\documentclass{aa}

\usepackage{graphicx}
\usepackage[space]{grffile}
\usepackage{latexsym}
\usepackage{amsfonts,amsmath,amssymb}
\usepackage{url}
\usepackage[utf8]{inputenc}
\usepackage{fancyref}
\usepackage{textcomp}
\usepackage{longtable}
\usepackage{multirow,booktabs}
\usepackage{subfigure}
\usepackage{natbib}
\usepackage{color}
\usepackage{gensymb}
\usepackage{enumitem}

\setlist[enumerate]{label*=\arabic*.}

\usepackage[backref,breaklinks,colorlinks,citecolor=blue]{hyperref}

\newcommand{\msol}{M$_\odot$}

\newcommand{\zsol}{Z$_\odot$}
\newcommand{\hubble}{{\it Hubble}}

\newcommand{\OVI}{[\hbox{{\rm O}\kern 0.1em{\sc vi}}]}
\newcommand{\Lalpha}{Ly$\alpha$}
\newcommand{\NV}{[\hbox{{\rm N}\kern 0.1em{\sc v}}]}
\newcommand{\CII}{[\hbox{{\rm C}\kern 0.1em{\sc ii}}]}
\newcommand{\SiIV}{[\hbox{{\rm Si}\kern 0.1em{\sc iv}}]}
\newcommand{\OIV}{[\hbox{{\rm O}\kern 0.1em{\sc iv}}]}
\newcommand{\NIV}{[\hbox{{\rm N}\kern 0.1em{\sc iv}}]}
\newcommand{\CIV}{\hbox{{\rm C}\kern 0.1em{\sc iv}}}
\newcommand{\HeII}{\hbox{{\rm He}\kern 0.1em{\sc ii}\kern 0.1em{$\lambda1640$}}}
\newcommand{\OIII}{\hbox{{\rm O}\kern 0.1em{\sc iii}}]}
\newcommand{\NIII}{[\hbox{{\rm N}\kern 0.1em{\sc iii}}]}
\newcommand{\AlIII}{\hbox{{\rm Al}\kern 0.1em{\sc iii}}}
\newcommand{\SiIII}{\hbox{{\rm Si}\kern 0.1em{\sc iii}}]}
\newcommand{\CIII}{\hbox{{\rm C}\kern 0.1em{\sc iii}]}}
\newcommand{\NeIV}{[\hbox{{\rm Ne}\kern 0.1em{\sc iv}}]}
\newcommand{\MgII}{\hbox{{\rm Mg}\kern 0.1em{\sc ii}}}

\newcommand{\HeIIoptical}{\hbox{{\rm He}\kern 0.1em{\sc ii}\kern 0.1em{$\lambda4686$} }}
\newcommand{\He}{\hbox{{\rm He}\kern 0.1em{\sc ii}}}
\newcommand{\CIIIL}{\hbox{{\rm C}\kern 0.1em{\sc iii}]\kern 0.1em{$\lambda1907,\lambda1909$}}}

\newcommand{\Hep}{He$^+$}
\newcommand{\Cpp}{C$^{++}$}
\newcommand{\Opp}{O$^{++}$}

\newcommand{\Hbeta}{H$\beta$}
\newcommand{\SII}{[\hbox{{\rm S}\kern 0.1em{\sc ii}}]}
\newcommand{\NII}{[\hbox{{\rm N}\kern 0.1em{\sc ii}}]}
\newcommand{\OII}{[\hbox{{\rm O}\kern 0.1em{\sc ii}}]}

\newcommand{\MgI}{\hbox{{\rm Mg}\kern 0.1em{\sc i}}}
\newcommand{\FeII}{\hbox{{\rm Fe}\kern 0.1em{\sc ii}}}
\newcommand{\HII}{{\ion{H}{ii}}}

\newcommand{\OI}{\hbox{{\rm O}\kern 0.1em{\sc i}}}
\newcommand{\NeII}{[\hbox{{\rm Ne}\kern 0.1em{\sc ii}}] }
\newcommand{\NaI}{[\hbox{{\rm Na}\kern 0.1em{\sc i}}] }
\newcommand{\NeIII}{[\hbox{{\rm Ne}\kern 0.1em{\sc iii}}] }

\newcommand{\Av}{$A(V)$}
\newcommand{\AvSED}{$A(V)_{SED}$}
\newcommand{\AvBeta}{$A(V)_\beta$}

\newcommand{\mosaic}{\emph{mosaic}}
\newcommand{\udf}{\emph{udf-10}}

\begin{document}

\bibpunct{(}{)}{;}{a}{}{,} 

\title{Exploring \HeII\ emission line properties at z$\sim2-4$}

\author{Themiya Nanayakkara\inst{1,*} \and 
Jarle Brinchmann\inst{1,2} \and 
Leindert Boogaard\inst{1} \and 
Rychard Bouwens\inst{1} \and 
Sebastiano Cantalupo\inst{3} \and 
Anna Feltre\inst{4,5} \and 
Wolfram Kollatschny\inst{6} \and 
Raffaella Anna Marino\inst{3} \and 
Michael Maseda\inst{1} \and 
Jorryt Matthee\inst{3} \and 
Mieke Paalvast\inst{1} \and 
Johan Richard\inst{4} \and 
Anne Verhamme\inst{7}}

\institute{Leiden Observatory, Leiden University, PO Box 9513, 2300
RA Leiden, The Netherlands.\thanks{themiyananayakkara@gmail.com}
\and Instituto de Astrofisica e Ciencias do Espaco, Universidade do Porto, CAUP, Rua das Estrelas, 4150-762 Porto, Portugal.
\and ETH Zurich, Department of Physics, HIT J31.5, Wolfgang-Pauli-Strasse 27 8093 Zurich, Switzerland.
\and Univ. Lyon, Univ. Lyon1, ENS de Lyon, CNRS, Centre de Recherche Astrophysique de Lyon (CRAL) UMR 5574, 69230 Saint-Genis-Laval, France.
\and Scuola Internazionale Superiore di Studi Avanzati (SISSA), Via Bonomea 265, I-34136, Trieste, Italy.
\and Institut f\"ur Astrophysik, Universit\"at G\"ottingen, Friedrich-Hund Platz 1, D-37077 G\"ottingen, Germany.
\and Observatoire de Gen\'eve, Universit\'e de Gen\'eve, 51 Ch. des Maillettes, 1290 Versoix, Switzerland.}

\date{5 November 2018 / 14 February 2019}

\abstract{
Deep optical spectroscopic surveys of galaxies provide us a unique opportunity to investigate  rest-frame ultra-violet (UV) emission line properties of galaxies at $z\sim2-4.5$. Here we combine VLT/MUSE Guaranteed Time Observations of the \emph{Hubble} Deep Field South, Ultra Deep Field, COSMOS, and several quasar fields with other publicly available data from VLT/VIMOS and VLT/FORS2 to construct a catalogue of \HeII\  emitters at $z\gtrsim2$. 
The deepest areas of our MUSE pointings reach a  $3\sigma$ line flux limit of $3.1\times10^{-19}$ erg s$^{-1}$ cm$^{-2}$.
After discarding broad line active galactic nuclei we find 13 \HeII\ detections from MUSE with a median M$_\mathrm{UV}=-20.1$ and 21 tentative \HeII\ detections from other public surveys.
Excluding \Lalpha, all except two galaxies in our sample show at least one other rest-UV emission line, with \CIIIL\ being the most prominent.
We use multi-wavelength data available in the \emph{Hubble} legacy fields to derive basic galaxy properties of our sample via spectral energy distribution fitting techniques. 
Taking advantage of the high quality spectra obtained by MUSE ($\sim10-30$h of exposure time per pointing), we use  photo-ionisation models to study the rest-UV emission line diagnostics of the \HeII\ emitters.
Line ratios of our sample can be reproduced by moderately sub-solar photo-ionisation models, however, we find that including effects of binary stars lead to degeneracies in most free parameters.   
Even after considering extra ionising photons produced by extreme sub-solar metallicity binary stellar models, photo-ionisation models are unable to reproduce rest-frame \HeII\ equivalent widths ($\sim0.2-10$ \AA), thus additional mechanisms are necessary in models to match the observed \HeII\ properties. 
}


\maketitle

\section{Introduction}
\label{sec:intro}

The transition of a chemically simple Universe to a complex and diverse structure was driven by the first generation of metal free stars (pop-III stars) which were formed within the first few million years of the Big Bang. 
In the current cosmological evolution framework, pop-III stars formed as individual stars or within the first (proto) galaxies produced high amounts of UV photons (UV ionising continuum) contributing to the re-ionization of the Universe and thereby, ending the cosmic ‘dark ages’ \citep{Tumlinson2000,Tumlinson2001, Barkana2001,Bromm2011,Wise2012,Wise2014}. 
Additionally, these stars generated the first supernovae in the Universe, which drove the cosmic chemical evolution process by synthesizing metals (elements heavier than He) and enriching the inter-galactic medium \citep[IGM; eg.,][]{Cooke2011}.

The existence of pop-III stars is yet to be observationally confirmed and numerous attempts are being made to explore the existence of such stars in the early Universe via current ground and space based telescopes. 
Narrow band \Lalpha\ surveys \citep{Hu2004,Tapken2006,Murayama2007,Ouchi2017} or Lyman Break techniques \citep{Steidel2003,Bouwens2010,McLure2011,Garcia-Vergara2017,Ono2017} observe galaxies at $z\sim2-8$ to make photometric pre-selections of high-$z$ galaxies.
These candidates are followed up spectroscopically to obtain multiple emission lines to explore stellar population and interstellar medium (ISM) conditions to confirm/refute the existence of pop-III stars \cite[e.g.,][]{Cassata2013,Sobral2015}. 
With large samples of high-$z$ galaxies, candidates for galaxies containing a significant population of Pop III stars can be selected due to the presence of strong \Lalpha\ and \He\ in the absence of other prominent emission lines. This can be interpreted as existence of pristine metal poor stellar populations \citep{Tumlinson2003,Raiter2010,Sobral2015}.

The absence of metals in primordial gas might result in higher stellar masses for pop-III stars \citep{Jeans1902,Bromm2004} leading to an extreme top heavy initial mass function \citep[IMF, e.g.,][]{Schaerer2002}. 
A stellar population with such an IMF will have relatively large numbers of very hot stars which produce \Hep\ ionizing photons with energies $> 54.4$ eV ($\lambda<228$ \AA). 
The resulting strong \He\ has been proposed as an indication of the presence of pop-III stars. 
This interpretation is however challenging in the face of other processes that can produce \Hep\ ionising photons. 
Additionally, the short life-time of $\sim1$Myr of pop-III systems and resulting ISM/IGM  pollution by pair-instability supernovae \citep{Heger2002}, uncertainties in photometric calibrations, presence of active galactic nuclei (AGN), pristine cold mode gas accretion to galaxies, limited understanding of high-redshift stellar populations and ISM contribute further to the complexity of detecting and identifying pop-III host systems \citep{Fardal2001,Yang2006,Sobral2015,Agarwal2016,Bowler2017,Matthee2017,Shibuya2017,Sobral2018b}.

In order to make compelling constraints of stellar populations in the presence of strong \He\ emission and link with pop-III hosts, a comprehensive understanding of the \He\ emission mechanisms is required.
The origin of \He\ emission, which is produced by cascading re-combination of He$^{++}$, has been explored extensively, however, the exact nature of physical mechanisms required to power the high ionization sources is still under debate \citep[e.g.,][]{Shirazi2012,Senchyna2017}. 
The shape of the \He\ profile has been attributed to different mechanisms that may contribute to the ionising photons.

Wolf-Rayet (W-R) stars are a long known source of HeII ionising photons in galaxies in the local Universe, which are hydrogen stripped massive evolved stars with high surface temperatures and high mass loss rates driven by strong and dense stellar winds \citep{Allen1976}.
Broad \He\ features are expected to originate in the thick winds of W-R stars  and are not recombination features. 
However W-R stars are also extremely hot and do produce photons with energies $>54$ eV, allowing some nebular \HeII\ emission. 
Therefore, in addition to nebular \HeII\ emission ($E_{photon}>54$ eV),  W-R stars and galaxies with W-R stars \citep[WR galaxies,][]{Osterbrock1982} show strong broad \He\ features ($E_{photon}>28$ eV) along with strong C or N emission lines with P-Cygni profiles \citep{Crowther2007}. 
Traditional stellar population models only produce nebular \He\ when there is an abundance of W-R stars \citep{Shirazi2012}, and therefore is limited to high metallicity stellar populations. 
At lower metallicities, the abundance of W-R stars decrease and observed \He\ profiles become narrower \citep[e.g.,][]{Senchyna2017}. 
Systems with strong nebular  \He\ emission in the absence of other W-R features require additional mechanisms that could produce high energy photons at lower metallicities.

The lack of W-R features in strong \He\ emitters in local low mass and metal poor galaxies have led to multiple theories that could power the \He\ emission and a new generation of stellar population and photo-ionization models attempt to quantify the effects of such mechanisms \citep[e.g.,][]{Gutkin2016,Eldridge2017}. 
Increase in stellar rotation, quasi homogeneous evolution (QHE), and production of stripped stars and  X-ray binaries  driven by binary interactions increase the surface temperatures of stars resulting in a higher  \Hep\ ionizing photon production efficiency. \citep{Garnett1991,Eldridge2008,Eldridge2012a,Miralles-Caballero2016,Stanway2016,Casares2017,Eldridge2017,Gotberg2017,Smith2017}.   
In addition to stars, fast radiative shocks and pre-shock and compressed post-shock regions of slower radiative shocks have been suggested as possible mechanisms to produce \Hep\ ionizing photons \citep{Allen2008,Izotov2012}, however, the abundance of such shocks as a function of metallicity is unclear. 
Post-Asymptotic Giant Branch (AGB) stars become a dominant mechanisms of ionizing radiation at low star-formation rates (SFRs), however, whether the observed \He\ emission can be attributed to such stars, specially at lower metallicities \citep{Shirazi2012,Senchyna2017} is questionable.

Ground and space based instruments have been used to observe rest-frame UV/optical features of local \citep[e.g.,][]{Kehrig2015,Senchyna2017} and high-redshift \citep[e.g.,][]{Cassata2013,Steidel2016,Berg2018} galaxies to examine possible origins for \He. 
In order to determine the origin of \He\ and link them to mechanisms that could be arisen from pop-III stellar systems, observations should be done in young, low-metallicity, highly star-forming systems which can give rise to a diverse range of exotic phenomena capable of producing high energy ionizing photons.
The Universe at $z\sim2-4$ was reaching the peak of the cosmic star-formation rate density \citep{Madau2014}, where the systems were highly star-forming and evolving rapidly giving rise to a diverse range of physical and chemical properties \citep[e.g.,][]{Steidel2014,Steidel2016,Kacprzak2015,Kacprzak2016,Sanders2015b,Sanders2015,Wirth2015,Kewley2016,Strom2017,Nanayakkara2017}.   
At $z\sim2-4$, the redshifted \HeII\ along with other prominent rest-UV features can be observed via optical spectroscopy.

In order to accurately identify systems that harbour pop-III stellar populations, observational signatures which can indicate differences in stellar and ISM metallicity independent of other physical conditions of galaxies in the early Universe are required. 
To constrain stellar population/ISM properties, high signal-to-noise (S/N) spectra ($\gtrsim20$) of galaxies with multiple emission/absorption lines in rest-frame UV/optical regions are required. 
Previous studies that investigated rest-UV properties of galaxies have been limited to either a single galaxy \citep{Erb2010,Vanzella2016,Patricio2017,Berg2018}, low-resolution observations of individual systems \citep{Cassata2013}, or to a single stacked spectrum of $\sim30-800$ galaxies at moderate resolution \citep{Shapley2003,Steidel2016,Nakajima2018,Rigby2018b}.

Surveys conduced using recently commissioned sensitive multiplex instruments in 8-10m class telescopes are instrumental to obtain samples of galaxy spectra ranging various physical and chemical compositions. 
Here, we use deep spectroscopic data obtained via the guaranteed time observations (GTO) of the Multi Unit Spectroscopic Explorer (MUSE) consortium to study properties of \HeII\ emitters at $z\sim2-4$ in individual and stacked galaxies. We complement our study by using deep photometric/spectroscopic data obtained by other public surveys.

The paper is arranged in the following way: In Section \ref{sec:sample_selection} we explain the sample selection, dust correction, and emission line fitting procedure of our sample used in this study. In Section \ref{sec:analysis} we perform spectrophotometric comparisons to our sample and in Section \ref{sec:model_comp}, we compare emission line ratios of our sample with photo-ionization models. We provide a brief discussion of the results of this study in Section \ref{sec:discussion} and outline our conclusions and future work in Section \ref{sec:conclusions}. 
Unless otherwise stated, we assume a \citet{Chabrier2003} IMF and a cosmology with  H$_{0}= 70$ km/s/Mpc, $\Omega_\Lambda=0.7$ and $\Omega_m= 0.3$. All magnitudes are expressed using the AB system \citep{Oke1983}.


\section{Sample Selection and characterization}
\label{sec:sample_selection}

In this section, we describe the \HeII\ sample selection procedure, dust corrections, and emission line fitting method used in this study.  
In general, we select all galaxies with redshift detections and visually inspect the spectra to determine the spectra for presence of sky lines and residual calibration issues and fit emission lines using a custom-built tool to obtain the systematic redshifts and line fluxes. 
We first briefly describe all deep MUSE GTO surveys explored and present a summary in Table \ref{tab:summary_table_muse}.

\subsection{\HeII\ detections}
\label{sec:sample_selection_muse}

MUSE  \citep{Bacon2010} is a second generation panoramic integral field spectrograph on the Very Large Telescope (VLT) operational since 2014. The instrument covers a field of view (FoV) of $1'\times1'$ with a $0.2''$ sampling in medium spectral resolution of $R\sim3000$.

MUSE \HeII\ detections are selected from three legacy fields, the Ultra Deep Field \citep[UDF,][]{Beckwith2006,Bacon2017}, the Hubble Deep Field South \citep[HUDF,][]{Williams1996,Bacon2015}, and the Cosmic Evolution Survey \citep[COSMOS,][]{Scoville2007} field along with the MUSE Extended quasar catalogue fields \citep{Marino2018}, all obtained as a part of the guaranteed time observations (GTO) awarded to the MUSE consortium. 
The MUSE spectra in our sample covers a nominal wavelength range of $\sim$4800--9300 \AA, implying that \HeII\ can be detected between $z \sim1.93-4.67$. 
Next we describe the sample selection from these fields.

\subsubsection{MUSE Ultra Deep Field}
\label{sec:sample_selection_muse_udf}

The current MUSE UDF coverage includes two distinct observing depths observed in good seeing conditions with a full width at half maximum (FWHM) of $\sim 0.6''$ at 7750\AA. 
The $3\times3$ arcmin$^2$ medium deep field (henceforth referred to as the \mosaic) has a depth of $\sim10$ hours obtained with a position angle (PA) of $-42\degree$. 
A further $1'\times1'$ region with a PA of $0\degree$  was selected within the \mosaic\ to be exposed for an additional $\sim21$ hours. 
The final deep region, henceforth referred to as \udf, comprise $\sim31$ hours of exposure time.    

The MUSE UDF catalogue used for this work includes 1574 galaxies with spectroscopic detections \citep{Inami2017}.
We select galaxies with a secure spectroscopic redshifts ({\tt CONFID$>$1}) between $z \sim1.93-4.67$. 
With these selection cuts we are left with 553 galaxies in the UDF out of which, 26 are flagged as merged\footnote{See \citet{Inami2017} for details} ({\tt MERGED=1}).
We visually inspect all 553 spectra and select high quality spectra to fit for \HeII\ features using our custom built line fitting tool (see Section \ref{sec:emission_line_measurements}).

Within UDF we identify nine unique galaxies with \HeII\ emission. 
Two galaxies are classified as AGN \citep[MUSE UDF AGN are flagged from the][Chandra Deep Field South catalogue]{Luo2017} and show strong broad \HeII\ features and four of the remaining galaxies show \CIII\ in emission. 
One galaxy shows a broad \HeII\ feature with a \HeII\ full-width at the half-maximum (FWHM) of 1068 km/s. We remove this galaxy from our sample because we are primarily interested in the narrow \HeII\ component of galaxies and it is a clear outlier in terms of \HeII\ FWHM compared to the rest of the sample (see Section \ref{sec:uncerternities_snr}).

\subsubsection{MUSE Hubble Deep Field South}
\label{sec:sample_selection_muse_hdfs}

MUSE HDFS observations were obtained in 2014 during the commissioning of MUSE and all data products are publicly available \citep{Bacon2015}. 
However, we use an updated version of the MUSE data reduction pipeline \citep[CubExtractor package;][Cantalupo et al., in prep]{Borisova2016} with improved flat fielding and sky subtraction to generate a modified version of the MUSE data cube for our analysis. 

The MUSE HDFS catalogue contains 139 secure spectroscopic redshifts ({\tt CONFID$>$1}) and 48 galaxies that fall within the spectral range for \HeII\ detection with MUSE were investigated. 
Using a similar procedure to UDF, we identify three galaxies with \HeII\ emission in the HDFS. Two galaxies have \CIII\ spectral coverage and show prominent \CIII\ emission. One galaxy shows \CIV\ absorption, while another shows indications for \CIV\ emission features.

\subsubsection{MUSE Groups catalogue}
\label{sec:sample_selection_muse_groups}

The MUSE groups GTO program targets galaxy groups (PI: T. Contini)  identified by the zCOSMOS survey \citep{Knobel2012}. So far 11 galaxy groups have been observed by the MUSE consortium with varying depths \citep[][]{Epinat2018}. 
For our analysis we selected five fields with exposure times greater than 2 hours, namely COSMOS-GR 114 (2.2 hours), VVDS-GR 189 (2.25 hours),  COSMOS-GR 34 (5.25 hours), COSMOS-GR 84 (5.25 hours), and COSMOS-GR 30 (9.75 hours).  
The seeing conditions of the fields vary between $0.5''-0.7''$.

Without imposing a redshift quality cut, we selected galaxies that lie within the spectral range for \HeII\ detection with MUSE. A total of 104 galaxy spectra were investigated to select one galaxy with \HeII\ signatures from our fitting tool. 
The galaxy is selected from COSMOS-GR 30, the deepest pointing of the MUSE COSMOS group catalogue and shows \CIII\ and \CIV\ in emission.

\subsubsection{MUSE Quasar fields}
\label{sec:sample_selection_muse_qso}

The MUSE extended quasar catalogue maps the cool gas distribution in the $z\sim3$ Universe by observing \Lalpha\ emission in the neighborhood of high redshift quasars at $z>3$ \citep{Marino2018}. 
In total, the catalogue contains 22 fields with varying exposure times from one hour to 20 hours. 

For our analysis, we select the three deepest quasar fields, J2321 (9.0 hours), UM287 (9.0 hours), and Q0422 (20.0 hours). We select all 49 galaxies with secure spectroscopic redshifts ({\tt CONFID$>$1}) within the spectral range for \HeII\ detection with MUSE. 
Visual inspection of the selected galaxies showed no \HeII\ features in the UM287 and J2321 fields. 
In Q0422 our fitting tool identified three galaxies with  \HeII\ features and one possible AGN with strong and broad \HeII, \CIII,\CIV, and \Lalpha. We further analyze the \CIII, \CIV, and \HeII\ line ratios following \citet{Feltre2016} diagnostics and find that line ratios are more likely to be powered via an AGN.

\begin{table}
\caption{Summary of MUSE GTO surveys explored. [Section \ref{sec:sample_selection_muse}]}
\label{tab:summary_table_muse}
\centering
\begin{tabular}{l l r r r  }
\hline\hline
Field name & FOV\tablefootmark{a} & Exposure  & Nc\tablefootmark{b}  & Ns\tablefootmark{c}  \\
		   & 					  & Time (h)  & 					 & 						 \\
\hline\hline
UDF10					& $1\arcmin\times1\arcmin$    & 31.00     &  122 &  0  \\
UDF MOSAIC				& $3\arcmin\times3\arcmin$    & 10.00     &  431 &  6  \\
HDFS					& $1\arcmin\times1\arcmin$    & 27.00     &  139 &  3  \\
Groups COSMOS  30		& $1\arcmin\times1\arcmin$    & 9.75      &   35 &  1  \\
Groups COSMOS  34		& $1\arcmin\times1\arcmin$    & 5.25      &    7 &  0  \\
Groups COSMOS  84		& $1\arcmin\times1\arcmin$    & 5.25      &   13 &  0  \\
Groups COSMOS 114		& $1\arcmin\times1\arcmin$    & 2.20      &    2 &  0  \\
Groups VVDS   189	    & $1\arcmin\times1\arcmin$    & 2.25      &    1 &  0  \\
Quasar J2321			& $1\arcmin\times1\arcmin$    & 9.00      &   13 &  0  \\
Quasar Q0422			& $1\arcmin\times1\arcmin$    & 20.00     &   25 &  3  \\
Quasar UM287			& $1\arcmin\times1\arcmin$    & 9.00      &   11 &  0  \\
\hline
\end{tabular}
\raggedright
\tablefoottext{a}{Field of view.\\}
\tablefoottext{b}{Number of galaxies selected to visually investigate for \HeII\ emission.\\}
\tablefoottext{c}{Number of galaxies selected to be analyzed in this study.}
\end{table}

\subsubsection{Other surveys explored}
\label{sec:other_surveys}

In addition to the MUSE GTO surveys, we also examine data from other public optical spectroscopic surveys to identify \HeII\ line emitters.
The GOODS FORS2 \citep{Vanzella2008}, GOODS VIMOS \citep{Balestra2010}, K20 \citep{Cimatti2002}, VANDELS \citep{Grogin2011}, VIPERS \citep{Garilli2014}, VUDS \citep{LeFevre2015}, VVDS \citep{LeFevre2013}, and zCOSMOS bright \citep{Lilly2007} surveys are utilized for this purpose. 
All publicly available data have a lower spectral resolution compared to MUSE and thus the spectra may suffer from blending between narrow and broad \HeII\ components. 
However, such surveys provide a wealth of spectra to investigate the \HeII\ and other UV nebular line properties of galaxies and are also suitable to be followed up with higher resolution spectrographs. 
In Appendix \ref{appendix:other_surveys} we provide a brief description of the examined surveys and provide a summary of the \HeII\ detections in Table \ref{tab:summary_table_others}.

\subsection{Dust corrections}
\label{sec:dust_corr}

Dust corrections are crucial to obtain accurate estimates of rest-UV emission line features of galaxies. 
The total-to-selective extinction ($k(\lambda)$) of a galaxy depends crucially on the physical nature of the dust grains and is a strong function  of wavelength.
Galactic and extra-galactic studies show a non-linear systematic increase in $k(\lambda)$ with decreasing wavelength \citep[e.g.,][]{Cardelli1989,Calzetti1994,Reddy2015}. 
UV dust extinction, is further complicated by the presence of a high UV absorption region at 2175 \AA\ \citep[UV absorption ``bump''][]{Mathis1990,Buat2011}, however, its origin is not yet well understood \citep[e.g.,][]{Calzetti2001,Zagury2017,Narayanan2018}. For our analysis, we use the dust obscuration law parametrized by \citet{Calzetti2000}, which is defined redwards of 1200 \AA. 
In Section \ref{sec:uncerternities_dust} we analyze how different dust laws affect our analysis.

The  MUSE UDF, HDFS, and COSMOS groups fields contain multi-wavelength photometric coverage from \hubble\  legacy fields, which we use to match with the MUSE observations \citep[e.g.,][]{Inami2017}. 
We use FAST \citep{Kriek2009} to match  synthetic stellar populations from \citet{Bruzual2003} models to the observed photometry using a $\chi^2$  fitting algorithm to derive best-fit stellar masses, ages, star formation timescales, and dust contents of galaxies.
FAST does not include models of the nebular emission from the photoionized-gas in addition to the continuum emission from stars. Even though the emission line contamination for stellar mass estimates have shown to be negligible for $z\sim1-3$ star-forming galaxies \citep{Pacifici2015}, the mass of galaxies in the presence of strong [\OIII$\lambda5007$ EW, which are likely in strong \HeII\ emitters, may be over-estimated.

Photometry used for SED fitting does not contain data redward of the \hubble\ F160W filter, thus lacks near-infra red coverage to better constrain degeneracies between derived parameters \citep{Conroy2013}.
The MUSE quasar catalogues do not contain HST photometry, and therefore we use the rest-UV continuum slope ($\beta$) parameterized by a power law of the form:
\begin{equation}
\label{eq:beta_def}
f_\lambda \propto \lambda^\beta
\end{equation}
where $f_\lambda$ is the observed flux at rest-frame wavelength $\lambda$, to obtain an estimation of the total dust extinction.

From \citet{Meurer1999}, we relate the UV slope $\beta$ to the total extinction magnitude at 1600 \AA\ ($A(1600)$) as:
\begin{equation}
\label{eq:A1600_to_beta}
A(1600) = 4.43 + 1.99\beta.
\end{equation} 
\citet{Meurer1999} demonstrated that the relationship in Equation \ref{eq:A1600_to_beta} is consistent with ionizing stellar population model expectations in dust free scenarios and with the \citet{Calzetti1994} extinction law within `reasonable' scatter \citep[also see][]{Reddy2018}. 
Therefore, we expect our spectroscopically derived \Av\ from $\beta$ (\AvBeta) to be consistent with photometrically derived FAST $A(V)$ (\AvSED) values within statistical uncertainty.

To validate our assumption, we use MUSE UDF data to investigate the relationship between \AvBeta\ and \AvSED. 
We use all galaxies in the MUSE UDF catalogue with a {\tt CONFID=3} and $z=2.2-4.7$ corresponding to galaxies with spectroscopic coverage between rest-frame $1500-1700$\AA. 
Using these criteria a total of 59 galaxies are selected from the UDF catalogue out of which we remove 23 galaxies that have weak continuum detections measured from the MUSE spectra (S/N$\lesssim1-2$) and 2 galaxies that have no stellar mass estimates from FAST. 
We divide the remaining 34 galaxies depending on their S/N level of the continuum into two bins by selecting galaxies with high (S/N$\gtrsim3$) and low (S/N$<3$) S/N.

For galaxies in these two bins, we mask out regions with rest-UV features as defined by Table 2 in \citet{Calzetti1994} and compute the inverse-variance weighted rest-UV power-law  spectral slope between the wavelength range of $1300-1900$\AA\ ($1600\pm300$\AA) using the power law function in the {\tt python LMFIT\footnote{http://lmfit.github.io/lmfit-py/}} module. 
We then convert $\beta$ to $A(1600)$ using Equation \ref{eq:A1600_to_beta}, and then use the \citet{Calzetti2000} dust attenuation law to compute the \AvBeta\ as follows:
\begin{equation}
A(V)_\beta = A(1600) * R_v / k(1600)
\end{equation}
where $R_v(=4.05)$ is the total attenuation and $k(1600)(=9.97)$ is the star-burst reddening curve at 1600 \AA.

\begin{figure}
\includegraphics[trim = 10 10 10 0, clip, scale=0.60]{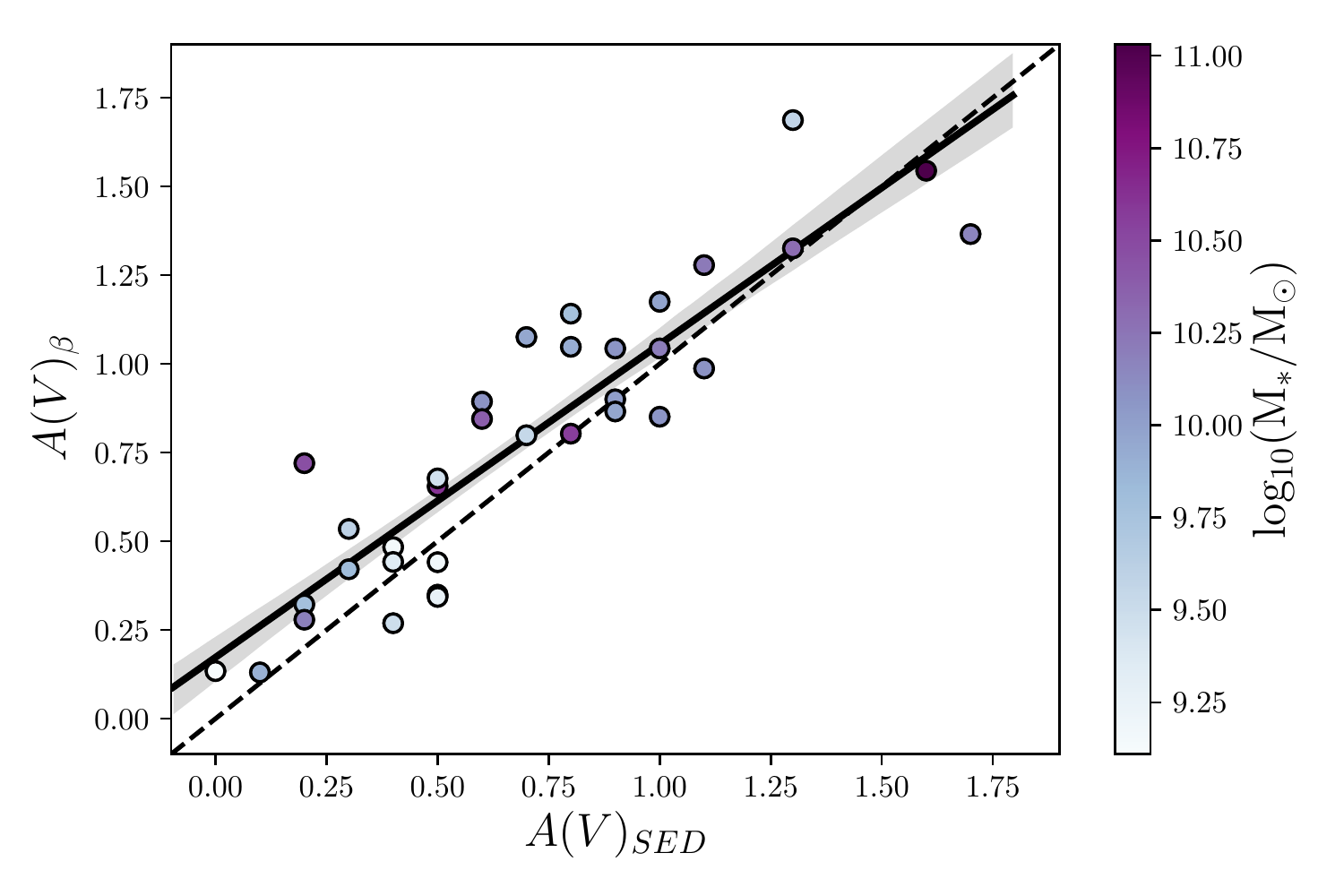}
\caption{Comparison between \AvBeta\ and \AvSED. Galaxies from MUSE UDF with {\tt CONFID=3} and $z=2.2-4.7$ with continuum S/N$\gtrsim3$ are shown here and are colour coded by their mass. The one to one line is shown as a black dashed line. The solid black line shows the linear regression model fit to the data along with its $1\sigma$ uncertainty (computed using 1000 bootstrap resamples) shaded in gray. [Section \ref{sec:dust_corr}]  
\label{fig:AvSED_AvBeta_comp}
}
\end{figure}

Figure \ref{fig:AvSED_AvBeta_comp} shows the relationship between \AvBeta\ and \AvSED\ for the MUSE UDF galaxies. \AvBeta\ shows good agreement with SED derived extinction values.  
In general lower stellar mass systems show low amounts of dust extinction. 
We conclude that the UV continuum slopes provide a reasonable estimate of the dust corrections required for galaxies (in comparison to estimates from SED fitting using FAST), which we use to calculate the dust extinction of galaxies in the MUSE quasar fields. All \AvBeta$<0$ is assigned an \AvBeta=0. 
\AvSED\ values are used to correct for dust extinction in all other galaxies. 
Dust corrections for the observed spectra are performed as follows:
\begin{equation}
\label{eq:dust_correction}
f(\lambda)_{int} = f(\lambda)_{obs} 10^{0.4   A(V)  k(\lambda)/R_v}	
\end{equation}
where $f(\lambda)_{int}$ and $f(\lambda)_{obs}$ are the intrinsic and observed flux at wavelength $\lambda$, \Av\ is the attenuation by dust, $k(\lambda)$ is the star-burst reddening curve from \citet{Calzetti2000}. 
$50\%$ of our galaxies have \Av=0 and for the rest we assume that the UV continuum suffers the same attenuation as the emission lines. Since the UV continuum in these actively star forming galaxies is to a large extent originating from the stars responsible for the emission lines, this seems reasonable but we will return to discuss this assumption in Section \ref{sec:uncerternities_dust}.

\subsection{Emission line measurements}
\label{sec:emission_line_measurements}

The line flux measurements of the MUSE surveys are performed using {\tt PLATEFIT} \citep{Tremonti2004,Brinchmann2008}, which uses model galaxy templates from \citet{Bruzual2003} to fit the continuum of the observed spectra at a predefined redshift and compute the line fluxes of each expected emission line using a single Gaussian fit. 
The redshift of the galaxies were determined as described by \citet{Inami2017}. 
We find  the \HeII\ profiles of our sample to be in general broader compared to the other observed rest-UV emission lines such as \CIII, which could be driven by multiple mechanisms that power \HeII\ compared to other nebular lines explored in this analysis (see Section \ref{sec:origion_of_HeII}). 
To accurately quantify the \HeII\ flux of our observed spectra we use a custom built fitting tool to perform the line fits and obtain the emission line fluxes and equivalent widths allowing greater flexibility ($\pm1.25$ \AA) in line centre and line widths.

Emission lines are fit allowing the line centre and line-width to vary as free parameters. 
Except for \HeII\ all other lines are fit such that line centre and line-width are fixed to a common best-fit value using {\tt python LMFIT} routine, however, given \HeII\ profiles are broader, we allow greater flexibility in the fitting parameters for \HeII. 
In Section \ref{sec:uncerternities_snr}, we further discuss and quantify the effects of allowing greater freedom for fitting parameters for \HeII\ compared to other lines.

The procedure we use to fit the lines and compute the equivalent-width (EW) is as follows:
\begin{enumerate}
\item We first manually inspect all spectra to identify galaxies with significant offsets between the \Lalpha\ redshift and the systemic redshift obtained via \CIII\ and \HeII\ emission lines. We modify the redshift of these galaxies to match the systemic redshift.
\item  We exclude $\pm\ 20\times$ muse wavelength sampling ($\sim20 \times 1.25$ \AA) around the rest-UV emission line regions of the spectra. 
\item  We define a continuum by calculating a running median within a window of 300 pixels, excluding  masked regions.
\item The emission line fluxes are calculated as follows:
	\begin{enumerate}
		\item Gaussian fits are performed on the continuum subtracted spectra.  
		\item The flux of each emission line is computed by integrating the best-fit Gaussian within $5\sigma$ of the determined line centre and line-width.
		\item Line flux errors are computed by integrating the error spectrum within the same $5\sigma\ \Delta \lambda$ Gaussian fit performed on the emission line. 
	\end{enumerate}
\item The EW is calculated similarly using the same Gaussian parameters and the continuum level. For each emission line, if the continuum is lower than the $1\sigma$ error spectrum, the $1\sigma$ error level is considered as the continuum to derive a lower limit to the EW. The error in the measured EW is computed by bootstrap resampling the spectrum, where each pixel is resampled using a random number parametrized by a gaussian function with mean at the flux value of the pixel in the observed spectrum and standard deviation by the corresponding value from the error spectrum.   
\end{enumerate}

Measured properties of the observed emission lines are presented in Tables \ref{tab:line_fluxes_muse_1}, \ref{tab:line_fluxes_muse_2}, and \ref{tab:line_ews_muse}.
We divide the MUSE sample in two categories, depending on whether the galaxy shows broad AGN like features, and we exclude them from our analysis. 
Since the \Hep\ ionization potential is higher (54.4 eV) than the C$^{+}$ ionization potential (24.38 eV), it is plausible for galaxies to only show \CIII\ nebular emission.  
However, C$^{++}$ ionization potential is 47.89 eV and resulting \CIV\ emission suffers from strong stellar wind absorption, thus only a handful of galaxy spectra show strong \CIV\ nebular emission in the absence of AGN activity.
All except two galaxies show \CIII\ in emission. One of the galaxies with no \CIII\ in emission shows a prominent \CIV\ emission feature, which suggests hard ionizing fields. 
This implies a higher electron temperature and, therefore, more prominent higher energy collisionally excited lines than in sources with less hard radiation field. 
We note that in the deepest MUSE pointings (UDF10 and HDFS), out of 17 \CIII\ emitters presented by \citet{Maseda2017}, only one galaxy (HDFS 87) is found to have a confident \HeII\ detection.

\begin{table*}
\caption{Summary of the MUSE \HeII\ sample. [Section \ref{sec:emission_line_measurements}]
\label{tab:line_fluxes_muse_1}}
\centering
\begin{tabular}{l l l l l l l r r r r r    }
\hline\hline
ID & RA  & Dec    &  Field  &  $z$    & Av     & M$_{\mathrm{UV}}$   & $\Delta$M$_{\mathrm{UV}}$  & \multicolumn{3}{c}{\HeII} 			 \\
   & 	 & 		  & 		& 		  & 	   & 					 &                            & Flux & Error &FWHM 					  \\
\hline\hline
1024 & 03: 32: 31 & $-$27:  47: 25 & UDF  	 & 2.87 & 0.7 & $-$21.08 & 0.02 & 177 & 52 & 5 \\
1036 & 03: 32: 43 & $-$27:  47: 11 & UDF  	 & 2.69 & 0.5 & $-$20.75 & 0.02 & 142 & 53 & 4 \\
1045 & 03: 32: 33 & $-$27:  48: 14 & UDF  	 & 2.61 & 0.4 & $-$20.57 & 0.03 & 156 & 58 & 4 \\
1079 & 03: 32: 37 & $-$27:  47: 56 & UDF 	 & 2.68 & 0.7 & $-$20.35 & 0.04 & 290 & 91 &11 \\
1273 & 03: 32: 35 & $-$27:  46: 17 & UDF  	 & 2.17 & 0.0 & $-$19.35 & 0.06 & 217 & 79 & 5 \\
3621 & 03: 32: 39 & $-$27:  48: 54 & UDF  	 & 3.07 & 0.0 & $-$19.34 & --   & 213 & 45 & 6 \\
87   & 22: 32: 55 & $-$60:  33: 42 & HDFS  	 & 2.67 & 0.0 & $-$19.29 & 0.02 &  59 & 12 & 4 \\
109  & 22: 32: 56 & $-$60:  34: 12 & HDFS  	 & 2.2  & 0.5 & $-$18.89 & 0.02 &  54 & 13 & 3 \\
144  & 22: 32: 59 & $-$60:  34: 00 & HDFS  	 & 4.02 & 0.0 & $-$19.62 & 0.04 &  48 & 12 & 4 \\
97   & 10: 00: 34 & $+$02:  03: 58 & cgr30 	 & 2.11 & 0.5 & $-$18.82 & 0.10 & 306 & 55 & 5 \\
39   & 04: 22: 01 & $-$38:  37: 04 & q0421 	 & 3.96 & 0.0 & $-$19.67 & 0.06 & 153 & 33 & 7 \\
84   & 04: 22: 01 & $-$38:  37: 21 & q0421 	 & 3.1  & 0.0 & $-$18.90 & --   & 161 & 29 & 4 \\
161  & 04: 22: 02 & $-$38:  37: 20 & q0421 	 & 3.1  & 0.5 & $-$18.85 & --   & 318 & 39 & 9 \\
\hline
{\bf AGN} & & & & & & & & & &   \\
1051 & 03: 32: 43 & $-$27: 47: 03 & UDF      & 3.19  & -- & -- & -- & -- & -- & -- \\
1056 & 03: 32: 40 & $-$27: 48: 51 & UDF      & 3.07  & -- & -- & -- & -- & -- & -- \\
78   & 04: 22: 02 & $-$38: 37: 18 & q0421    & 3.10  & -- & -- & -- & -- & -- & -- \\
\hline
\end{tabular}
\tablefoot{All line fluxes are in $\mathrm{1\times10^{-20} erg/s/cm^{2}}$ and FWHM in \AA. Line fluxes of non-detected lines are shown as $3\sigma$ upper limits with their corresponding error shown as --. 
If the UV magnitude is not detected above the $1\sigma$ noise (see Section \ref{sec:observed_sample}), the corresposing error spectrum is used to compute an upper limit to the magnitude and the error in magnitude is given as --.
$^{\bf a}$ We note UDF 3621 is at a separation of $\sim3.''$ (28 kpc at $z~3$) and $dz=0.004$ of X-ray confirmed AGN, for which we do detect \Lalpha, \CIV, and \HeII. It is also $<1''$ and $dz=0.001$ away from another \Lalpha\ emitter for which no \HeII\ is detected. It is therefore plausible that the \HeII\ ionisation could be due to the AGN \citep[e.g.,][]{Cantalupo2019}, and not from internal sources in the galaxy.} 
\end{table*}


\section{MUSE \HeII\ sample analysis}
\label{sec:analysis}

\begin{table*}
\caption{Continuation of Table \ref{tab:line_fluxes_muse_1}. [Section \ref{sec:emission_line_measurements}]
\label{tab:line_fluxes_muse_2}}
\centering
\begin{tabular}{ l r r r r r r r r r r r r r     }
\hline\hline
		   {ID}                  &
           \multicolumn{3}{c}{\CIII1907}       &
           \multicolumn{2}{c}{\CIII1909}       &
           \multicolumn{2}{c}{\OIII1661}       &
           \multicolumn{2}{c}{\OIII1666}       &
           \multicolumn{2}{c}{\SiIII1883}      &
           \multicolumn{2}{c}{\SiIII1892}      \\
           {}                          & 
           {Flux}                      &
           {Error} 					   &
           {FWHM}       		       &
           {Flux}                      &
           {Error} 					   &
           {Flux}                      &
           {Error} 					   &
           {Flux}                      &
           {Error} 					   &
           {Flux}                      &
           {Error} 					   &
           {Flux}                      &
           {Error} 					   \\
\hline\hline
1024  & 308  & 49 & 6   & 206   & 42  & 151  & --  & 161        & 48   & 238    & --	& 317  & -- \\
1036  & 436  & 36 & 4   & 300   & 41  & 155  & --  & 230        & 54   & 167    & 55 	& 184  & -- \\
1045  & 388  & 49 & 4   & 211   & 54  & 186  & --  & 200        & 56   & 141    & 38 	& 129  & 44 \\
1079  & 111  & -- & 4   & 111   & --  & 162  & --  &  81        & --   &  64    & --	& 118  & -- \\
1273  & 402  & 48 & 3   & 271   & 47  & 165  & --  & 195        & 56   & 141    & 55 	& 153  & -- \\
3621  & 252  & -- & 4   & 212   & --  & 106  & --  & 105        & --   & 122  	& --	& 122  & -- \\
  87  &  78  & 11 & 3   &  33   & 11  &  33  & --  &  49   		& 11   &  32  	& 10 	&  47  & 13 \\
 109  &  71  & 12 & 3   &  63   & 12  &  38  & --  &  72   		& 13   &  57    & 12 	&  36  & -- \\
 144  &  --  & -- & --  & --    & --  &  37  & 11  & 150   		& 21   & --  	& --	&  --  & -- \\
  97  & 369  & 50 & 3   & 258   & 62  & 148  & 43  & 284   		& 44   & 131  	& --	& 134  & -- \\
  39  &  --  & -- & --  & --    & --  &  71  & --  &  66   		& --   & --  	& --	&  --  & -- \\
  84  & 174  & 47 & 3   &  72   & 23  &  66  & --  &  54   		& --   & 393  	& 59 	& 156  & 40 \\
 161  & 143  & -- & 3   &  62   & --  &  64  & --  &  58   		& 19   & 181  	& --	& 130  & -- \\
\hline
{\bf AGN} & & & & & & &  \\
1051 & -- & -- & -- & -- & -- & --	& --& --& --& --& -- & -- & -- \\
1056 & -- & -- & -- & -- & -- & --	& --& --& --& --& -- & -- & -- \\
78   & -- & -- & -- & -- & -- & --	& --& --& --& --& -- & -- & -- \\
\hline
\end{tabular}
\tablefoot{All line fluxes are in $\mathrm{1\times10^{-20} erg/s/cm^{2}}$ and FWHM in \AA. Line fluxes of non-detected lines are shown as $3\sigma$ upper limits with their corresponding error shown as --.} 
\end{table*}

\subsection{The observed sample}
\label{sec:observed_sample}

In total we have obtained 13 high quality \HeII\ emission line detections from the MUSE GTO surveys. 
In addition we have three galaxies which either show broad \CIV\ and/or \CIII\ emission or are flagged as AGN \citep{Inami2017}, which we have removed from our sample.
The spectra of our full \HeII\ sample are shown by Figures \ref{fig:HeII_spectra_1} and \ref{fig:HeII_spectra_2}. As is evident, our sample spans a large variety in spectral shape and emission line profiles. 
We define S/N $>2.5$ as a line flux detection, and three galaxies in our sample fall between S/N of $2.5-3.0$.
We additionally perform a false detection test for these three galaxies by forcing our line fitting algorithm to fit a line iteratively at random blueward of \HeII\ between 1580\AA -- 1620\AA. 100 such iterations show no false detections. 

\begin{figure*}
\includegraphics[trim = 28 50 10 20, clip, scale=1.0]{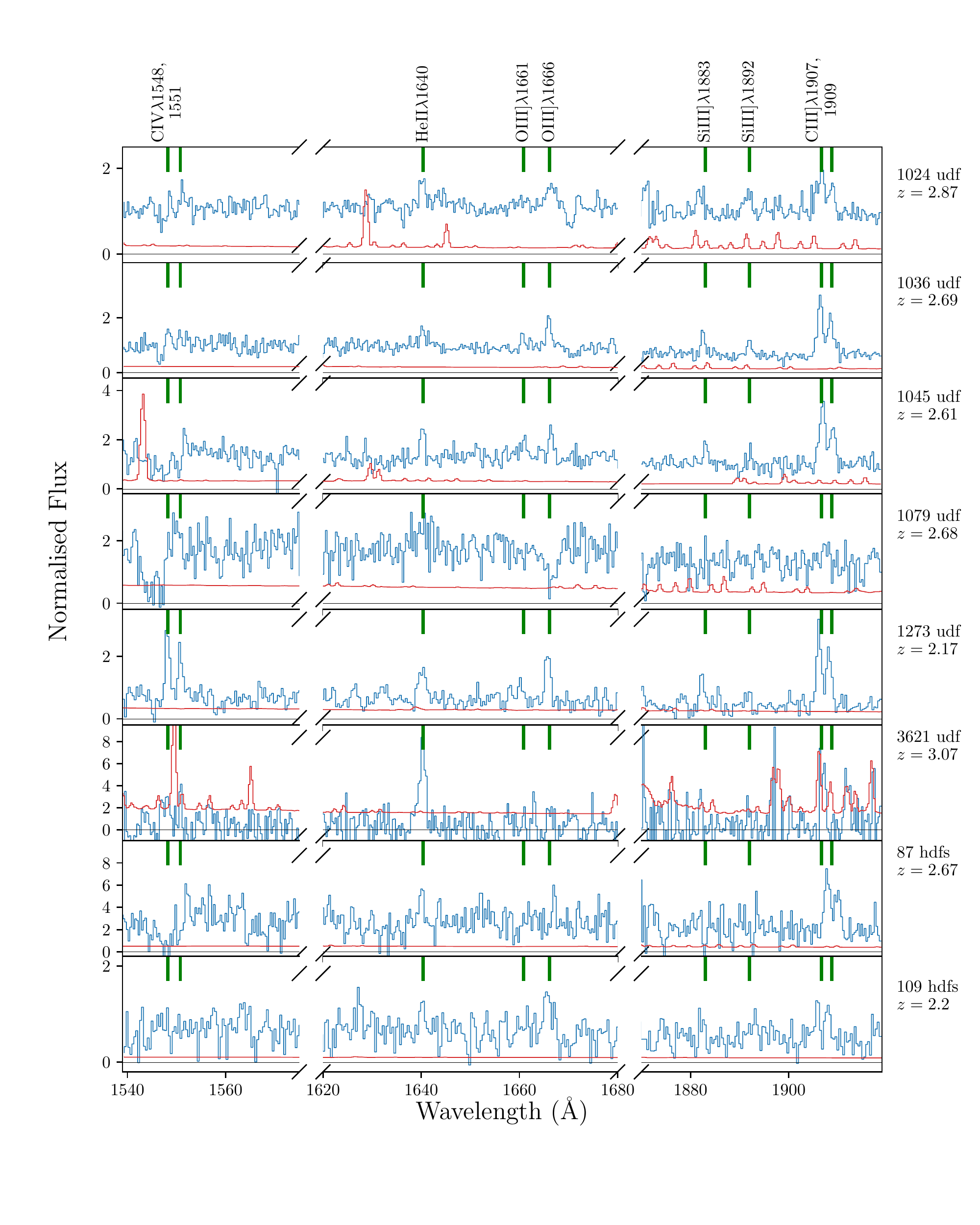}
\caption{Spectra (blue) of the MUSE \HeII\ detections with their respective noise spectrum (red). All spectra are shown at their rest-frame wavelength and are normalized at $\sim1600$\AA. The ID, field, and the spectroscopic redshift of each target is shown in the panels. The green vertical lines indicate selected rest-UV emission/absorption features. [Section \ref{sec:observed_sample}]
\label{fig:HeII_spectra_1}
}
\end{figure*}

\begin{figure*}[h!]
\includegraphics[trim = 10 50 0 20, clip, scale=1.0]{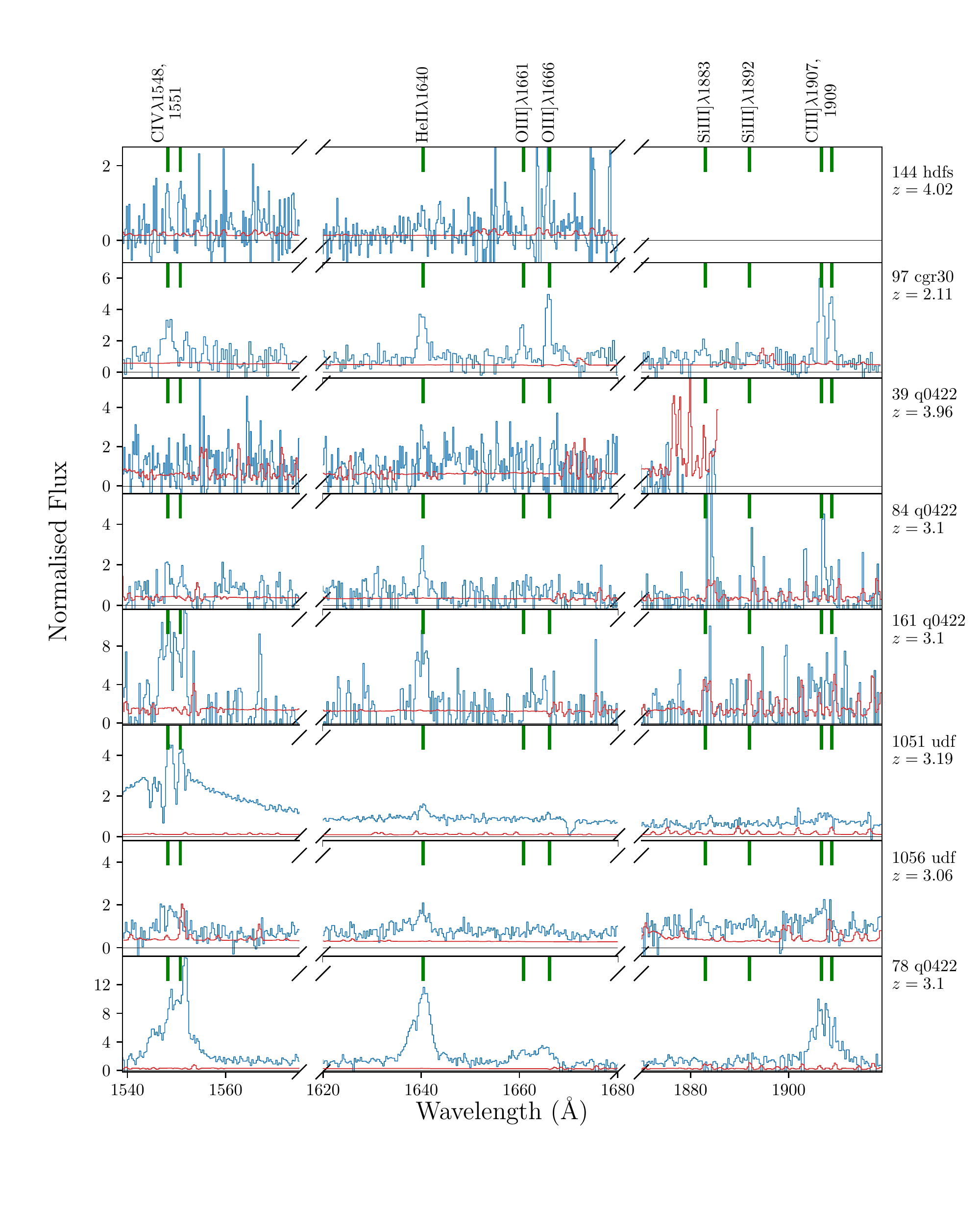}
\caption{Continuation of Figure \ref{fig:HeII_spectra_1}. The last three panels show the spectra dominated by AGN activity. [Section \ref{sec:observed_sample}]
\label{fig:HeII_spectra_2}
}
\end{figure*}

\begin{figure}
\includegraphics[scale=0.75]{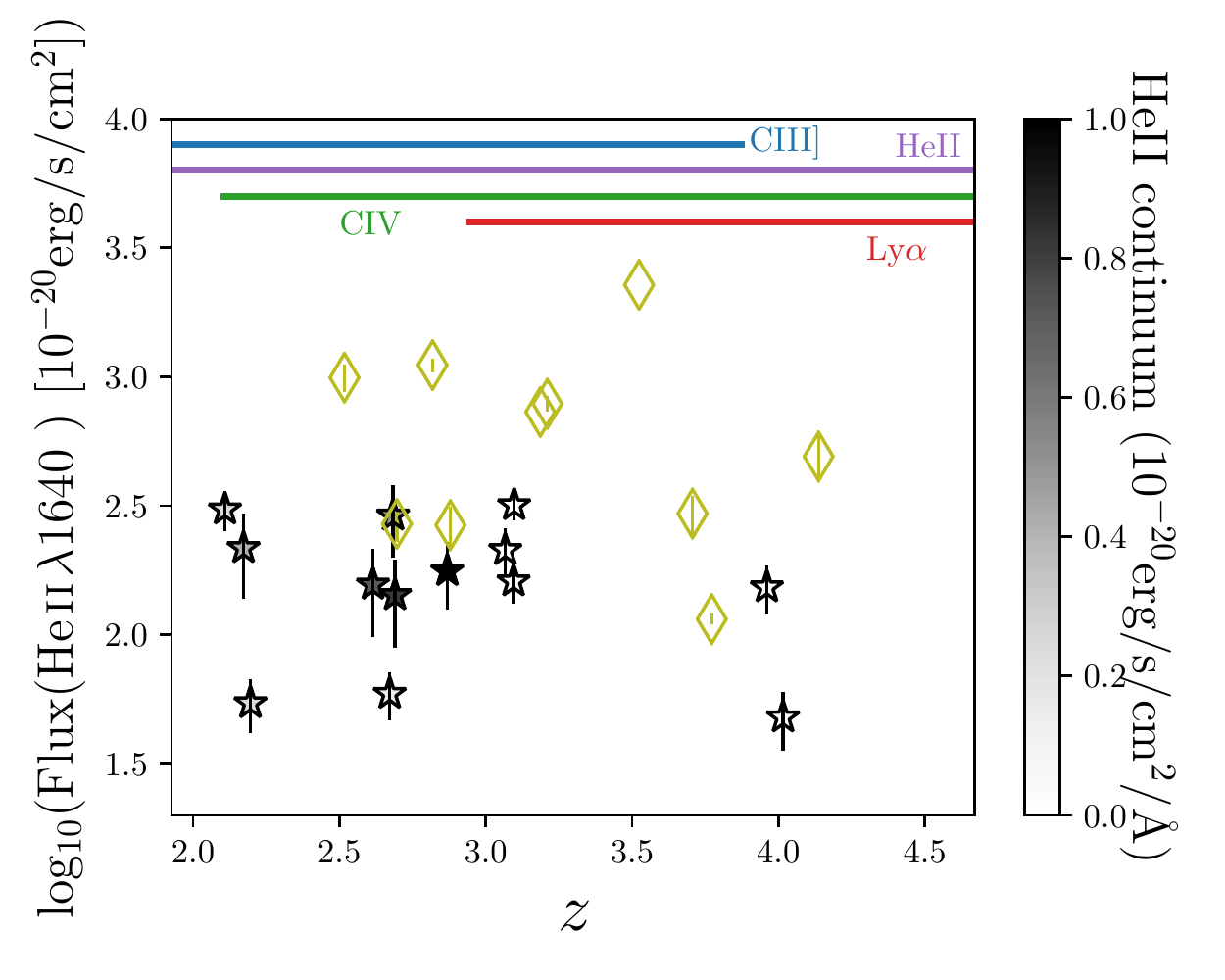}
\includegraphics[trim = 8 0 0 0, clip, scale=0.80]{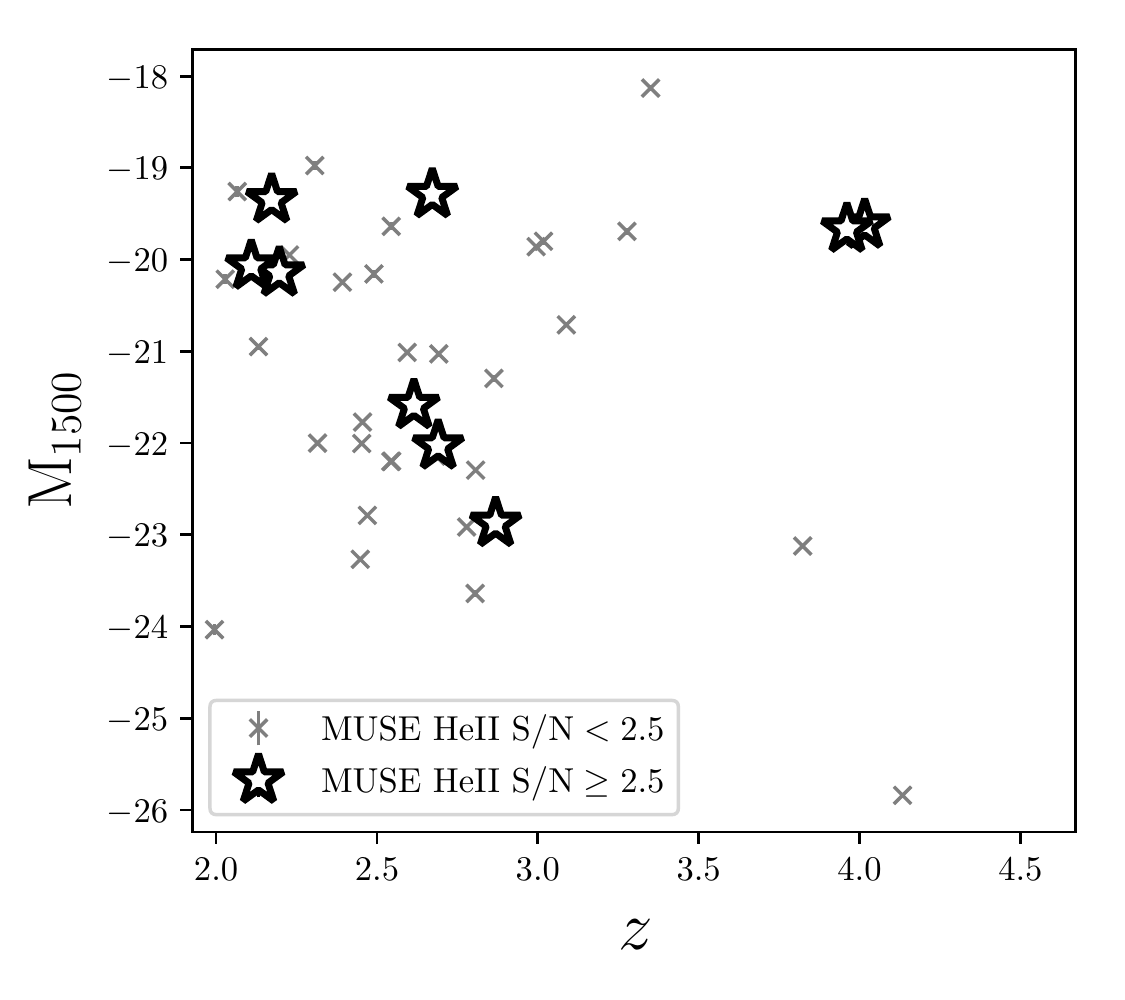}
\caption{{\bf Top:} Here we show the \HeII\ flux as a function of redshift. \HeII\ detections from MUSE are shown as stars and are colour coded depending on their median continuum flux at $\sim1640$ \AA. 
\HeII\ detections from other surveys within the plot range are shown by diamonds.
The redshift dependent MUSE wavelength coverage of a few prominent rest-UV features are shown in the top of the panel.  
{\bf Bottom:} M$_{1500}$ as a function of redshift for the MUSE \HeII\ detected and \HeII\ non-detected (set B, see Section \ref{sec:spec_stacking}) galaxies. [Section \ref{sec:observed_sample}]
\label{fig:HeII_flux_comparisions}
}
\end{figure}

In Figure \ref{fig:HeII_flux_comparisions}, we examine the \HeII\ flux distribution of our sample as a function of redshift and continuum S/N. 
It is evident from the figure that MUSE achieves better flux limits of \HeII\ compared to other surveys. 
We further show the absolute UV magnitude of the MUSE \HeII\ detected and MUSE \HeII\ coverage (set B, see Section \ref{sec:spec_stacking}) galaxies as a function of redshift. UV magnitudes are computed from rest-frame dust corrected (following \citealt{Calzetti2000} attenuation curve) MUSE spectra using a box-car filter between $1500\pm100$\AA.
We opt to use the MUSE spectra to compensate for limitations in rest-UV photometric coverage between our fields. 
Only galaxies with UV magnitude detected above $1\sigma$ noise between  $1500\pm100$\AA\ are selected for this analysis. 
The corresponding magnitude errors are computed using 100 bootstrap iterations of the spectra where the normalized median absolute deviation ($\mathrm{\sigma_{NMAD}=1.48\ |x_{i}-median(x)|}$) of the bootstrapped UV magnitudes are considered as the error.   
There is no statistically significant difference in absolute UV magnitude between MUSE \HeII\ detected and \HeII\ non-detected galaxies and a simple two sample K-S test for the two samples gives a Ks statistic of 0.40 and a P value of 0.17, thus we cannot reject the null hypothesis that the two independent samples are drawn from the same continuous distribution.

\subsection{Spectral Stacking}
\label{sec:spec_stacking}


Driven by observational constraints, spectral stacking techniques are commonly used to obtain high S/N UV rest-frame spectra of high redshift galaxies \citep[e.g.,][]{Shapley2003,Steidel2016}. 
While it provides strong constraints on the \emph{average} properties of observed galaxies, stacking of galaxies without any prior information about them may not constrain the  observed diversity of galaxies and could result in strong systematic biases.
For our analysis, we divided our sample of \HeII\ detected and non-detected galaxies in mass and redshift bins in order to mitigate any biases that may arise by having a large range of galaxy masses/redshifts in a single stack.

We define Set A (N=13) as the stack of all galaxies with \HeII\ detections. Set B (N=46) are all galaxies with no \HeII\ detections in the individual spectra and contains all galaxies with {\tt CONFID=3} (secure redshift, determined by multiple features) redshift quality classification between $1.93<z<4.67$ but with galaxies in set A removed. 
Each bin is then divided into three mass and redshift bins.
Since the MUSE quasar catalogue does not contain photometric information to constrain the stellar masses, galaxies in this field are not used for the mass stacks.

\begin{figure*}
\includegraphics[trim = 10 10 10 0, clip, scale=1.0]{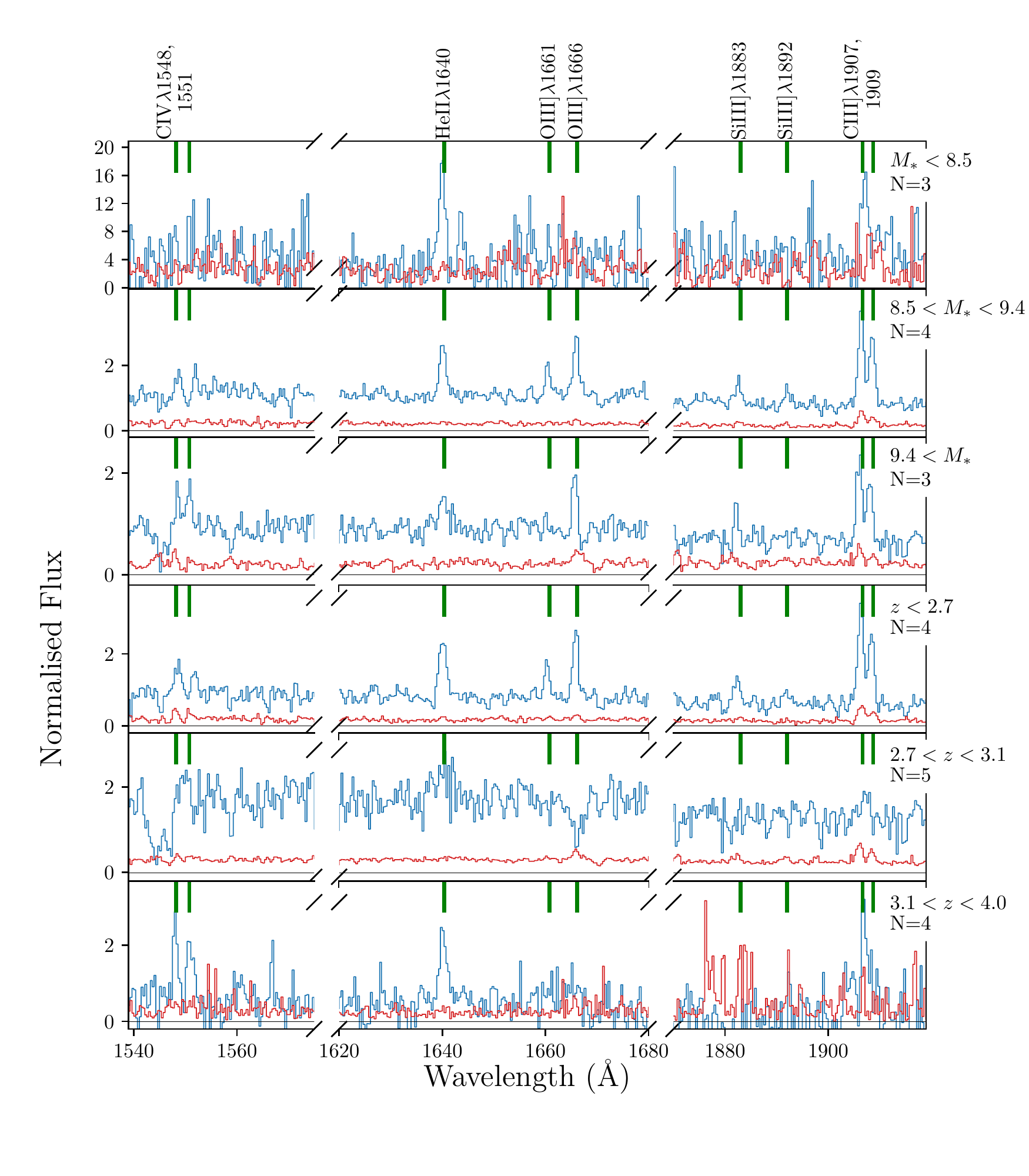}
\caption{Similar to Figure \ref{fig:HeII_spectra_1} but for the mass and redshift binned stacked spectra of the MUSE \HeII\ detected sample. The corresponding bin parameters are shown in each panel. [Section \ref{sec:spec_stacking}] 
\label{fig:similar_to_figure_figheii_spectra_1_but_for_the_mass_and_redshift_binned_stacked_spectra_of_the_muse_heii_detected_sample_the_corresponding_bin_parameters_are_shown_in_each_panel_figheii_detected_stacked}
\label{fig:HeII_detected_stacked}
}
\end{figure*}

\begin{figure*}
\includegraphics[trim = 10 10 10 0, clip, scale=1.0]{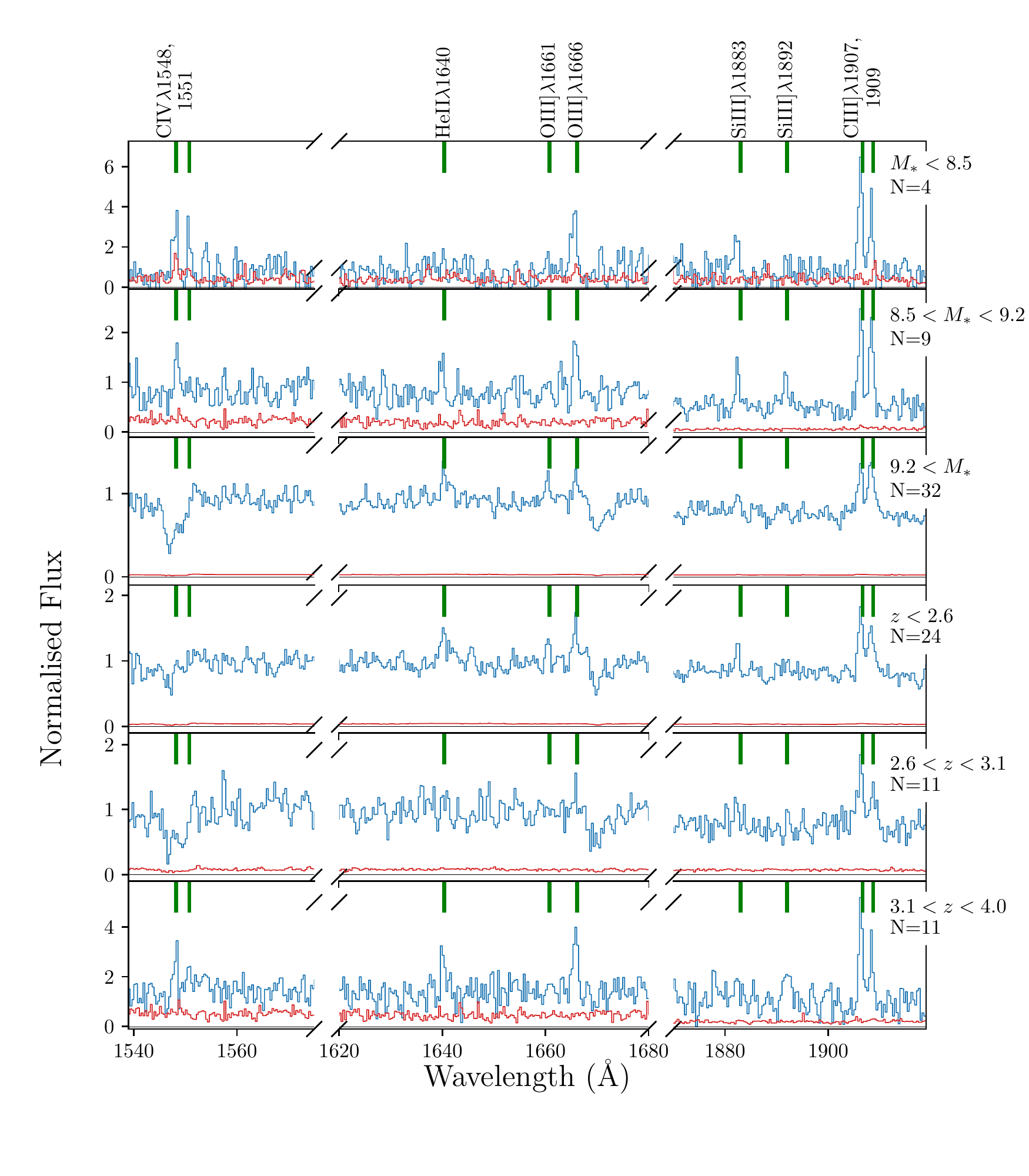}
\caption{Similar to Figure \ref{fig:HeII_detected_stacked} but for the stacks of \HeII\ undetected sample. [Section \ref{sec:spec_stacking}]
\label{fig:HeII_undetected_stacked}
}
\end{figure*}

We first measure the systematic redshift of galaxies by excluding \Lalpha\ from the redshift fitting procedure.
Then we resample the rest-frame spectra onto a regular grid between $1400-2700$ \AA\ with a sampling of 0.367 \AA\ corresponding to the native resolution of MUSE at $z=2.5$ in the rest-frame.
The final stacked spectra are calculated via median stacking and fitted using the method described above with the errors determined using 1000 bootstrap repetitions. We quote uncertainties using $\sigma_{NMAD}$.
We show our sample of stacked spectra in Figures \ref{fig:HeII_detected_stacked} (set A) and \ref{fig:HeII_undetected_stacked} (set B).

\subsection{Comparison with \citet{Gutkin2016} photo-ionization modeling}
\label{sec:model_comp}

The nature of rest-frame UV emission lines that originate from the ISM is driven by the properties of stars that heat up the ISM and the physical/chemical conditions of the ISM itself. 
Therefore, by making simplifying assumptions about the stellar populations, geometry of the ionization regions, and physics and chemistry of dust and ISM, the observed rest-UV emission line ratios can be used to infer average properties of the ISM and underlying stellar populations of the observed galaxies.

In this section we use photo-ionization models by \citet{Gutkin2016} to infer the average ISM conditions of galaxies in our sample. The \citet{Gutkin2016} models are based on the new generation of \citet{Bruzual2003} stellar population models and uses the photo-ionization model {\tt CLOUDY} \citep[c13.03,][]{Ferland2013} to model emission lines of \HII\ regions by self-consistently accounting for the influence of gas phase and interstellar abundances. 
The wide range of interstellar parameters spanned by these models makes them ideally suited for comparisons to the observed line ratios of our sample for which we expect properties clearly different from the average population of local star-forming galaxies \citep[e.g.,][]{Erb2010}.
We use the following emission lines for our analysis: \HeII, \CIII=(\CIII$\lambda1907$+\CIII$\lambda1909$), \OIII(=\OIII$\lambda1661$+\OIII$\lambda1666$), \SiIII(=\SiIII$\lambda1883$+\SiIII$\lambda1892$). 
For each emission line ratio diagnostic, we select a subsample of galaxies with S/N$\geq3$ for the emission lines considered in that specific diagnostic.  
A further analysis of the \citet{Gutkin2016} rest-UV emission line ratios discussed in our study is presented in Appendix \ref{appendix:model_comp_models}.


\subsubsection{Individual detections}
\label{sec:model_comp_HeII_individual}

\begin{figure*}
\includegraphics[trim = 10 10 10 0, clip, scale=0.625]{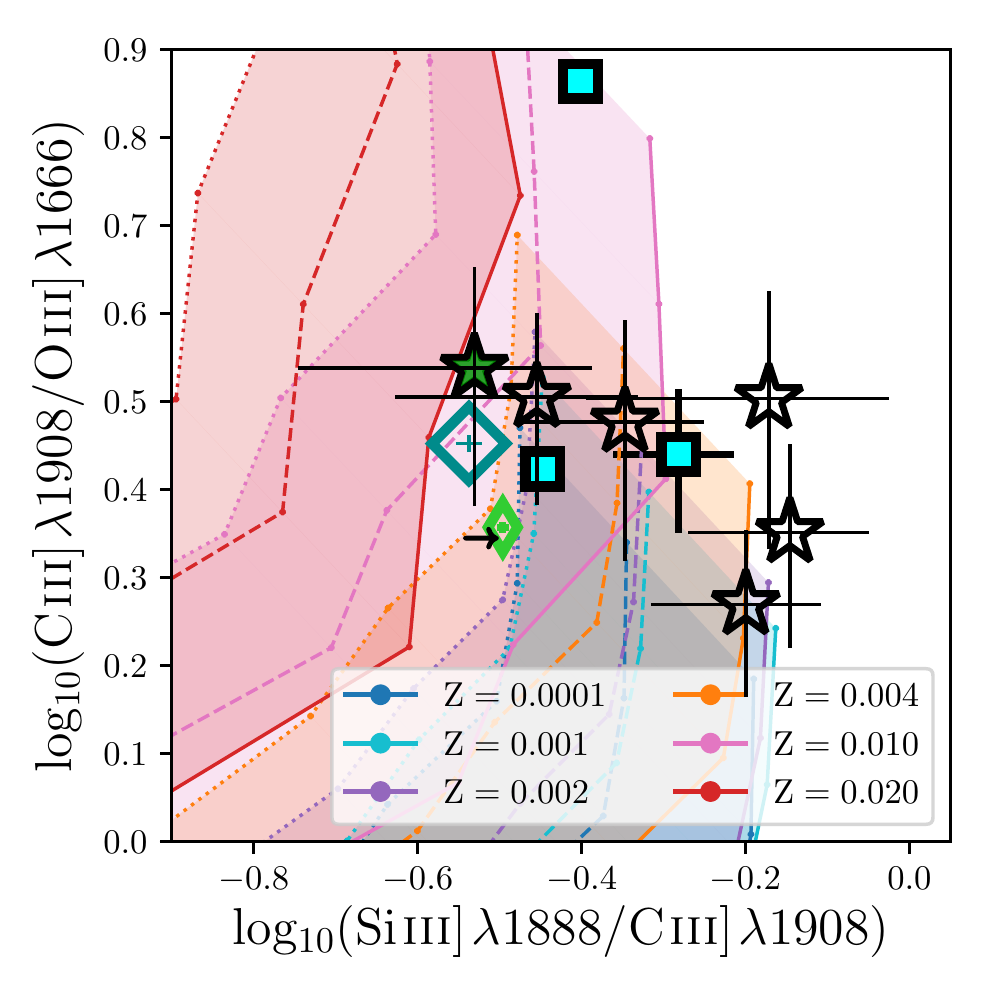}
\includegraphics[trim = 10 10 10 0, clip, scale=0.625]{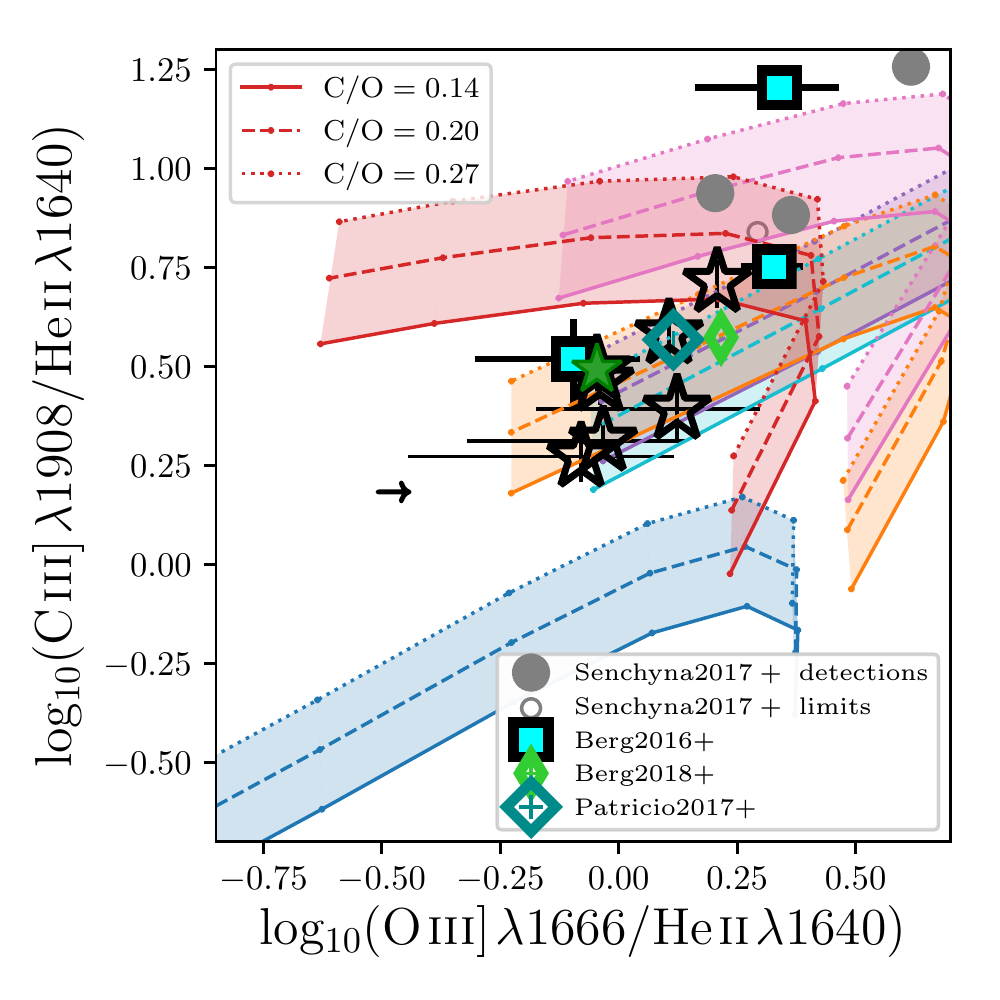}
\includegraphics[trim = 10 10 10 0, clip, scale=0.625]{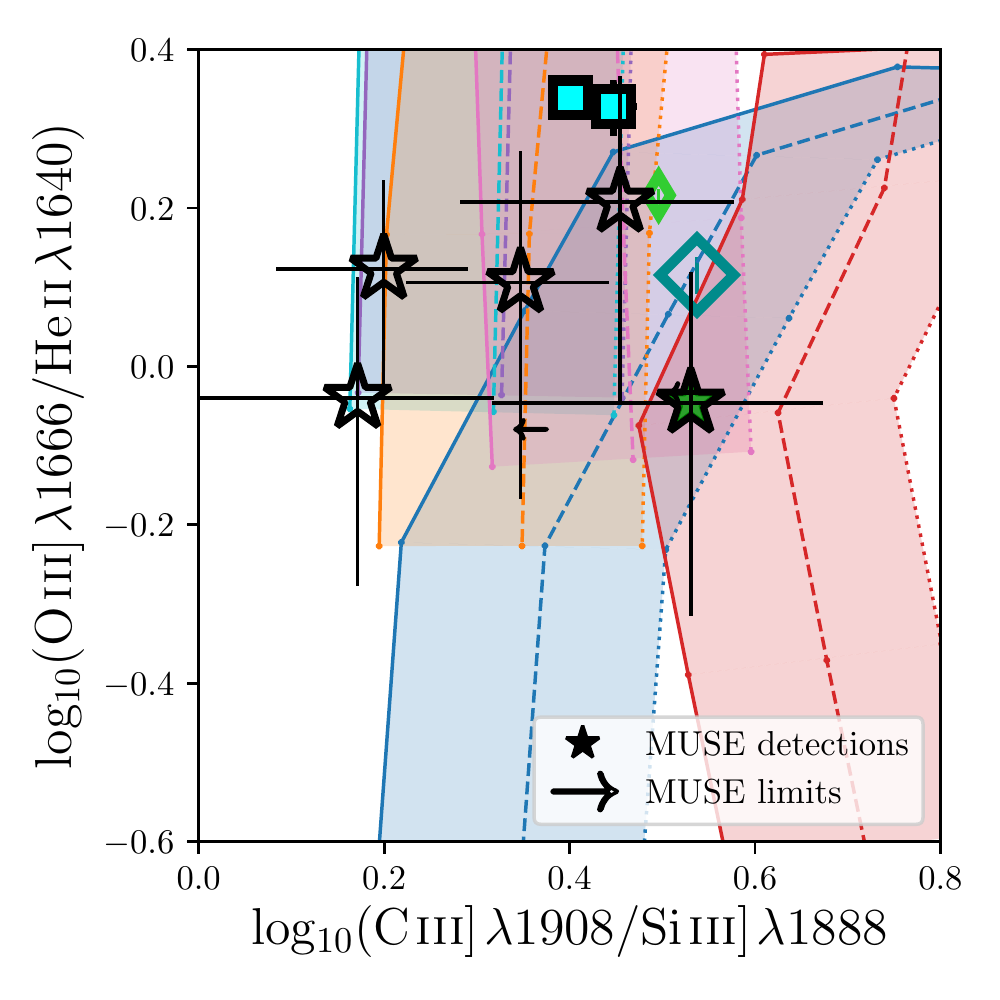}
\caption{Rest-frame UV emission line ratios of the MUSE \HeII\ sample. 
{\bf Left:} \CIII/\OIII vs \SiIII/\CIII\ ratios. Individual galaxies with S/N$>2.5$ for all four emission lines are shown as stars. Limits are shown as arrows. The tracks are from \citet{Gutkin2016} models which are powered by star-formation. Each set of tracks with same colour show three C/O ratios and the region between the minimum and maximum C/O tracks are shaded by the same colour. 
From top to bottom the ionization parameter increases. 
Where available line ratios from \citet{Patricio2017}, \citet{Senchyna2017}, and \citet{Berg2018} are shown for comparison. MUSE line ratios of the Lyman continuum emitting candidate from \cite{Naidu2017} is shown by filled the green star.
{\bf Centre:} Similar to the left panel but \CIII/\HeII\ vs \OIII/\HeII\ emission line ratios, where detections are defined as galaxies with S/N$>2.5$ for all three emission lines. 
{\bf Right:} Similar to left but \OIII/\HeII vs \CIII/\SiIII\ emission line ratios where detections are defined as galaxies with S/N$>2.5$ for all four emission lines. [Section \ref{sec:model_comp_HeII_individual}] 
\label{fig:line_ratios_gutkin}
}
\end{figure*}

In order to probe the general ISM properties of our \HeII\ detections and investigate whether we can constrain the dominant ionizing source, in this section we explore the observed distribution of emission line ratios of the individual galaxies and make comparisons with \citet{Gutkin2016} photo-ionisation models.  
In Figure \ref{fig:line_ratios_gutkin} we show three selected line ratio diagrams. 
Due to the wavelength coverage of MUSE and detection thresholds of our observations, not all galaxies with \HeII\ are detected with the full suite of rest-UV emission lines considered in the models. 
Therefore, in each panel, we select all galaxies for which the considered emission lines would fall within the wavelength range of MUSE and divide them into two bins depending on their S/N,  where S/N$\geq2.5$ are considered as MUSE detections and galaxies which do not make the cut for at least one of the emission lines are considered as MUSE limits. 
The error level is constrained by the noise spectrum and we consider the 3$\sigma$ error level as the upper limit to the line flux for emission lines that fail the S/N cut. 
Given the degeneracy between model parameters and observational constraints driven by weak line detections, quantitative predictions about specific ISM conditions of our sample cannot be inferred within the current scope of our work and thus, we refrain from inferring best-fit model values on a per galaxy basis.

In the \CIII/\OIII\  vs \SiIII/\CIII\ line ratio diagram, all galaxies with MUSE line detections fall within reasonable limits of the  \citet{Gutkin2016} models. 
With the existing data we cannot place constraints on the metallicity but most model tracks require an ionisation parameter ($U_{s}$)$\gtrsim-2$. 
In MUSE data, the weakest emission line in this line ratio diagnostic is \SiIII, thus observed \SiIII/\CIII\ ratios of the MUSE limits should be considered as upper limits. Therefore, MUSE limits would prefer lower metallicity, lower ionisation parameter models. 
Additionally, it is evident from  Figure \ref{fig:line_ratios_gutkin} that MUSE detected emission line ratios agree well with emission line ratios obtained for the  \citet{Berg2018} and \citet{Patricio2017} lensed galaxies at $z\sim2$ and $z\sim3.5$, respectively.  
The line ratios of most of the \citet{Berg2016a} $z\sim0$  low metallicity dwarf galaxies are also consistent with those measured in our MUSE sample

The \CIII/\HeII\ vs \OIII/\HeII\ diagnostic diagram has been suggested as a rest-UV emission line diagnostic for the separation of AGN and stellar ionising sources \citep[e.g.,][however also see \citealt{Xiao2018}]{Feltre2016}, and all our galaxies in the MUSE detected sample occupy the region where the emission lines can be powered purely by star-formation processes. 
In this diagnostic diagram MUSE galaxies occupy a region preferred by sub-solar metallicity tracks ($\sim1/5$th to $\sim1/100$th) with low ionisation parameters in conflict with the \CIII/\OIII\  vs \SiIII/\CIII\ line ratio diagram.   
Higher metallicities can be accommodated but would require C/O ratios lower than the typical C/O ratios ($\sim0.15-1.30$) observed in high-$z$ galaxies \citep{Shapley2003,Erb2010,Steidel2016}. 
This would require either relatively low fraction of mass loss and ISM enrichment from massive stars for a given metallicity \citep{Henry2000}  or a longer time-delay in the production of carbon by lower mass stars compare to oxygen \citep[][also see \citealt{Akerman2004,Erb2010}]{Chiappini2003}, which is primarily produced by massive stars.  
Here the MUSE limits are driven by weak \OIII\ emission line and thus \OIII/\HeII\ limits should be considered as upper limits.
The change of  \CIII/\HeII\ vs \OIII/\HeII\ line ratios as a function of $U_{s}$ is not linear (see Appendix \ref{appendix:model_comp_models}) and thus we cannot make any constraints about the expected ISM conditions of the limits in this line ratio diagnostic. 
The high-$z$ lensed galaxies from \citet{Berg2018} and \citet{Patricio2017} occupy a similar region to MUSE detections. 
We also show the $z\sim0$ sample from \citet{Senchyna2017} which clearly requires higher metallicity models to explain the emission line ratios.
Low metallicity $z\sim0$ dwarf galaxies from \citet{Berg2016a} also on average prefers higher metallicity models compared to the high-$z$ samples.

Driven by the close proximity of the line wavelengths of \OIII/\HeII\ and \CIII/\SiIII, we select the \OIII/\HeII\ vs \CIII/\SiIII\ line ratio diagram to measure the photo-ionisation properties of our sample relatively independent of dust attenuation.
MUSE detected galaxies in our sample favour models with solar to sub-solar (down to $\sim1/200$th) metallicities. However, at lower metallicity, the different stellar tracks cannot be distinguished from each other.
At fixed metallicity, this line ratio diagnostic is ideal to constrain the C/O ratios of galaxies.
Due to multiple effects, metallicity shows a complex relationship with semi/forbidden emission line ratios. 
For example, \citet{Jaskot2016} show that at lower metallicities where C abundance is lower, counterintuitively the \CIII\ flux is enhanced due to the harder ionizing SED and higher gas temperature increasing the \CIII\ collisional excitation rate. 
Similar to the other line ratio diagrams, \citet{Berg2018} and \citet{Patricio2017} lensed galaxies occupy a similar parameter to our MUSE detections, however, \citet{Berg2016a} sample shows higher \OIII/\HeII\ ratios compared to the high-$z$ samples. 
MUSE limits within the plot range are driven by weak \SiIII\ lines and therefore, the \CIII/\SiIII\ ratio should be considered as a lower limit.

Analysis of individual emission line ratios of the MUSE \HeII\ sample in multiple line ratio diagnostics does show, in general, good agreement with the line ratio space occupied by the \citet{Gutkin2016} models. 
As aforementioned, we refrain from inferring best-fit model values on a per galaxy basis due to modeling and observational constraints. 
Additionally, Lyman continuum leakage results in high-energy ionising photons to escape the dusty molecular clouds without being converted to lower energy photons as assumed by the photo-ionisation models.  
This results in extra complications for comparisons between observed line ratios with model predictions. 
We have one Lyman continuum leaking candidate \citep{Naidu2017} in our sample, which we have highlighted in Figure \ref{fig:line_ratios_gutkin} (green star). 
The emission line ratios of this galaxy does not stand out relative to the rest of the sample, but given the estimated high escape fraction ($f_{esc}\sim60\%$) the parameters inferred from the \citet{Gutkin2016} models (which assume no escape) are expected to be biased. Since to first order Lyman continuum escape implies reduced Balmer line fluxes, we would typically infer higher U and/or lower Z values than the intrinsic values.


\subsubsection{Stacked sample}
\label{sec:model_comp_stacked}

To improve the S/N in the weak lines, we now turn to the stacked spectra discussed in Section  \ref{sec:model_comp_HeII_individual}.
As shown by Figures \ref{fig:HeII_detected_stacked} and \ref{fig:HeII_undetected_stacked}, continuum normalized \HeII\ detected stacked spectra show a trend between \HeII\ emission line strength and stellar mass, with the lowest mass galaxies stacked sample showing the strongest \HeII\ emission compared to the continuum level. 
The higher stellar mass systems show broader \HeII\ profiles which could be linked to increased stellar contribution to the \HeII\ emission. 
The stacks of \HeII\ non-detected galaxies also show weak \HeII\ emission, thus, it is possible that some galaxies show weak \HeII\ emission which is below the MUSE detection limit for individual objects. 
There is no strong redshift evolution for \HeII\ detected sample, however, high redshift stacks of \HeII\ undetected galaxies show weak narrow \HeII\ features.

\begin{figure*}
\includegraphics[trim = 10 10 10 0, clip, scale=0.6252]{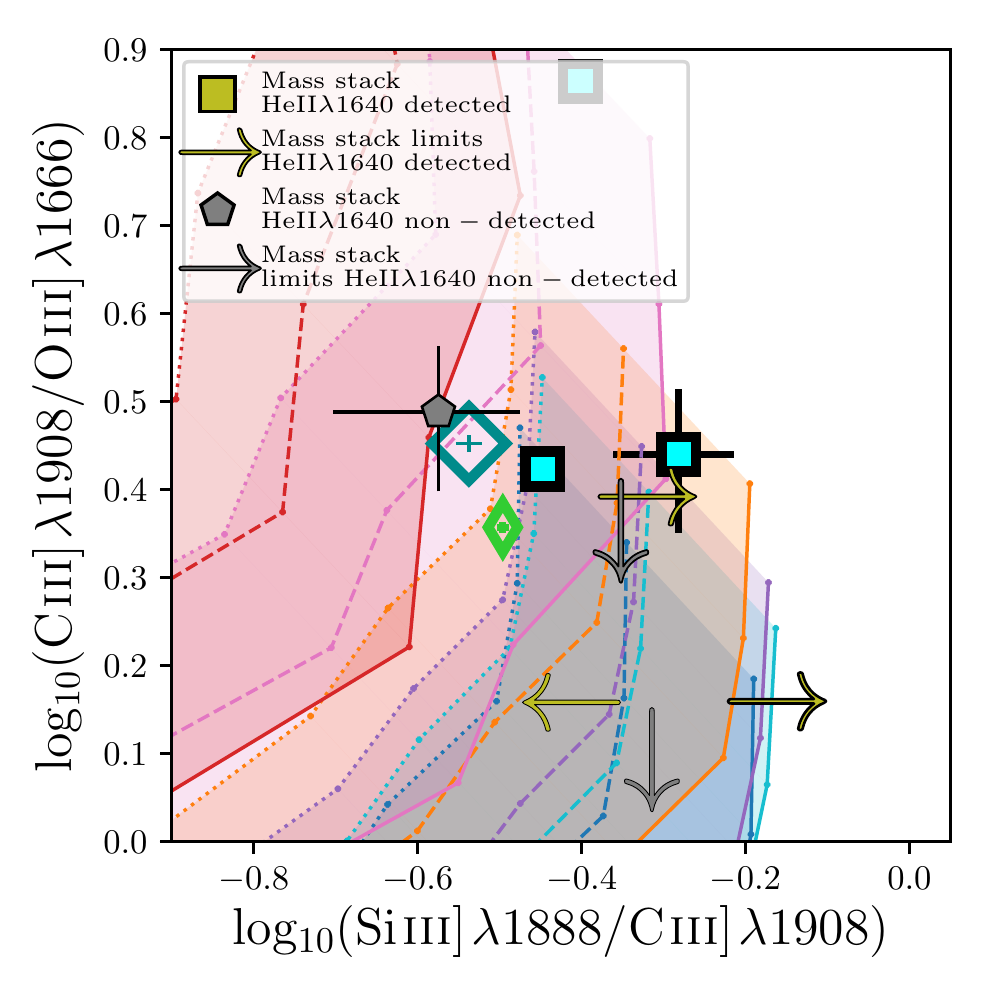}
\includegraphics[trim = 10 10 10 0, clip, scale=0.6252]{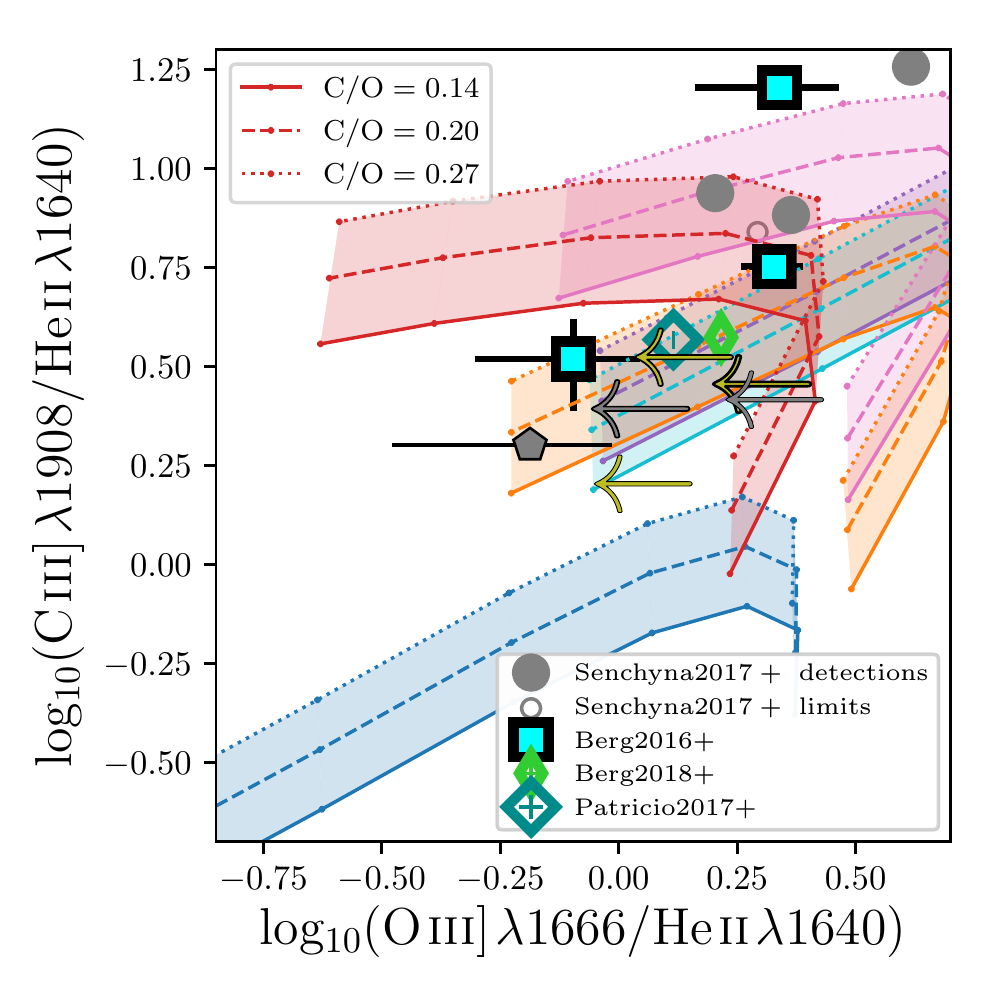}
\includegraphics[trim = 10 10 10 0, clip, scale=0.6252]{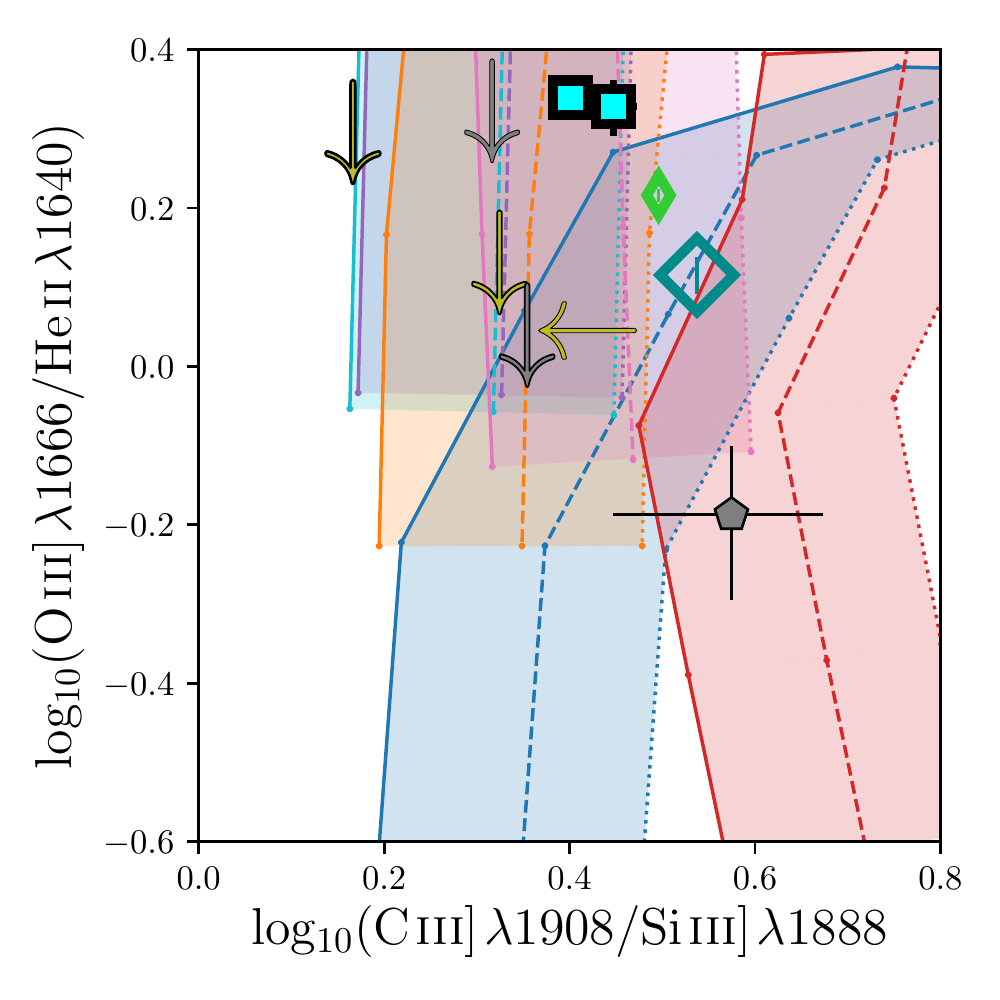}
\includegraphics[trim = 10 10 10 0, clip, scale=0.6252]{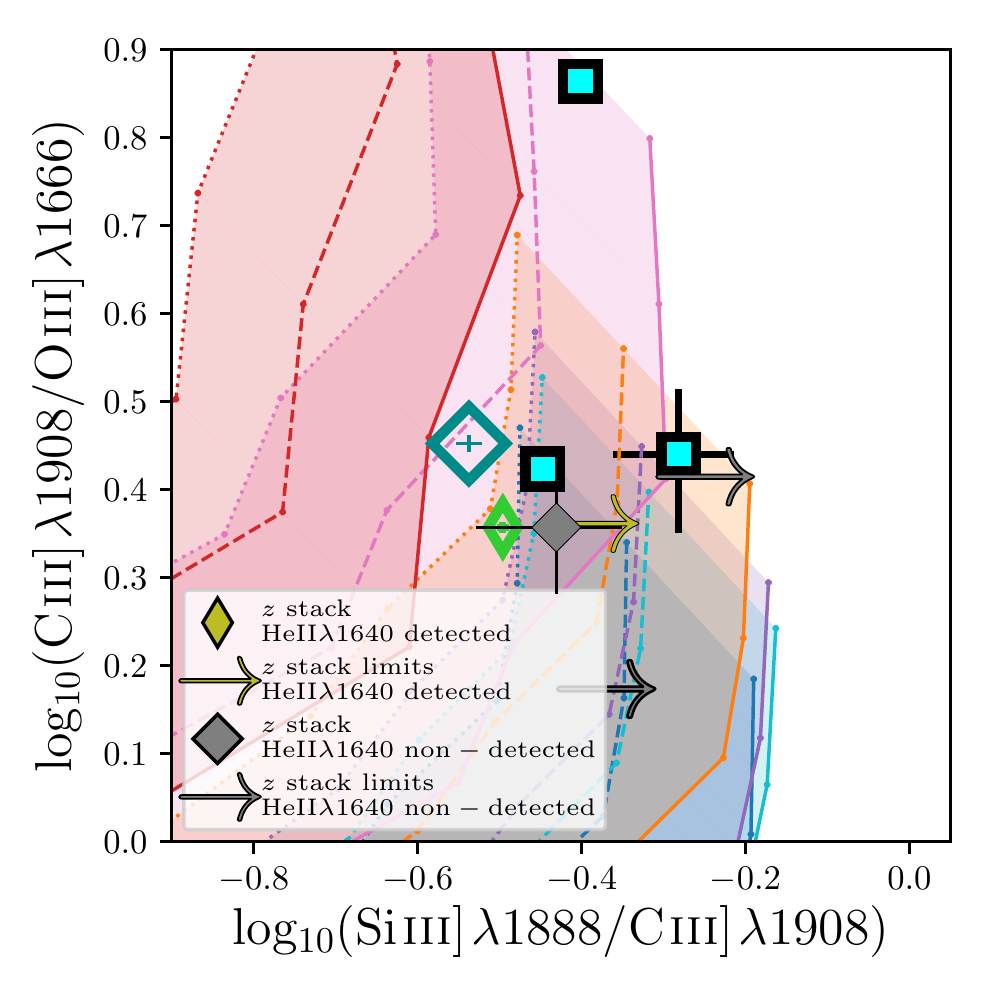}
\includegraphics[trim = 10 10 10 0, clip, scale=0.6252]{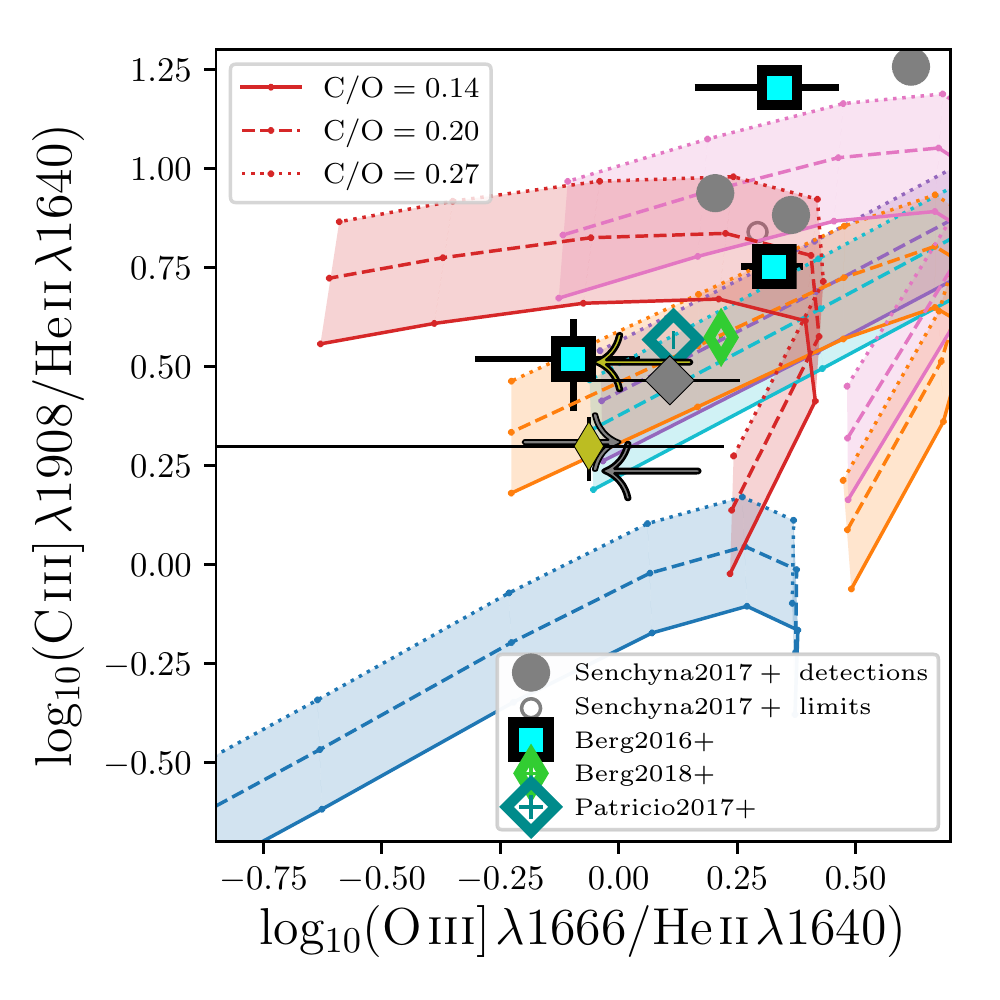}
\includegraphics[trim = 10 10 10 0, clip, scale=0.6252]{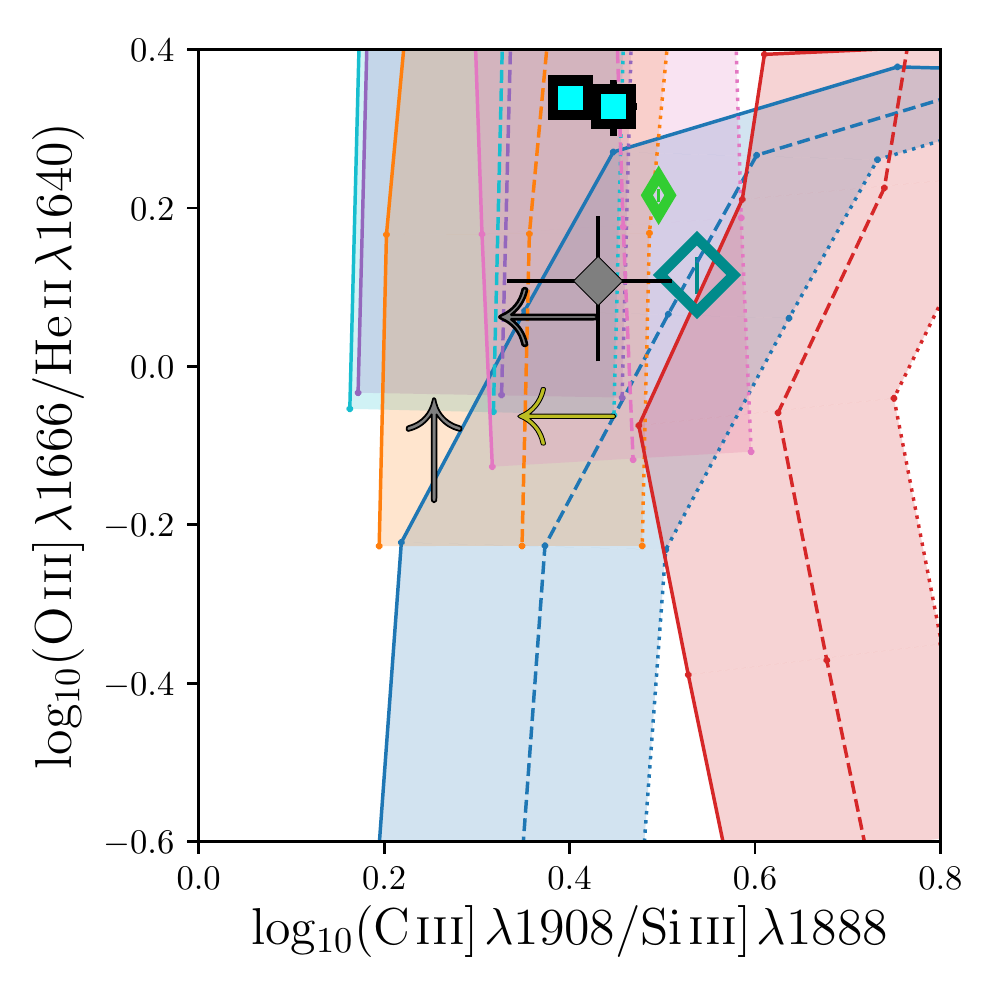}
\caption{Rest-frame UV emission line ratios of the MUSE stacked galaxies compared with \citet{Gutkin2016} models. Panels from left to right are similar to Figure \ref{fig:line_ratios_gutkin}. 
Galaxies are stacked in mass and redshift bins, with line width of the markers increasing with mass and redshift. 
Limits resemble stacks with emission lines (considered in each panel) lower than the 3$\sigma$ error limit. For such stacks, $3\sigma$ error is used as the respective line flux.   
{\bf Top:} MUSE stacked sample for stellar mass bins: $log_{10}({M_{*}/M_{\odot}})<9.5$, $9.5<log_{10}({M_{*}/M_{\odot}})<10.0$ , $log_{10}({M_{*}/M_{\odot}})>10.0$.   
{\bf Bottom:} MUSE stacked sample for redshift bins: $z<2.5$, $2.5<z<3$ , $z>3$. [Section \ref{sec:model_comp_stacked}]    
\label{fig:line_ratios_gutkin_stacked}
}
\end{figure*}

We show the emission line ratios of the \HeII\ detected stacked sample in Figure \ref{fig:line_ratios_gutkin_stacked}. In all three line ratio diagrams, the stacked galaxies with line detections occupy a similar region to the individual galaxies shown in Figure \ref{fig:line_ratios_gutkin}.  
The low S/N of \SiIII\ and \OIII\ line fluxes of the stacked sample refrain us from making strong constraints with emission line ratio diagnostics.

The \CIII/\OIII\ vs \SiIII/\CIII\ line ratios of the MUSE stacked detections do not show any trend with either stellar mass or redshift. 
Driven by the weak \OIII\ emission line, the \CIII/\HeII vs \OIII/\HeII\ line ratios of the moderate-low mass bins show a preference for sub-solar models with low ionisation parameter. 
As aforementioned, higher metallicity tracks with lower C/O ratios than what is illustrated in the figure could also explain the emission line ratios of these bins.  
The higher redshift stacks also show a similar preference. 
Low mass and high redshift systems have been shown to have lower gas phase \citep[e.g.,][]{Sanders2015,Kacprzak2015} and stellar metallicities \citep[e.g.,][]{Steidel2016} compared to local galaxies, and thus such a trend is expected. 
The stacked galaxy sample show no clear trend with either stellar mass or redshift in the \OIII/\HeII vs \CIII/\SiIII\ line ratio distribution.

We perform a similar analysis on all galaxies where we are unable to detect a narrow \HeII\ emission line. Though individual galaxies do not show such features, once stacked, specially the lower mass and higher redshift stacks show narrow \HeII\ emission. 
The \CIII/\OIII vs \SiIII/\CIII\ emission line ratios of these galaxies also do not show any trend with redshift, but marginally prefer models with higher metallicities or C/O ratios, compared to the \HeII\ detected sample.

\subsection{Comparison with BPASS \citet{Xiao2018} models}
\label{sec:model_comp_HeII_Xiao}

\begin{figure}
\includegraphics[trim = 10 10 10 0, clip, scale=0.9]{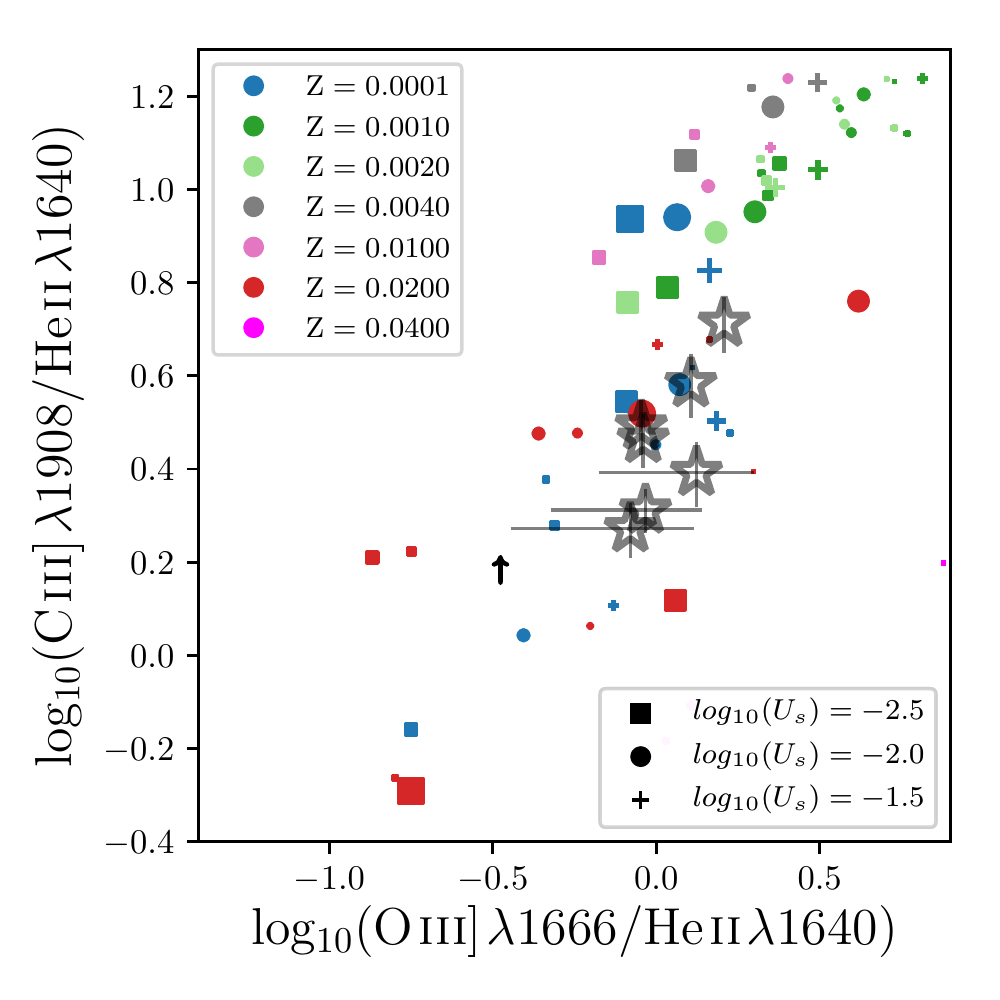}
\caption{Rest-frame UV emission line ratios of the MUSE \HeII\ sample compared with the model line ratios computed by \citet{Xiao2018} using BPASS binary stellar population models. Here we show the \CIII/\HeII vs \OIII/\HeII\ line ratios for the MUSE \HeII\ detected sample. Individual galaxies with S/N$\geq2.5$ for all three emission lines are shown as stars. Galaxies which fail the S/N cut are shown as arrows. 
The symbol size is proportional to the age from the onset of the star formation burst between  $t=1$Myr (smallest) and $t=50$Myr.
Models are computed with $log_{10}(n_H)=1.0$ with varying $U_s$ between --2.5 and --1.5. BPASS \zsol=0.02.
[Section \ref{sec:model_comp_HeII_Xiao}] 
\label{fig:line_ratios_xiao}
}
\end{figure}

The \citet{Gutkin2016} photo-ionisation models are built on an updated version of the \citet{Bruzual2003} stellar population models (Charlot \& Bruzual, in preparation), which considers stars up to 350\msol\ in a range of metallicities. However, these models do not account for any effects of stellar rotation nor effects of stars interacting with each other, i.e. binary stars. 
However, the Universe contains many binary stars. In the Galaxy, $\sim50\%$ of O stars have shown to be in binary systems \citep[e.g.,][]{Langer2012,Sana2012,Sana2013} and stellar population analysis of local massive star clusters in $z\sim0$ galaxies have shown the need to consider interactions between binary stars to accurately predict the observed photometry \citep{Wofford2016}. 
Additionally, modeling of rest-UV and optical spectra of galaxies at $z\sim2$ find that models that include binaries perform better than the single star models considered \citep{Steidel2016,Strom2017,Nanayakkara2017,Berg2018}. 
In this section, we use photo-ionisation models by \citet{Xiao2018} to explore the effects of including binary star interactions in our rest-UV emission line/EW analysis of the \HeII\ emitters.


\subsubsection{Comparison of observed line ratios}
\label{sec:model_comp_HeII_lr_gutkin}

\citet{Xiao2018} use BPASSv2 \citep{Eldridge2017} stellar population models as the source for the ionizing continuum to  self consistently predict the nebular continuum and emission line flux using the photo-ionisation code {\tt CLOUDY}. These photo-ionisation models are generated as a function of time for a single stellar population with a constant SFH up to 100 Myr assuming a spherical ionization bound gas nebula with uniform hydrogen density. The models assume no dust and considers the nebular gas metallicity to be same as that of the stellar metallicity.    
The \citet{Xiao2018} models are run on two distinct BPASSv2 stellar population implementations: models with and without binary star interactions. 
Here we only analyze the binary stellar populations.
For a single star-burst, implementing the effects of binary evolution results in the ionizing continuum being harder for a prolonged period of time compared to a non interacting model with the same initial conditions.  
Binary interactions prolongs the life time and/or rejuvenates the  stars via gas accretion and rotational mixing enhanced by the angular momentum transfer, which results in efficient hydrogen burning within the stars \citep[e.g.][]{Stanway2016}. 
Additionally, binary interactions effectively remove the outer layers of the massive red super-giants resulting in a higher fraction of W-R stars and/or low-mass helium stars, specially at lower metallicities and at later times  ($>5$ Myr) in single burst stellar populations. 
Including such effects to the ionizing continuum causes the number of \Hep\ ionizing photons to increase (up to $\sim3$ orders of magnitude), at $t>10$ Myr for higher metallicities and $t\sim10$ Myr for lower metallicity models.
Therefore, considering the effects of binaries is crucial to probe mechanisms of \HeII\ production.

In Figure \ref{fig:line_ratios_xiao} we show the distribution of the observed \CIII/\HeII/ vs \OIII/\HeII\ line ratios of the MUSE \HeII\ sample with \citet{Xiao2018} models that include binary stellar populations. 
As discussed in Appendix \ref{appendix:model_comp_models}, at fixed ionisation parameter rest-UV emission line strengths of higher metallicity models have a strong dependence on hydrogen gas density, thus at $log_{10}(n_H)\leq1$, super solar metallicity models could also produce the observed line ratios but only at extreme ionisation parameters ($log_{10}(U)\geq-1.5$). 
If sub solar metallicity models (down to $\sim1/200$ \zsol) are to produce the observed line ratios, BPASS single stellar populations model require galaxies to harbor extremely young ($<10$ Myr) stellar populations.
One large uncertainty in \citet{Xiao2018} models is the negligence of dust depletion and dust physics in the photo-ionisation modeling.
Considering dust depletion will lead to depletion of metals from the gas phase which will further increase the parameter space of the models \citet[][also see discussion in Section \ref{sec:uncerternities_dust}]{Charlot2001,Brinchmann2013,Gutkin2016}.  
When considering BPASS models that only include single stellar populations, the observed line ratios in general can only be produced by solar metallicity models and are not shown in Figure \ref{fig:line_ratios_xiao}.

When binary stars are included, most parameters become degenerate with each other. Therefore, a variety of models ranging from \zsol\ to  $\sim1/200$ \zsol\ are able to reproduce the observed line ratios largely independent from photo-ionisation properties (also see Figure B2 of \citealt{Xiao2018}). 
However, the BPASS binary models rule out lower ionisation parameter models ($U_s\lesssim-2.5$) at every metallicity considered. 
Hence, we conclude that extra degeneracies introduced by including  effects of binary star interactions prohibit us from putting strong constraints on ISM conditions of our \HeII\ sample. 
Full spectral fitting analysis with higher S/N spectra of individual galaxies might allow stronger constraints on the binarity of the stellar populations enabling more detailed understanding of stellar and ISM conditions of \HeII\ emitters at high-$z$. 
However, this is outside the scope of the present paper. 
We further caution against direct comparison of emission line ratios between \citet{Gutkin2016} and \citet{Xiao2018} models due to significant differences in the underlying stellar population and photo-ionization modeling assumptions.

\begin{figure*}
\includegraphics[trim = 10 10 10 0, clip, scale=0.95]{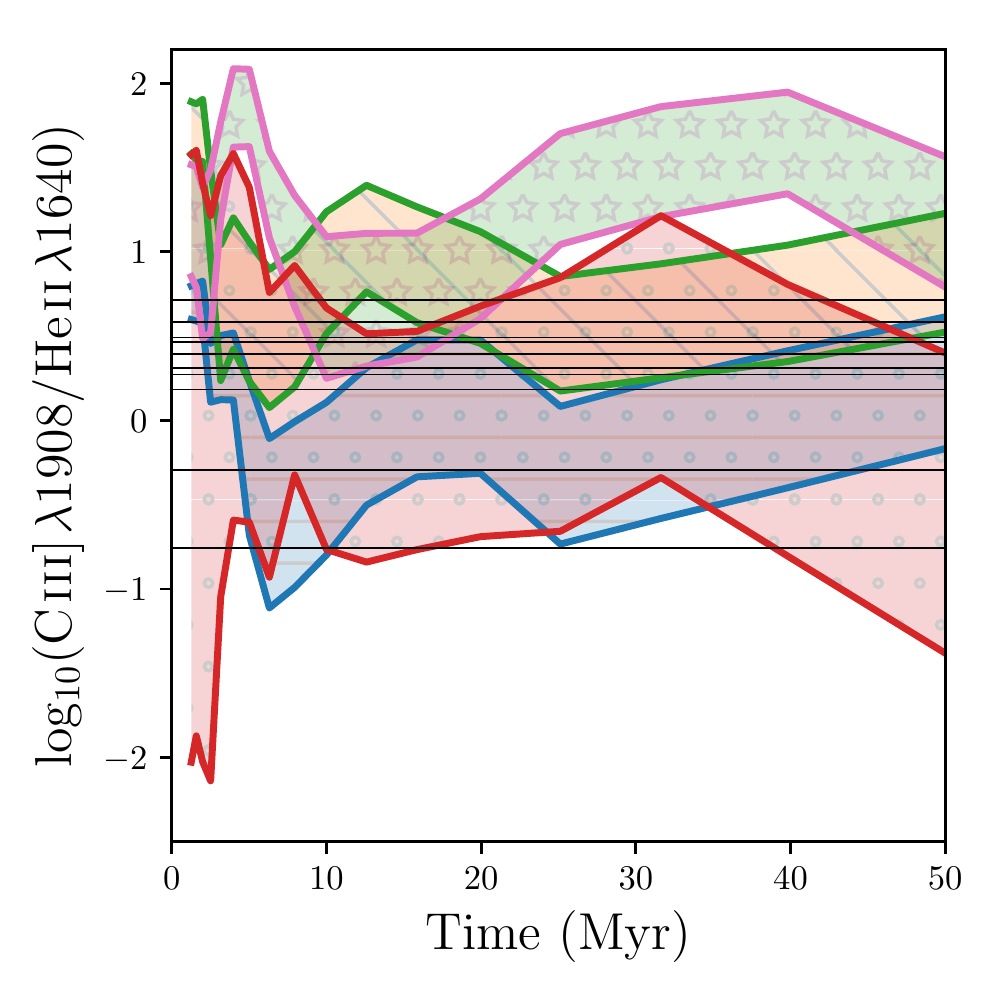}
\includegraphics[trim = 10 10 10 0, clip, scale=0.95]{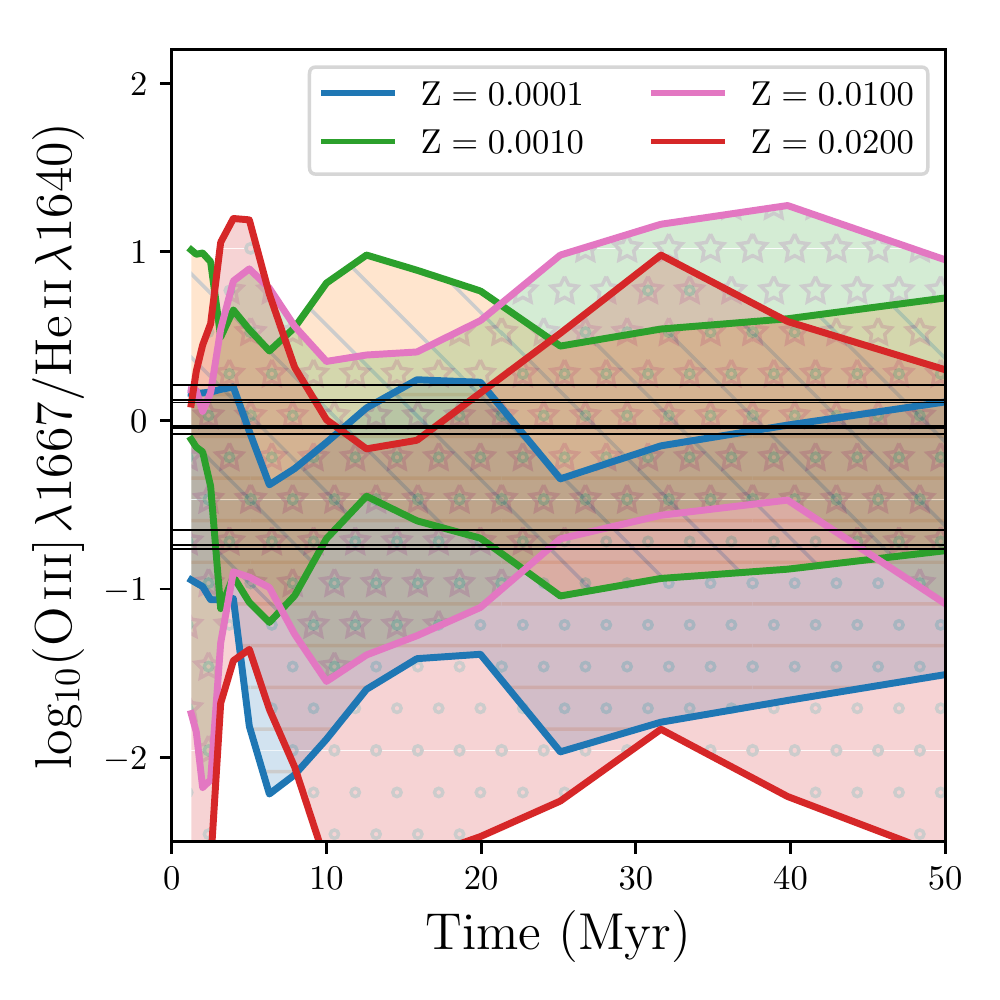}
\caption{ \citet{Xiao2018} rest-UV emission line ratio evolution as a function of time. 
Here we show  {\bf Left:} \CIII/\HeII\ vs time  and {\bf Right:} \OIII$\lambda1666$/\HeII\ vs time for the BPASS binary models computed with a $log_{10}(n_H)=1.0$ and $U_{s}=-1.5$ and $U_{s}=-3.5$ (upper and lower limits of each shaded region, respectively)  at different metallicities between 1 \zsol\ to 1/200th \zsol. 
We only show a limited set of model metallicities to enhance the clarity of the figure.  
The black horizontal lines show the line ratios of the MUSE \HeII\ sample. [Section \ref{sec:model_comp_HeII_Xiao}]
\label{fig:line_ratio_xiao_time_evolution}
}
\end{figure*}

In Figure \ref{fig:line_ratio_xiao_time_evolution}, we examine the  time evolution of  \CIII/\HeII\ and \OIII$\lambda1666$/\HeII\ emission line ratios in the \citet{Xiao2018} models. 
Models with lower $U_s$ always show lower emission line ratios in both \CIII/\HeII\ and \OIII$\lambda1666$/\HeII\ line ratios, with higher metallicity models in general showing a larger dependence of $U_s$. 
Lower metallicity models produce more \HeII\ flux, hence show lower line ratios compared to their higher metallicity models at earlier times. 
However, at later time the enhanced production of W-R stars in higher metallicity systems  decreases the emission line ratios. 
A mixture of QHE effects, ISM abundances, and W-R stars give rise to the complex variations in the time evolution of the models (also see Figure \ref{fig:bpass_model_predictions}).
Our observed emission line ratios can be produced by a variety of models relatively independent of the age within the first 100 Myr of the onset of the star-burst.


\subsubsection{Comparison of observed EWs }
\label{sec:model_comp_HeII_EW}

\begin{table*}
\caption{EWs of the MUSE \HeII\ sample used in this analysis. [Section \ref{sec:emission_line_measurements}, \ref{sec:model_comp_HeII_EW}]
\label{tab:line_ews_muse}}
\centering
\begin{tabular}{l r r r r r r r r r r r r r r r    }
\hline\hline
		   {ID}                                &
           \multicolumn{2}{c}{\HeII}     	   &   
           \multicolumn{2}{c}{\CIII1907}       &
           \multicolumn{2}{c}{\CIII1909}       &
           \multicolumn{2}{c}{\OIII1661}       &
           \multicolumn{2}{c}{\OIII1666}       &
           \multicolumn{2}{c}{\SiIII1883}      &
           \multicolumn{2}{c}{\SiIII1892}        \\
           {}                         		   & 
           {EW}                       		   &
           {$\Delta$EW} 			  		   &
           {EW}                       		   &
           {$\Delta$EW} 			  		   &
           {EW}                       		   &
           {$\Delta$EW} 			  		   &
           {EW}                       		   &
           {$\Delta$EW} 			  		   &
           {EW}                       		   &
           {$\Delta$EW} 			  		   &
           {EW}                       		   &
           {$\Delta$EW} 			  		   &
           {EW}                       		   &
           {$\Delta$EW} 			  		   \\
\hline \hline
1024   & 18.9 & 3.5 & 18.5 & 3.1 & 20.5 & 3.5 & 23.2 & 3.6 & 21.7 & 3.3 & 20.9 & 3.2 & 21.3 & 3.0 \\
1036   & 12.9 & 2.9 & 4.4 & 1.0 & 8.3 & 1.4 & 14.7 & 1.1 & 12.0 & 1.1 & 12.2 & 0.9 & 14.3 & 1.2 \\
1045   & 12.6 & 2.2 & 6.6 & 1.2 & 11.8 & 1.7 & 14.9 & 1.3 & 13.5 & 1.4 & 14.0 & 1.3 & 14.3 & 1.7 \\
1079   & 35.6 & 11.5 & 16.6 & 0.7 & 15.8 & 0.7 & 17.1 & 0.6 & 17.5 & 0.7 & 17.3 & 0.6 & 16.9 & 0.7 \\
1273   & 13.7 & 3.7 & 6.5 & 2.0 & 0.4 & 1.0 & 11.2 & 1.6 & 7.6 & 1.1 & 7.5 & 1.6 & 11.8 & 1.7 \\
3621   & 6.8  & -- & 9.3 & -- & 15.3 & -- & 11.5 & -- & 11.4 & -- & 16.6 & -- & 10.2 & -- \\
87     & 11.4 & 2.1 & 5.5 & 0.7 & 9.6 & 0.7 & 10.5 & 0.7 & 8.8 & 0.6 & 9.7 & 0.6 & 8.5 & 0.9 \\
109    & 11.0 & 1.0 & 8.7 & 0.7 & 9.4 & 0.7 & 12.9 & 0.8 & 10.1 & 0.6 & 9.9 & 1.0 & 12.2 & 0.9 \\
144    & 5.2  & 1.5 & -- & -- & -- & -- & 2.9 & 1.9 & 24.6 & 3.0 & -- & -- & -- & -- \\
97     & 5.5  & 2.3 & 14.0 & 2.7 & 5.3 & 4.4 & 6.1 & 1.5 & 2.0 & 2.0 & 7.2 & 2.2 & 11.9 & 2.0 \\
39     & 3.6  & -- & -- & -- & -- & -- & 4.7 & -- & 9.6 & -- & -- & -- & -- & -- \\
84     & 8.5  & 2.9 & 8.9 & -- & 3.4 & -- & 4.1 & 2.2 & 6.3 & 1.6 & 32.0 & -- & 14.2 & -- \\
161    & 28.3 & -- & 11.7 & -- & 1.5 & -- & 5.7 & -- & 2.7 & -- & 6.8 & -- & 12.6 & -- \\
\hline
\end{tabular}
\tablefoot{All EWs are in \AA. 
EW errors are obtained from bootstrap resampling of the spectrum (see Section \ref{sec:emission_line_measurements}) and account for the uncertainty in continuum fitting. If a line is not covered by the spectral range of muse, the EW is —. 
If a line is covered, but the continuum level around the considered line is below the error level, the $\Delta$EW is —.
In these cases, the EW is computed assuming continuum level = noise level and the EW presented should be considered as a lower limit.
} 
\end{table*}

Our analysis of emission line ratios demonstrates that the \citet{Xiao2018} models are able to reproduce the observed emission line ratios within the considered photo-ionisation parameter space.
Next we use \citet{Xiao2018} models to investigate if the observed \HeII\ EWs of the MUSE sample could be reproduced by BPASS models.

\begin{figure*}
\includegraphics[trim = 10 10 10 0, clip, scale=0.95]{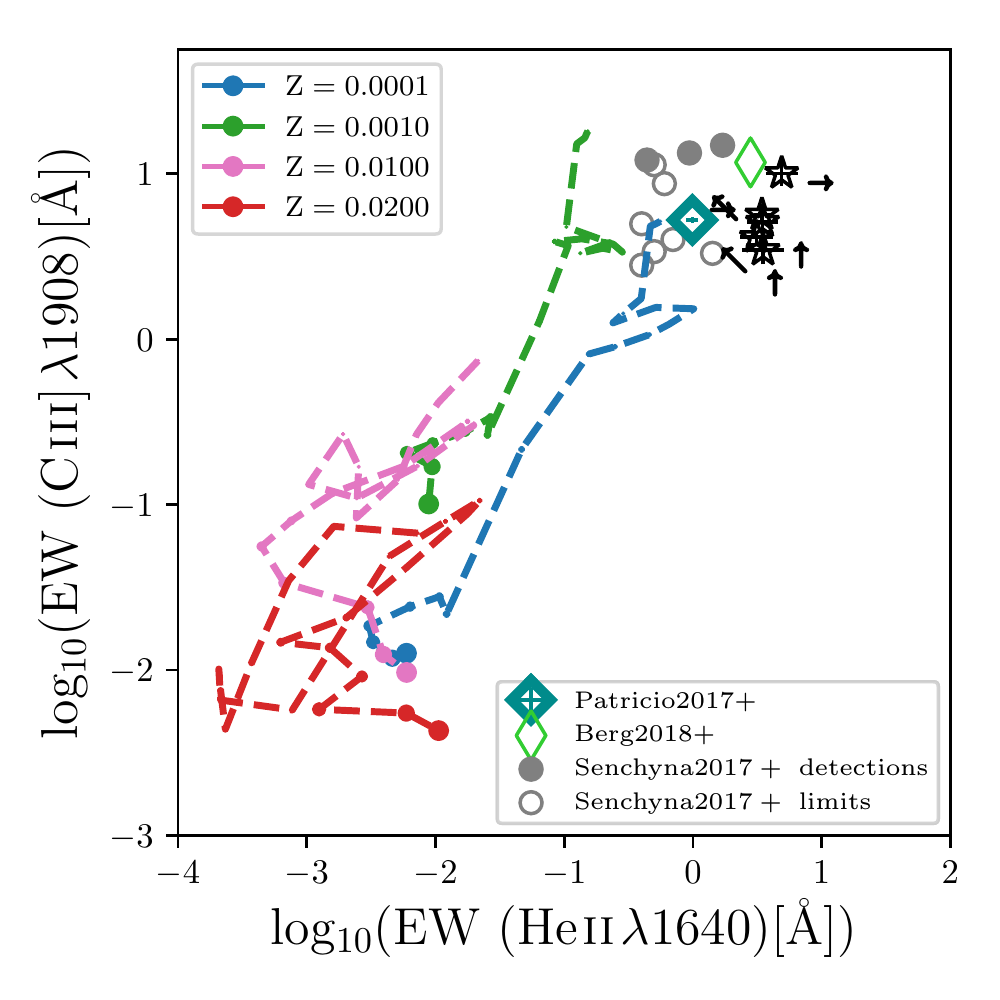}
\includegraphics[trim = 10 10 10 0, clip, scale=0.95]{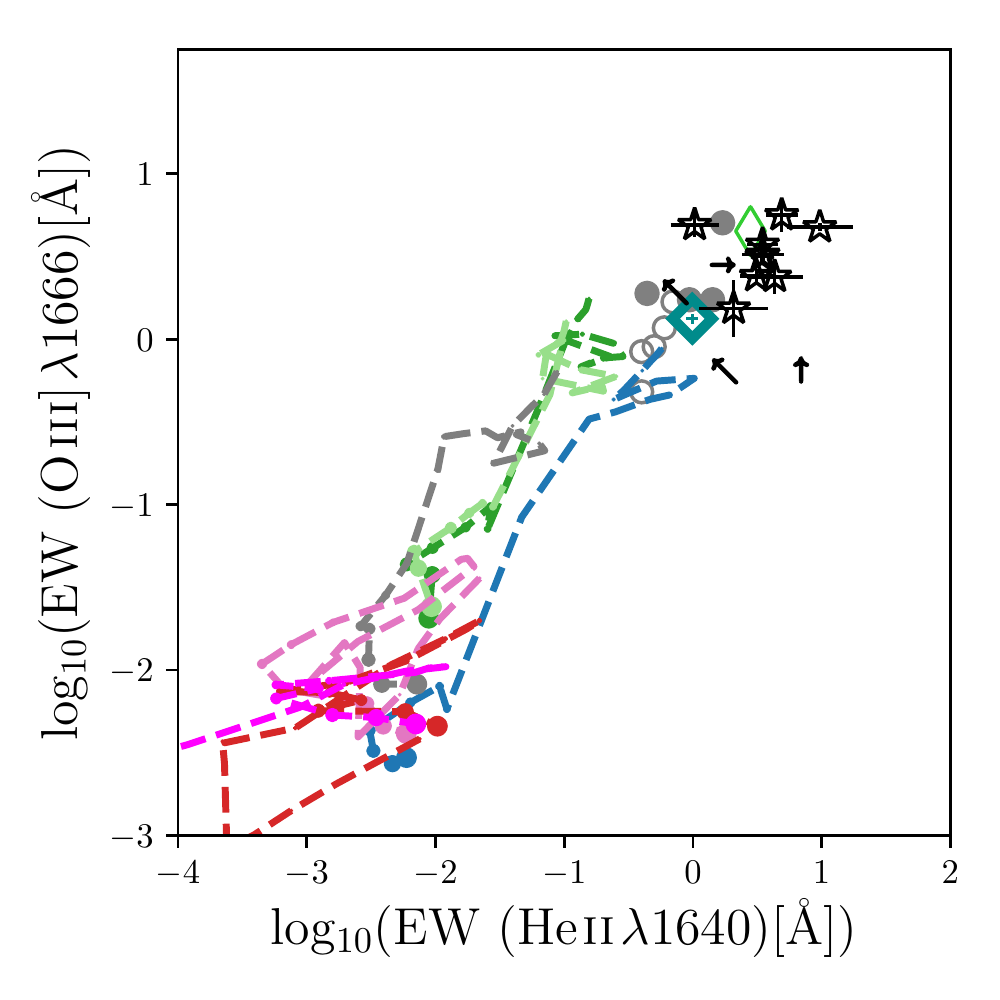}
\caption{ EW comparison of the MUSE \HeII\ sample using BPASS stellar population models. {\bf Left:} \CIII$\lambda1907$+\CIII$\lambda1909$ EW vs \HeII\ EW  and {\bf Right:} \OIII$\lambda1666$ EW vs \HeII\ EW are shown here. Galaxies with S/N$\geq2.5$ are shown by stars and others are shown as lower limits to the EW as triangles. We compare our observed EWs with model tracks from \citet{Xiao2018} BPASS binary tracks. Models are computed for a $log_{10}(n_H)=1.0$ and $U_{s}=-1.5$ at different metallicities between 2 \zsol\ to 1/200th \zsol. The size of the symbols increase with time. 
EWs from literature are also shown for comparison. [Section \ref{sec:model_comp_HeII_EW}]
\label{fig:ews_xiao}
}
\end{figure*}

We show the distribution of the \CIII\ EW vs \HeII\ EW and  \OIII$\lambda1666$ EW vs \HeII\ EW of the MUSE sample in Figure \ref{fig:ews_xiao}. 
Models are able to reproduce the \CIII\ EWs at very early times of the star-burst at high $U_{s}$ and low metallicities. 
However, the models are unable to reproduce the  \HeII\ and \OIII$\lambda1666$ EWs. 
This is in contrast to the ability of \citet{Xiao2018} models to reproduce observed rest-UV emission line ratios within the photo-ionisation model parameter space. 
Therefore, it is evident that the relative strength of \HeII\ compared to \CIII\ and \OIII\ is within the scope of model grids, however, the \HeII\ and \OIII\ flux to their respective rest-UV continuum at $\sim1640$ \AA\ and $\sim1666$ \AA\ is not.
Given the ionisation energy of  C$^{+}$ ($\sim24.38$eV) is relatively low compared to  \Hep, and O$^{+}$ ($\sim35.11$eV), it is likely that the lack of high energy ionisation photons drive the low \HeII\ and \OIII\ EWs in the \citet{Xiao2018} models at fixed C/O.

Next in Section \ref{sec:hep_production} we further discuss the ionisation photon production efficiency of the BPASS models.
We also note that spectro-photometric modeling by \citet{Berg2018} was able to model the \OIII\ doublet accurately but was unable to reproduce the \HeII\ emission. 
Therefore, additional constraints of the \emph{individual} stellar populations along with extra far-UV ionising photons are required to accurately predict the extra source of ionisation photons. 
\citet{Steidel2016} argue that core-collapse supernovae dominating at high-$z$ drives the ISM of $z\sim2$ galaxies to be O enriched with super-solar O/Fe \citep[also see][]{Matthee2018a}. 
Thus the stellar metallicity relevant to model the emission lines is lower than the gas phase metallicity, which results in an ionising spectrum that is harder resulting in a higher \OIII\ flux.

\subsection{Investigation of \Hep\ ionising photon production}
\label{sec:hep_production}

In this section we use the BPASS stellar population models to investigate their \Hep\ ionising photon production efficiencies and derive a simple calibration to investigate under what conditions the observed \HeII\ luminosities could be reproduced by the models. 

\begin{figure*}
\includegraphics[trim = 0 10 10 0, clip, scale=0.9]{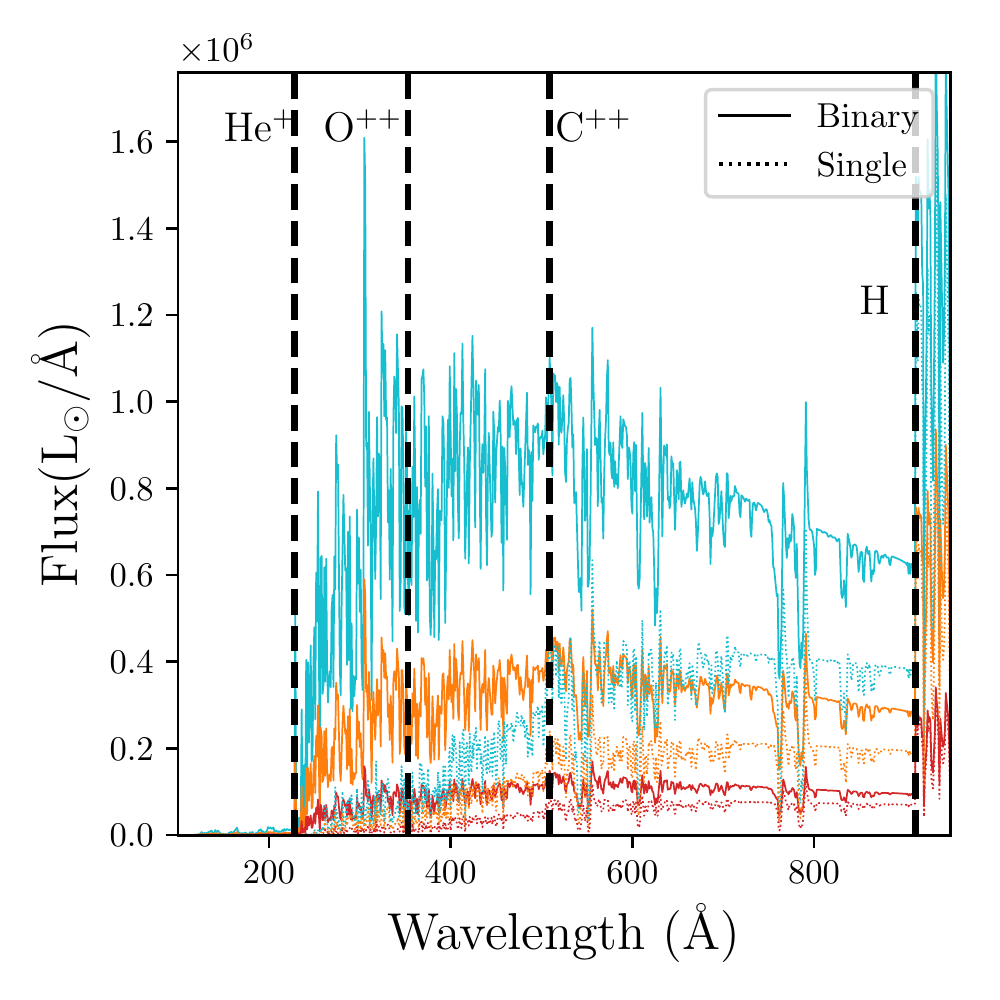}
\includegraphics[trim = 0 10 10 0, clip, scale=0.9]{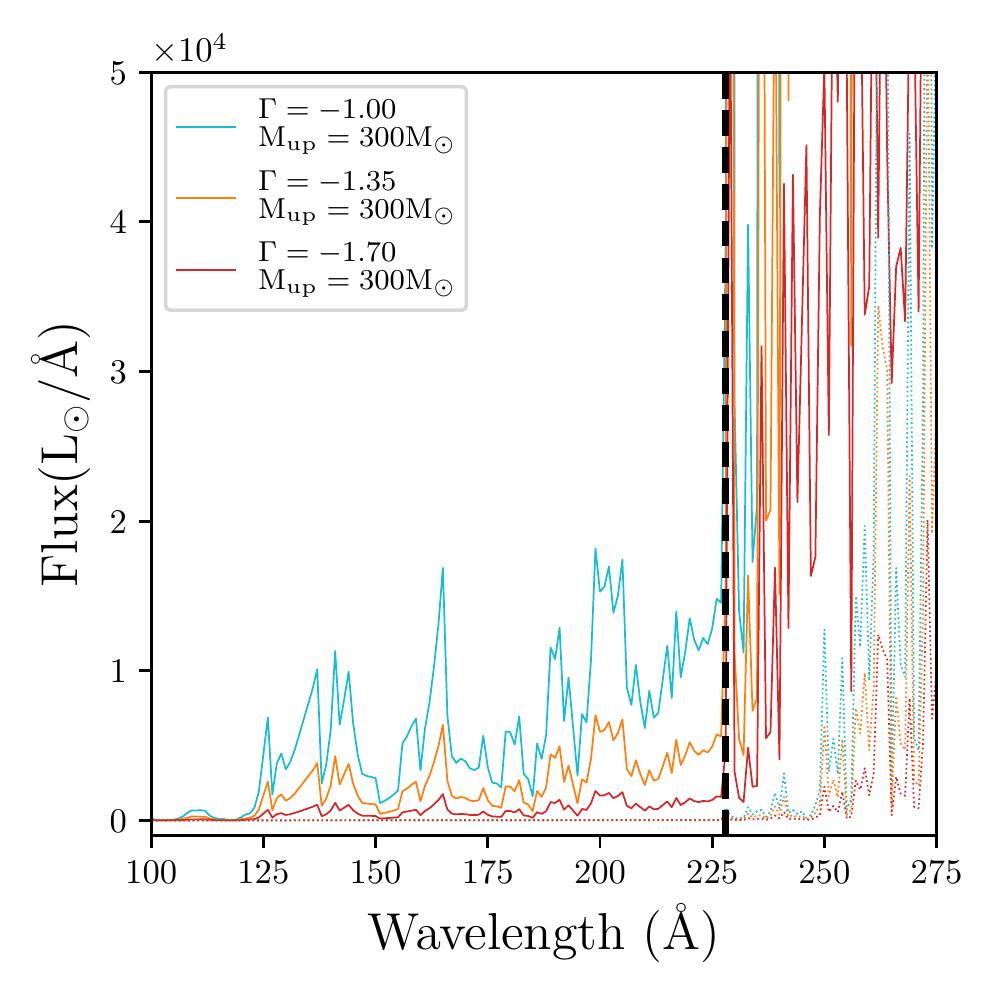}
\caption{{\bf Left:} Example BPASSv2.1 model spectra of a single burst stellar population after 5 Myr from the star-burst. Both single (dashed) and binary (continuous) model predictions are shown for different IMFs ($\Gamma=-1.0, -1.35, -1.70$) with a high mass IMF cutoff at 300\msol. The dashed black vertical lines mark $\mathrm{\lambda=228\AA, 353\AA,\ 508\AA,\ and\ 912\AA}$, below which \Hep, \Opp, \Cpp, and H  ionizing photons are produced. 
{\bf Right:} Zoomed in region $\lambda<275$\AA\, clearly showing the difference in flux around \Hep\ ionising limits. [Section \ref{sec:hep_production}]
\label{fig:bpass_Lyman_continuum_spectra}
}
\end{figure*}

In Figure \ref{fig:bpass_Lyman_continuum_spectra} we show the Lyman continuum spectra of the BPASS single and binary stellar models. Compared to single stellar populations, the effects of binary stellar evolution leads the Lyman continuum to increase substantially ($\times\gtrsim2$).
The Lyman continuum flux is driven by the young O and B stars and given their high temperatures, an increase in flux of $\sim400-600$ \AA\ is observed. At shorter wavelengths ($\lambda\lesssim300$\AA), the observed flux reduces rapidly, and hence between \Cpp\ and \Hep\ ionisation limits, the flux decreases by around one magnitude. 
However, we also note that our limited empirical constraints on far-UV spectroscopy of stars introduce additional uncertainties into stellar population modeling at this wavelength regime. 
Additionally, variations in the IMF also lead to an increase in Lyman continuum flux, which we discuss in Section \ref{sec:imf_variations}.

\begin{figure*}
\includegraphics[ scale=0.9]{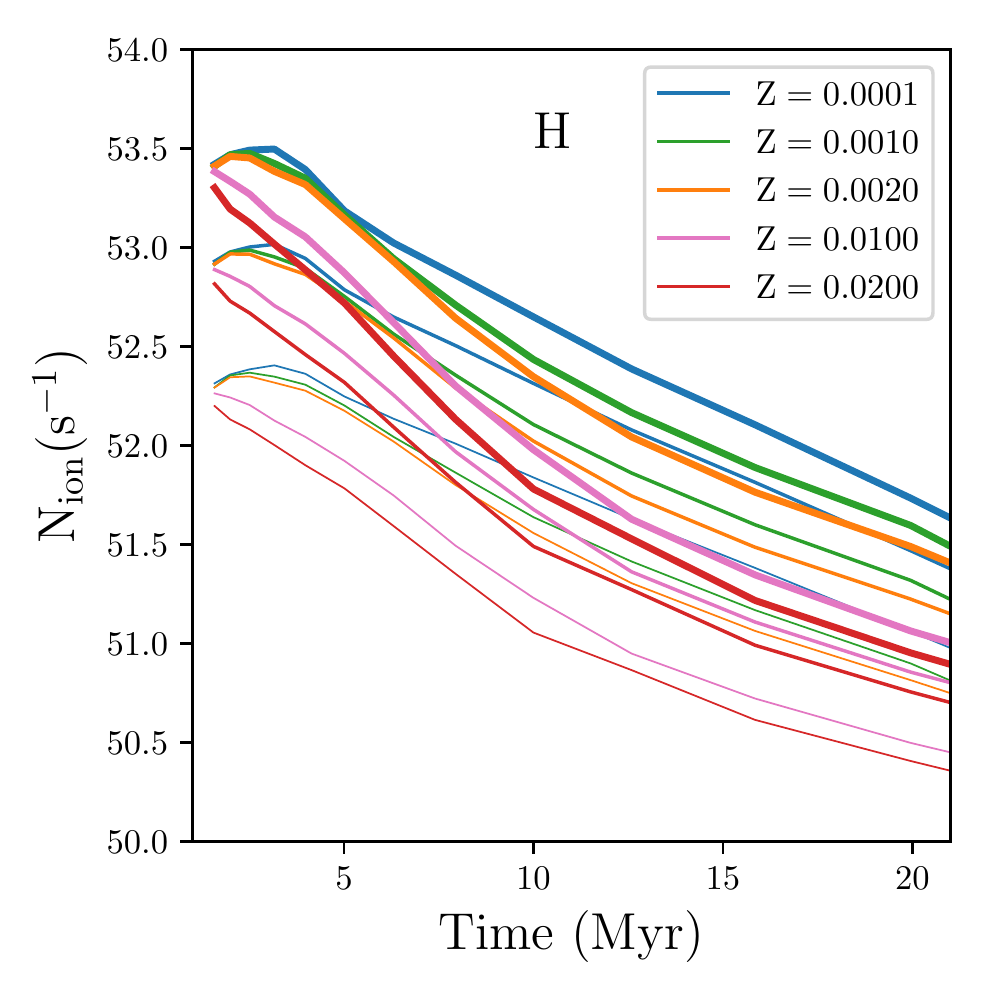}
\includegraphics[ scale=0.9]{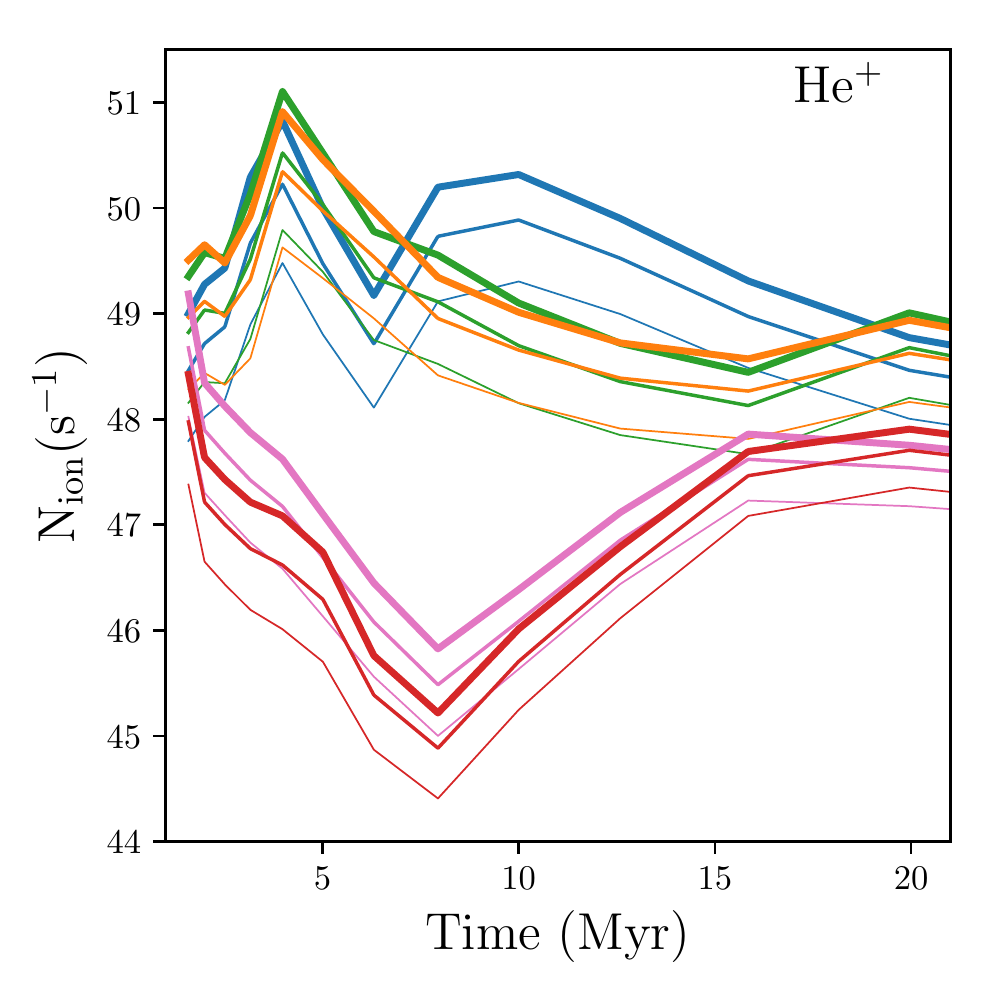}
\includegraphics[ scale=0.9]{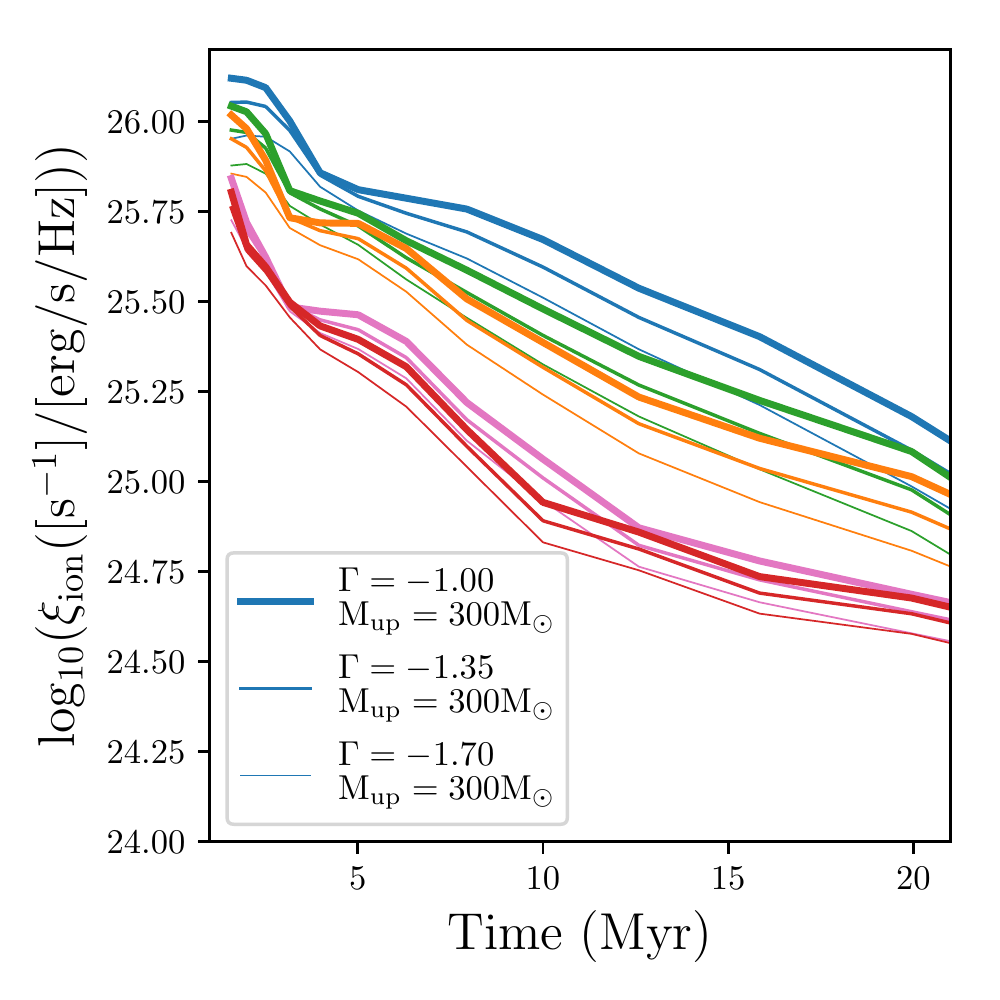}
\includegraphics[ scale=0.9]{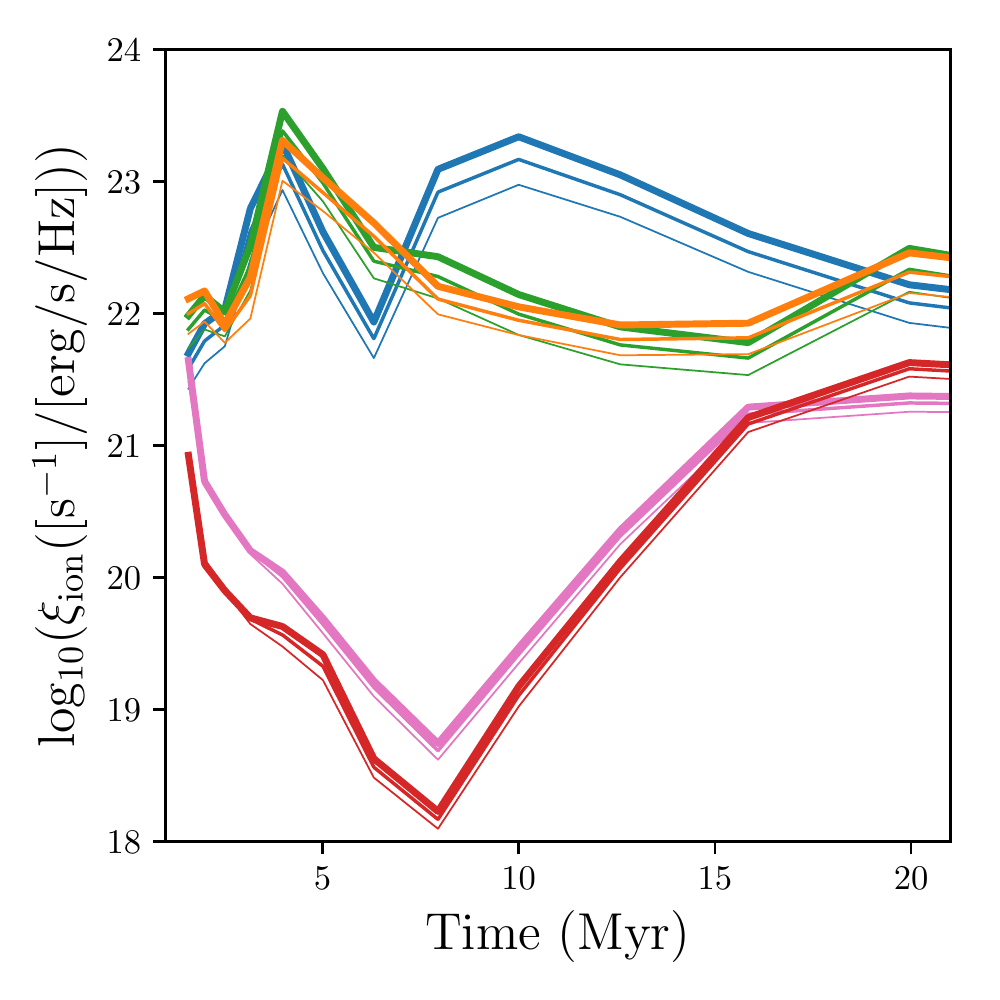}
\caption{{\bf Top panels:}  BPASS stellar population predictions for the evolution of the number of ionisation photons produced by a single instantaneous star-burst ($10^6$\msol) stellar population as a function of time. From {\bf left} to {\bf right} the panels show H and \Hep\ ionisation photons computed by integrating the spectra at $\mathrm{\lambda=228\ \AA\ and\ 912\ \AA}$, respectively.  
The models are computed at $Z=0.0001,0.001,0.002,0.01,0.02$ for binary models with different IMFs ($\Gamma=-1.0, -1.35, -1.70$) and IMF upper mass cutoffs (100 \msol\ and 300 \msol). 
{\bf Bottom panels:} Similar to the top panels but shows the evolution of $\xi_{ion}$ of H, and \Hep\ as a function of time. [Section \ref{sec:hep_production}]
\label{fig:bpass_model_predictions}
}
\end{figure*}

In Figure \ref{fig:bpass_model_predictions} we show the ionizing photon production efficiency of  BPASS models. 
For simplicity, we do not show the single stellar models in the figure, however, we note that binary models show a higher amount of photon production compared to their single stellar model counterparts. 
Thus, binary stellar evolution plays a vital role in producing ionizing photons for a prolonged time after a star-burst. 
We additionally investigate the time-evolution of $\xi_{ion}$ for H and \Hep\ in BPASS binary models. 
We define $\xi_{ion}$ for each element/ion as the Lyman continuum photon production efficiency above energies that could ionize the given element/ion which is computed as:
\begin{equation}
\label{eq:xi_ion}
\xi_{ion} = \frac{N(X)}{L_{UV}}
\end{equation}
where $N(X)$ is the ionizing photon production rate of the considered element/ion (in 1/s) and $L_{UV}$ is the luminosity at 1500\AA (in erg/s/Hz). 
Here we assume $f_{esc}=0$.
Both $N(X)$ as $\xi_{ion}$ are strongly sensitive to the metallicity, with lower metallicity models producing high values of $N(X)$ and $\xi_{ion}$.
As discussed is Section \ref{sec:model_comp_HeII_Xiao} (also see \citealt{Stanway2016,Eldridge2017,Xiao2018}), the two main effects of binaries with regard to production of ionising photons is to prolong the life time of massive O and B stars and enhance the production of W-R/Helium stars even at lower metallicities.

We further develop a simple prescription to investigate the difference in \HeII\ ionising photons between the observed data and the \citet{Xiao2018} model predictions.

\begin{figure*}
\includegraphics[ scale=0.60]{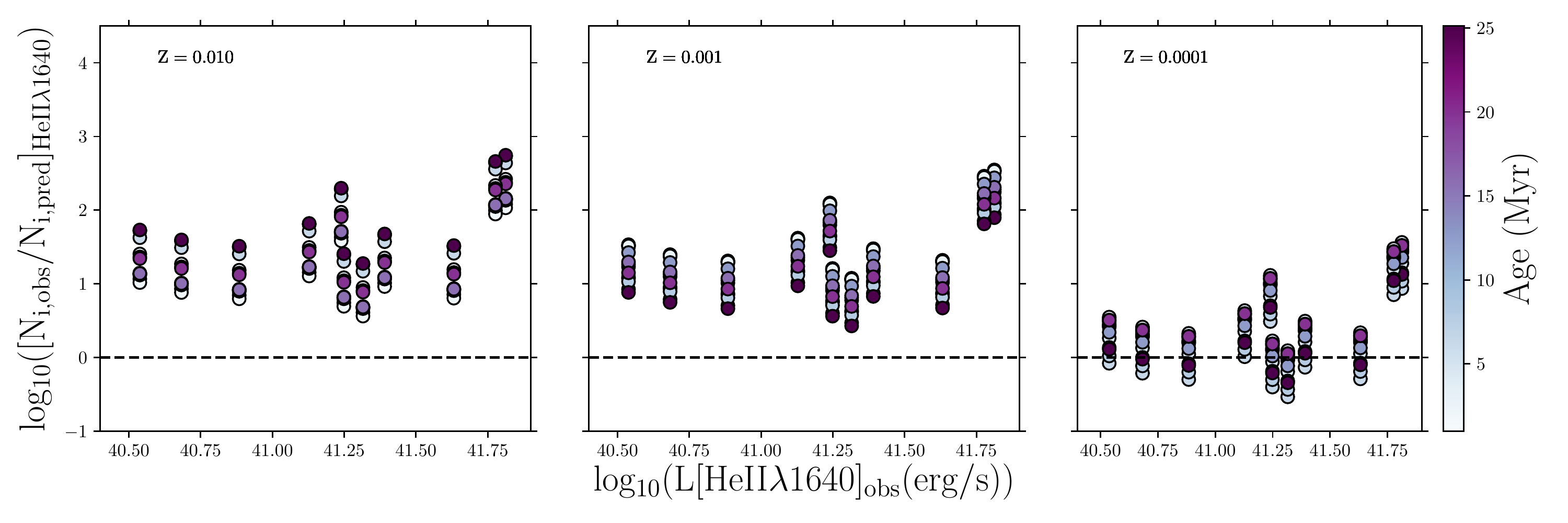}
\caption{The fraction of observed \Hep\ ionising photons compared to \citet{Xiao2018} model expectations as a function of observed \HeII\ luminosity of the MUSE \HeII\ sample. From {\bf left} to {\bf right}, we show the \Hep\ model predictions computed for three metallicities, $Z=0.01, 0.001, 0.0001$ with $log_{10}(n_H)=1.0$ and $U_{s}=-1.5$ for different times between $1-20$ Myr from the onset of the star-burst. The dashed horizontal line indicates y=0, where there is no difference between observations and model predictions. [Section \ref{sec:hep_production}] 
\label{fig:NHeII_def}
}
\end{figure*}

We compute a normalization constant (C), as:
\begin{equation}
\label{eq:cal_const}
C=\frac{L_{\CIII\ model}}{L_{\CIII\ data}}
\end{equation}
using the \CIII\ luminosities of the models and observed data. 
We use the calibration constant to compute the predicted \HeII\ luminosity from the models as, 
\begin{equation}
L_{\HeII pred}=\frac{L_{\HeII model}}{C}
\end{equation}
and obtain the approximate difference in \Hep\ ionising photons between observations and models assuming that $L_{\HeII}\propto N_{i,\HeII}$.  In Figure \ref{fig:NHeII_def} we show the fraction of observed \Hep\ ionising photons compared to the predictions from the models.
Only extreme sub-solar metallicities ($\sim1/200th$) are able to accurately predict the observed \Hep\ ionising photons. 
In Section \ref{sec:binary_effects} we discuss the mechanisms in binary models that drive extra production of ionising photons in binary stellar models and the role of metallicity in such models.

\section{Discussion}
\label{sec:discussion}

In this study we have presented a population of \HeII\ emitters from deep MUSE spectra obtained from a variety of spectroscopic surveys conducted by the MUSE consortium.  
By taking advantage of the other rest-UV emission lines with MUSE coverage, we have explored the stellar population/ISM properties of our sample.


\subsection{Uncertainties affecting our analysis}
\label{sec:uncertinities}

\subsubsection{Dust}
\label{sec:uncerternities_dust}

Our limited understanding of interstellar dust at high-redshift plays a role in our analysis of emission line properties in the rest-UV in many folds. 
Metal depletion and dust dissociation of galaxies play a role in the photo-ionisation models, with only a handful of models accounting for dust in chemical evolution models 
\citep[e.g.,][]{Gutkin2016,Gioannini2017}. Providing tight constraints for these parameters at high-$z$ requires a thorough understanding of element abundances, which is currently limited at high-$z$ due to observational constraints. We further discuss uncertainties related to this in Appendix \ref{appendix:dust_appendix}.

\begin{figure*}
\includegraphics[trim = 10 10 10 0, clip, scale=0.625]{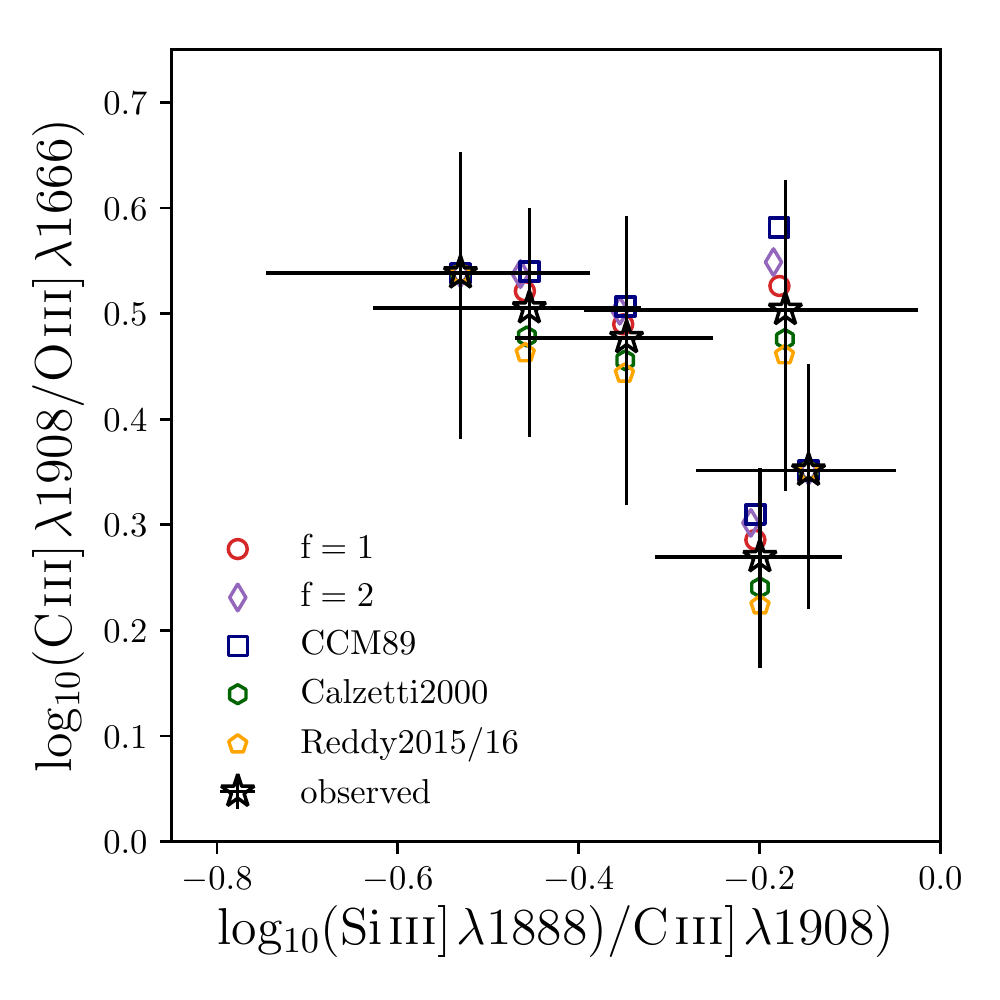}
\includegraphics[trim = 10 10 10 0, clip, scale=0.625]{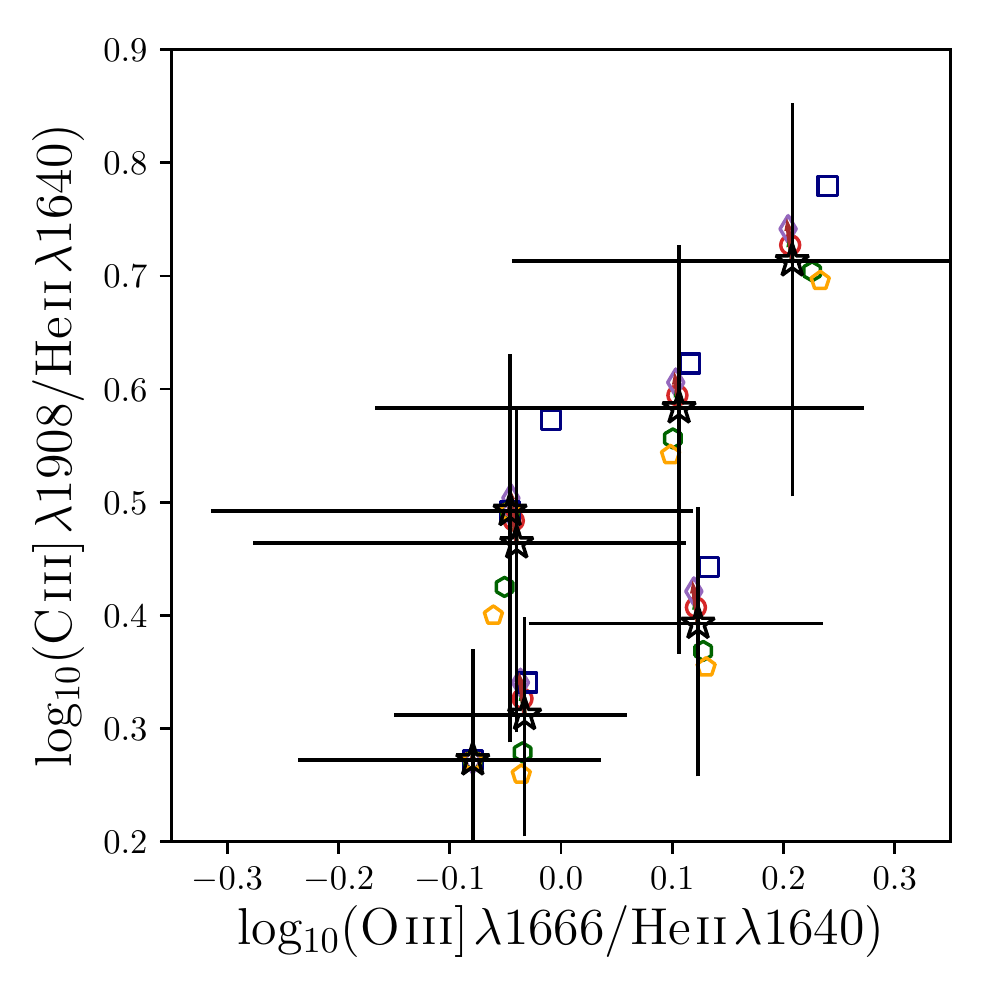}
\includegraphics[trim = 10 10 10 0, clip, scale=0.625]{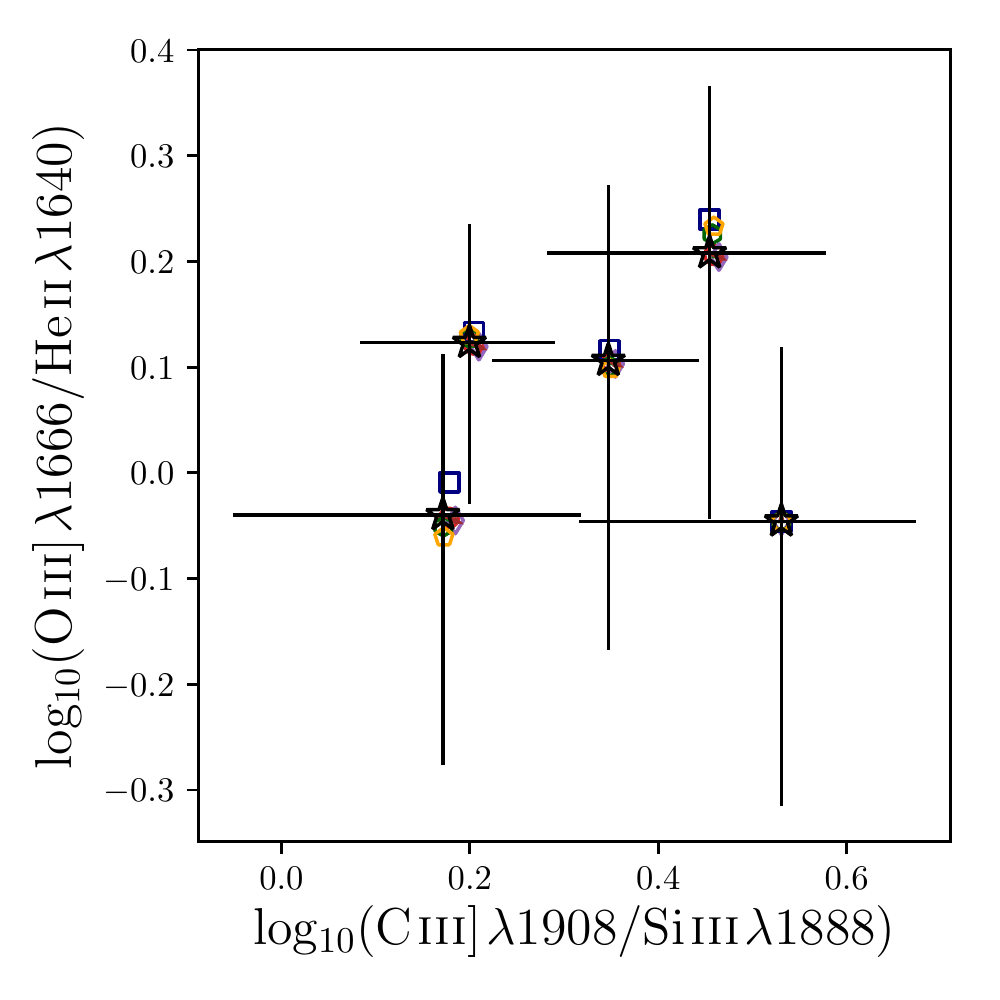}
\caption{Rest-frame UV emission line ratios of the MUSE \HeII\ sample computed with different dust laws. 
The panels are similar to Figure \ref{fig:line_ratios_gutkin}, however, \citet{Gutkin2016} models are removed for clarity. 
Only galaxies with S/N$\geq2.5$ for emission line considered in each panel are shown in the figure. 
Observed emission line ratios are shown by stars. $f=1$ and $f=2$ resemble dust corrections applied considering no extra attenuation and $\times2$ extra attenuation for ionized gas regions (computed using \citet{Cardelli1989} attenuation law) compared to stars (computed using \citet{Calzetti2000} attenuation law). We further show the distribution of the line ratios using \citet{Cardelli1989}, \citet{Calzetti2000}, and \citet{Reddy2015,Reddy2016} attenuation laws. [Section \ref{sec:uncerternities_dust}]
\label{fig:line_ratios_dust}
}
\end{figure*}

In addition to the parameters related to photo-ionisation modeling, dust attenuation of the observed spectra introduce additional complexities when interpreting observed emission lines. 
If nebular emission has systematically higher attenuation, line flux values will change significantly ($\sim5\%-80\%$), however, line ratio diagnostics will be significantly less impacted. 
We show this in Figure \ref{fig:line_ratios_dust} where we compare the observed emission line ratios with dust corrections applied using different attenuation laws and different extinction between stellar and ionized gas regions. 
Using the \citet{Calzetti2000} attenuation law for the continuum and \citet{Cardelli1989} attenuation law for the nebular emission lines, we derive dust corrected emission line flux ratios for our \HeII\ sample considering (i) no difference in extinction between stellar and ionized gas regions (ii) ionized gas regions are twice as extincted compared to stellar regions. Figure \ref{fig:line_ratios_dust} show that the change in emission line ratios between (i) and (ii) are quite modest and are within the error limits of the line fluxes. 
We further show the difference in dust corrected emission line flux ratios between \citet{Cardelli1989}, \citet{Calzetti2000}, and \citet{Reddy2015,Reddy2016}. Regardless of the attenuation law most galaxies lie within the line flux measurement errors. The significant outliers in \CIII/\OIII\ vs \SiIII/\CIII\ and  \CIII/\HeII\ vs \OIII/\HeII\ line ratios are primarily driven by the variations of the \HeII\ fit performed on the spectra once dust corrections are applied using different attenuation laws.

In this analysis we completely ignore the fact that the \Av\ values of our are sample are obtained through either SED fitting or $\beta$, which are calibrated to a certain dust attenuation law and stellar population models. Therefore, a more accurate treatment of dust require recalibration of attenuation laws with a variety of stellar population models \citep[e.g.,][]{Reddy2018,Theios2018} and is out of scope of this work. 
However, we show that to first order  for the rest-UV emission line ratios considered in our analysis, dust correction does not have a significant effect, and that only observed outliers are driven by variations introduced by wavelength dependent broadening of emission lines.

\subsubsection{S/N and line fitting}
\label{sec:uncerternities_snr}

The low S/N of the observed spectra of our sample affects our analysis via (i) uncertainties associated with the continuum fitting process and (ii) weak emission line strengths compared to the continuum level.  
In our analysis, we have completely ignored the uncertainties associated with the continuum fitting process. As shown by Figure \ref{fig:HeII_spectra_1} and \ref{fig:HeII_spectra_2}, the continuum levels of a majority of our galaxies are less than $3\sigma$ of the noise level. 
Therefore, despite the fact that visual identification of emission lines show clear features, line fluxes, which are measured by subtracting the continuum from the spectra may have larger uncertainties.

To quantify the low significance of the emission lines compared to the continuum level and uncertainties associated with the continuum fitting, we perform a bootstrap resampling analysis of the spectra. 
For each spectrum, we randomly resample each pixel flux value with a Gaussian distribution around $\pm\sigma$ of the error level of that pixel. We then refit the continuum and measure the line fluxes and perform this iteratively 100 times. 
We consider the median value of the line flux distribution as the line flux and the standard  deviation of the measured values as the associated error level of the line flux. Out of the 13 galaxies identified with \HeII\ detections, we find that though all \HeII\ line fluxes are measured at $>2\sigma$ five of the galaxies fail to make a S/N$\geq2.5$ cut for \HeII\ emission and only four galaxies are detected with S/N$\geq3$. 
Therefore, we conclude that the low S/N of data is a non-negligible uncertainty of our analysis and we require deeper integrations to constrain the continuum of galaxies with greater significance.

Additionally, the method that we implemented to obtain the \HeII\ fluxes may give rise to uncertainties associated with the emission line fitting algorithm. 
As we discussed in Section \ref{sec:emission_line_measurements}, we fit the \HeII\ emission line width using a single gaussian parametric fit allowing more freedom compared to the other emission lines. 
We opt for this approach in recognition of the fact that the \HeII\ can originate from a multitude of processes (see Section \ref{sec:origion_of_HeII}).
However, our photo-ionisation model comparisons assume that the nature of \HeII\ is purely nebular. 
Thus it is necessary to investigate how allowing more flexibility in the fit affects the \HeII\  flux measured on the galaxy spectra \citep{Brinchmann2008}.

\begin{figure*}
\includegraphics[trim = 0 0 0 0, clip, scale=0.9]{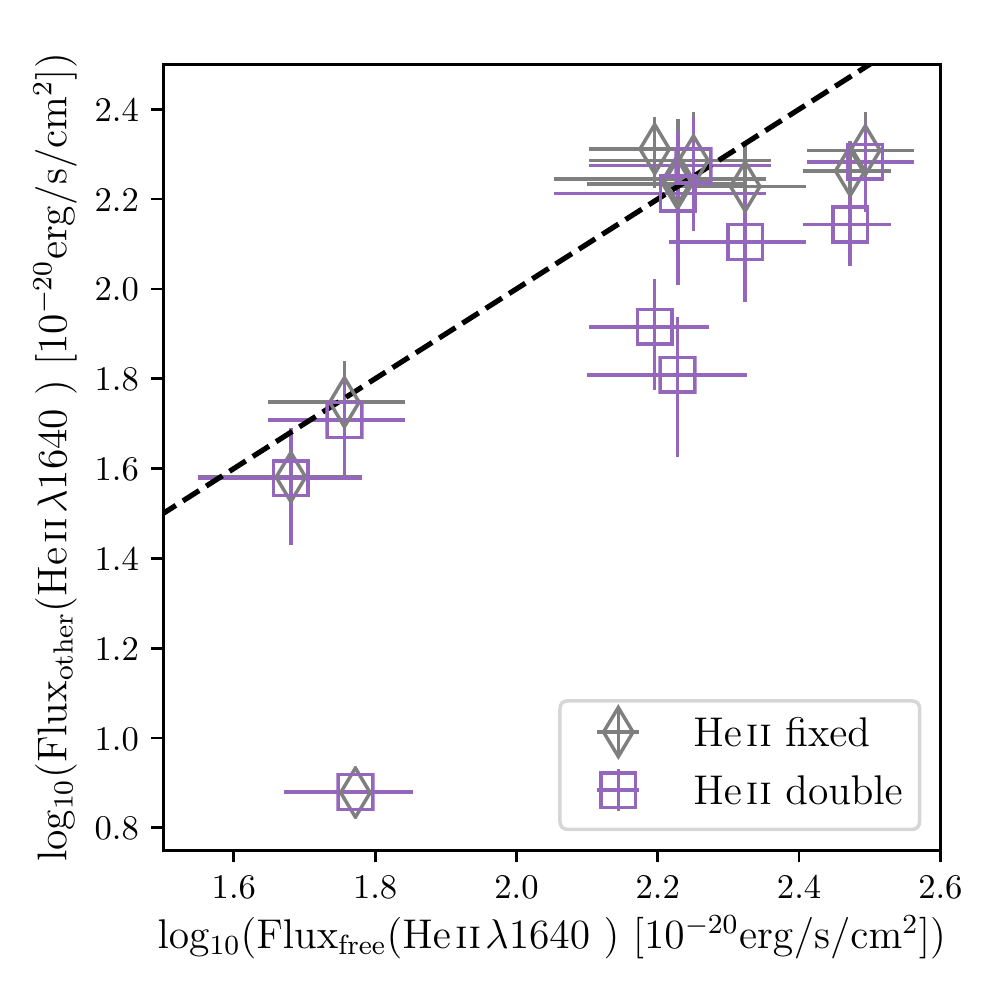}
\includegraphics[trim = 0 0 0 0, clip, scale=0.9]{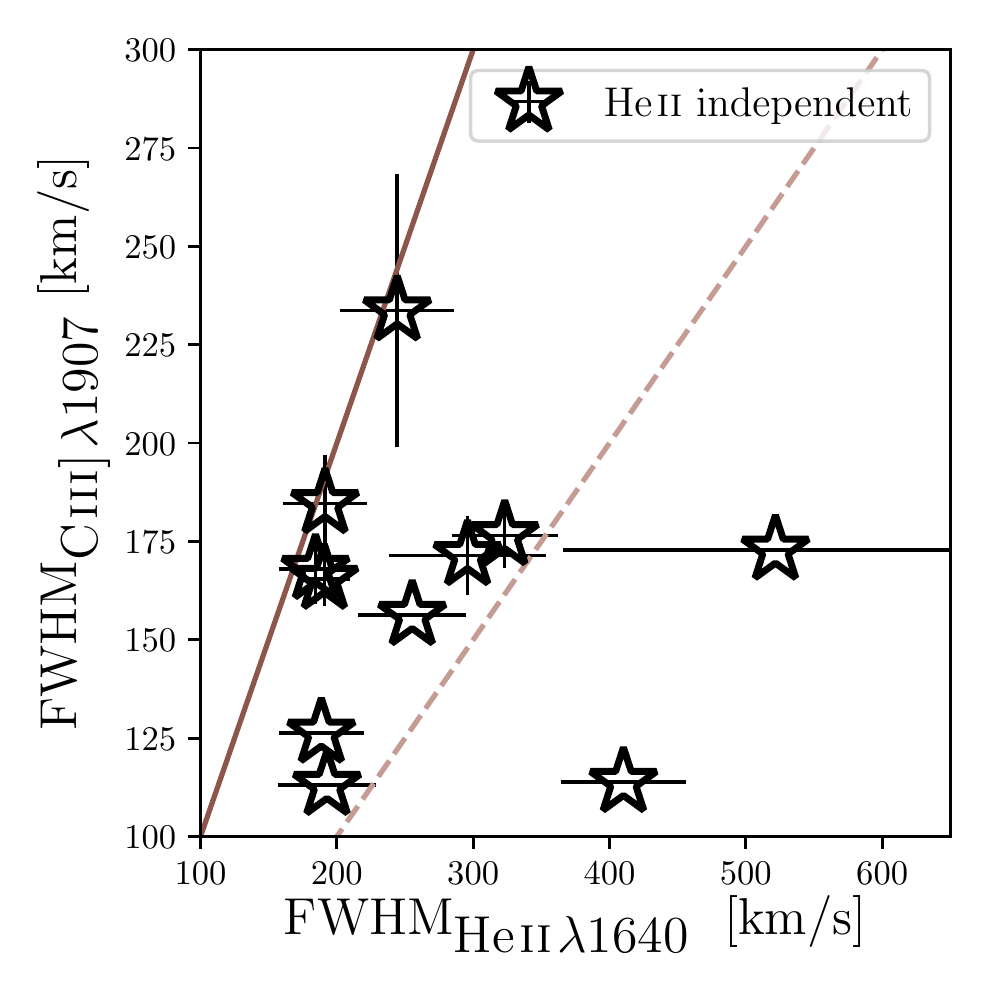}
\caption{ {\bf Left:} Comparison of \HeII\ line flux measurements obtained for different \HeII\ line flux parameterizations. We compare the \HeII\ line flux measurements of galaxies with a S/N $> 3$ obtained using an independently fit \HeII\ single gaussian fit  with measurements obtained by 1) fitting \HeII\ using a single gaussian with line width fixed with the other emission lines (diamonds) and 2) a double gaussian profile with the one component line width fixed with the other emission lines (stars). For the double gaussian fit, we only consider the nebular component for comparison.  
{\bf Right:} Comparison of FWHM of \CIII$\lambda1907$ with FWHM of \HeII\ fit independently of the other nebular emission lines. The solid lines denotes $x=y$ and the dashed denotes $x= 2y$. 
[Section \ref{sec:uncerternities_snr}]
\label{fig:line_fits_difference}
}
\end{figure*}

In Figure \ref{fig:line_fits_difference} we show a comparison of \HeII\ line flux measurements between different line fitting methods. We use the independently fit \HeII\ single gaussian fit as the base line and compare with measurements obtained by 1) fitting \HeII\ using a single gaussian with line centre and width fixed with the other emission lines and 2) a double gaussian profile with the one component line centre and width fixed with the other emission lines. 
For spectra with multiple emission line detections, once the \HeII\ line centre and width is fixed with the other emission lines, there is a tendency for \HeII\ flux  to be underestimated by  $\sim20\% \pm 27\%$ compared to the independently fit \HeII. 
Similarly, with a multi-gaussian fit, the difference in flux for the nebular component is much greater with an observed underestimation of flux $\sim36\%\pm24\%$. 
In all cases, the \HeII\ fit performed independently of the other emission lines performs better in obtaining a better fit to the observed emission line.
Ambiguities associated with the line fitting is an inherent uncertainty in our analysis, and only high S/N emission line detections of weaker rest-UV nebular emission line features will grant stronger constraints on the nebular component of the  \HeII\ features.

Considering \HeII\ line width to be independent of other nebular emission lines results in a systematic difference between line width velocities of  \HeII\ with the other nebular emission lines.  
In Figure \ref{fig:line_fits_difference} we compare the line FWHM of the gaussian fits of \CIII$\lambda1907$ and \HeII\ emission lines. 
All our galaxies show \HeII\ FWHM to be higher than that of \CIII$\lambda1907$.
Since \CIII$\lambda1907$ is purely driven by the nebular emission, the difference in line velocities suggest that \HeII\ may also have a contribution from a different source. Given low-S/N of our data, we are refrain from over interpreting this result but we discuss possible origins for a narrow stellar driven \HeII\ component in Section \ref{sec:WR-stars}.

\subsubsection{Stellar population models}
\label{sec:uncerternities_ssp_models}

Stellar population models used to infer ISM properties of our \HeII\ sample contributes to uncertainties in interpreting the observed emission line ratios. 
In Section \ref{sec:model_comp} we show that the \citet{Gutkin2016} models, which do not account for the effects of stellar rotation or binary stars but have self-consistent treatment of element abundances and depletion on to dust grains, show different emission line ratios compared to \citet{Xiao2018} models that incorporate effects of binary stellar evolution.

To obtain stronger constraints on the underlying stellar populations and the ISM, nebular emission lines should be jointly used with rest-UV/optical stellar and ISM (neutral + ionized, e.g.,  \citealt{Vidal-Garca2017}) absorption lines for comparison with predictions from the stellar population models. Recently there has been a number of advanced full spectral fitting algorithms for stellar population models developed to perform full spectro-photometric analysis of galaxies \citep[e.g.,][]{Chevallard2016,Leja2017}. 
In the local Universe, \citet{Senchyna2017} showed that at low metallicity full spectral fitting fails to accurately predict the observed \HeII\ features using models that does not include stellar rotation or binaries.

At high redshift, most rest-UV studies suffer strong observational constraints due to the low S/N of the continua of galaxy spectra. 
To overcome the low S/N, studies have attempted spectral stacking techniques and gravitational lensing to obtain rest-UV spectra with high-S/N. 
\citet{Steidel2016} used STARBURST99 \citep{Leitherer1999} and BPASSv2 models to obtain a best-fit spectral model for a stacked composite spectra at $z\sim2$. Their results demonstrated that the best fit models showed considerable difference between  STARBURST99 and BPASS for various spectral features, e.g., stellar wind features lacked any metallicity dependence in BPASS models. 
The analysis of \citet{Steidel2016} highlighted an important aspect with regard to \HeII: the observed \HeII\ feature of the stacked spectrum was completely attributed to a stellar origin from BPASS while STARBURST99 suggested a purely nebular feature arising from \HII\ regions.  
However, \citet{Steidel2016} did not investigate if the Lyman continuum photons from the STARBURST99 best fit SED were sufficient to produce the observed \HeII\ feature.
Using a gravitational lensed galaxy at $z\sim2$ \citet{Berg2018} showed that all the observed emission lines except \HeII\ can be best-fit by a BPASS stellar population model. 
We note that \citet{Berg2018} galaxy has similar rest-UV emission line ratios compared to galaxies in our sample (e.g., see Figure \ref{fig:line_ratios_gutkin}).  
The necessity for binary models to explain high-$z$ observed spectral features (e.g., \CIII\ EW) has also been demonstrated by \citet{Jaskot2016}, but with the caveat of being unable to reproduce the \HeIIoptical/\Hbeta\ ratios of $z\sim0$ galaxies.

Therefore, even the latest generation of \citet{Bruzual2003} or BPASS stellar population models are currently unable to accurately predict the observed \HeII\ features. 
Due to such complications in stellar population models, cross validation using a multitude of spectral diagnostics is imperative to make strong conclusions about the ISM and stellar conditions of high redshift galaxies.  
We discuss the effects of binaries on \HeII\ emission in further detail in Section \ref{sec:binary_effects} and defer a full spectral fitting analysis of stacked spectra from MUSE to a future study.


\subsection{The origin of \HeII}
\label{sec:origion_of_HeII}

Multiple mechanisms are currently being used to describe the origin of the \He\ emission line (refer to \citet{Shirazi2012} and \citet{Senchyna2017} for a detailed discussion). Here we explore whether we can rule in favour or against of any of such mechanisms. However, we note that the lack of rest-frame optical coverage of our sample hinders making strong conclusions of the origin of \He.


\paragraph{AGN:}\label{sec:AGN_contamination}
Rapidly accreting supermassive black holes release high energy photons to the surrounding environment \citep[e.g.,][]{Kormendy1995,Magorrian1998}, which contributed to strong rest-frame UV/optical nebular emission lines \citep{Feltre2016}. 
In our sample we identify three galaxies with possible strong AGN contribution based on X-ray detections and enhanced line-widths. All three of these galaxies show broad \CIV\ emission features that clearly distinguish them from stellar ionisation sources. 
We demonstrated in Figure \ref{fig:line_ratios_gutkin} that our \HeII\ sample emission line ratios  completely fall within the region powered by star-formation and additionally, the line ratios of our sample  does not fall within the AGN segment of \citet{Feltre2016}.
However, we do not rule out effects of sub-dominant AGN, which may still contribute to the ionization processes of our sample but to a lesser degree compared to stellar sources. 
Thus it is possible that at least some of the \HeII\ emission of our sample to be arisen by AGN. 

\paragraph{Radiative shocks:}\label{sec:shocks}
Radiative shocks in galaxies also contribute to ionizing photons capable of producing \HeII\ \citep[e.g.,][]{Thuan2005,Allen2008,Jaskot2016} but are only expected to be dominant at higher metallicities \citep{Shirazi2012}. However, metallicity correlations are yet to be tested with newer generation of stellar population models incorporating advanced treatment of W-R stellar evolution (Charlot \& Bruzual, in preparation), and binaries/stellar rotation (e.g., BPASS).  
Radiative shocks could also contribute to spatial offsets between \HeII\ emission and the continuum \citep{Thuan2005}

In order to accurately distinguish whether shocks (or even AGN) play a dominant role in producing ionizing photons we require multiple emission line diagnostics or high S/N emission lines with broad components.
Using emission line diagnostics alone to differentiate shocks/AGN from star-formation activity requires caution at high redshift, given significant differences in the ISM of high redshift galaxies compared to local star-forming galaxies \citep[e.g.,][]{Steidel2014,Kewley2016,Strom2017}. 
Recent studies demonstrate that the capability to decompose narrow and broad components of observed emission lines is correlated with the S/N \citep{Freeman2017}. 
Thus, at lower SNRs the contribution to an emission line from star-formation and shocks would become degenerate. 
If shocks are correlated with galactic outflows, at lower masses the outflow mass per SFR will be higher \citep{Muratov2015} and thus if SFR $\propto$ L(\HeII) \citep{Schaerer2003}, it is plausible for the strongest \HeII\ emitting low mass galaxies to have a larger shock contribution to \HeII. 
Given the \HeII\ S/N of our sample is $<10$, we are unable to perform a meaningful study on the individual \HeII\ detections to distinguish between star-formation and AGN/shocks and thus refrain from fitting multiple Gaussian components to \HeII\ to identify broader emission components. 
Additional rest-frame optical emission line diagnostics couple with high S/N data from MUSE/\emph{JWST} will be crucial to distinguish the contribution of shocks to the emitted spectrum.

\paragraph{Wolf-Rayet stars:}\label{sec:WR-stars} 
Strong winds driven by the powerful radiation pressure in the W-R stars results in characteristic broad emission lines, thus, star-forming galaxies with a significant population of W-R stars will show a composite of broad and narrow \He\ features \citep{Crowther2007}. 
At low metallicities stellar winds will be weaker, increasing the relative efficiency of stellar rotation and mass transfer between binaries \citep{Eldridge2009,Szecsi2015}.
Thus, W-R stars in metal poor galaxies produce \Hep\ ionizing photons without broad W-R features that are characteristic of high metallicity W-R stars \citep{Schmutz1992,Crowther2006,Grafener2015}. 
The nebular \He\ components in $z\sim0$ galaxies show a very prominent transition to high \HeIIoptical/\Hbeta\ ratios as a function of metallicity, with low metallicity systems requiring up to an order of magnitude higher \Hep\ photons \citep{Brinchmann2008,Shirazi2012,Senchyna2017}.
WC stars are formed in binary systems around more luminous O stars, but are hotter but bolometrically fainter than typical core-He burning W-R stars which contribute to high energy ionizing photons while being `observationally invisible' \citep{McClelland2016}.  
Additionally, very massive low-Z WNh stars (hydrogen rich WN stars) produce narrow \HeII\ emission features of $\gtrsim 300\  km/s$ with no other accompanying features \citep{Grafener2015}. 
In Figure \ref{fig:line_fits_difference}, we show that \HeII\ line widths of our sample are $\sim200-400\ km/s$ and thus it is plausible for a subset of our \HeII\ emitters to be powered by stellar emission. 
Here, the \HeII\ line-width is a direct proxy for the velocities of the stellar winds, which are considerably weaker at low-Z, thus moderate to high resolution spectra with high S/N are required to accurately extract nebular and stellar components of the \HeII\ emission \citep[e.g.,][]{Senchyna2017}. 
As we discuss in Section \ref{sec:uncerternities_snr}, we require high S/N data with multiple confident rest-UV emission line detections to accurately distinguish between different \HeII\ mechanisms that produce comparable emission line features. We note that we remove one galaxy from our \HeII\ sample (see Section \ref{sec:sample_selection_muse_udf}) due to its broad \HeII\ feature with a FWHM $>1000\ km/s$, which can also be explained by WN type stars at $\sim0.5$ \zsol\ \citep{Grafener2015}.
We can rule out high metallicity W-R stars with broader winds to be a strong contributor to our \HeII\ sample, however, we cannot completely rule out the presence of low metallicity WN type stars, which may become more prominent at lower metallicities once effects of binaries are considered.

\paragraph{Effects of binary interactions:}\label{sec:binary_effects}
Effects of binaries have shown to play a crucial role in increasing the ionising photon production in young stellar systems through multiple processes \citep{Eldridge2017} and may contribute to alleviate the  tension between models and data for \He\ emitters observed in low metallicity systems \citep{Shirazi2012,Senchyna2017}. 
In contrast to stellar population models that incorporate advanced treatments of stellar rotation  \citep{Leitherer2014}, driven by large degeneracies between effects of stellar rotation and binaries BPASSv2 stellar population models implement a simplified approach to consider effects of stellar rotation \citep{Eldridge2017}. 
Nonetheless, it is important to consider the effects of binaries and stellar rotation together, since it has been shown that rapid rotation in stars may only arise due to binary interactions \citep[e.g.,][]{deMink2013}.  
At $Z\leq0.004$, BPASS models could generate up to $>0.1$ dex more ionizing photons from effects of QHE alone (see Figure 6 of \citet{Stanway2016}, also see Figure \ref{fig:bpass_model_predictions} of this paper). A physically motivated gradual transition of QHE effects as a function of metallicity may provide further constraints to balance the production of \Hep\ photons with observed W-R features.

Regarding the production of \He\ ionizing photons, studies have shown that even BPASS models with effects of binaries do not produce sufficient \Hep\ ionizing photons to consistently model \He\ with other observed emission lines  \citep{Jaskot2016,Berg2018}. 
In Figure \ref{fig:bpass_model_predictions} we show that compared to the H ionising photons, \Hep\ ionising photon production rate is only mildly sensitive to the age of the stellar population within $\sim20$ Myr from the star-burst. This is driven by the contribution of older W-R stars after the first few million years from the onset of the star-burst. 
The effect of QHE can be clearly seen here, where the lower metallicity models produce significantly higher amounts of ionizing photons compared to models with $Z>0.004$. 
Therefore, in BPASS the contribution to \Hep\ ionising photons is primarily driven by the extra production of W-R stars at lower masses/metallicities due to binary interactions and is insufficient to match with observed correlations. 
Compared to \Hep\ at all times the \Cpp\ ionising photon production rate is higher (also see Figure \ref{fig:bpass_Lyman_continuum_spectra}), and as we show in our analysis, BPASS models accurately predict observed \CIII\ EWs but not higher energy \OIII\ and \HeII\ EWs. 
The lack of \Hep\ ionizing photons in BPASS could be driven by the white dwarf treatment where BPASS asymptotic giant branch (AGB) stellar models do not produce white dwarfs in the single star mode, while in the binary mode the stars do not evolve up to the AGB phase driven due to effects of binary interactions before the second dredge-up \citep{Eldridge2017}. 
More advanced treatment of white-dwarf production within BPASS and consideration of emission from white-dwarf accretion disks \citep{Woods2016} may contribute to an enhancement of \Hep\ ionizing photons within BPASS to account for the deficiency compared to observations.

\paragraph{X-ray binaries:}\label{sec:x_ray_binaries} 
have shown to produce hard ionizing photons capable of ionizing \Hep\ \citep[e.g.,][]{Garnett1991}. 
A black hole or neutron star/pulsar that is accreting material from their companion O/B star will undergo heating by the strong X-ray ionizing photons produced during accretion. 
The strength of the X-binary binary is expected to be determined by the mass transfer rate of the secondary star, the magnetic field strength of the compact source, and by the X-ray luminosity which contributes to the heating of the accretion disk \citep{Casares2017}.   
At fixed SFR lower metallicity systems are found to have high X-ray luminosity, thus at lower metallicities X-ray binaries may play a larger role in producing \Hep\ ionizing photons \citep{Brorby2016,Schaerer2019}. 
Low redshift rest-UV studies have either not shown any strong evidence for X-ray point source detections in strong \He\ emitters or show spatial offsets between \HeII\ detections and the X-ray sources \citep{Thuan2005,Kehrig2011,Kehrig2015,Shirazi2012,Senchyna2017}.    
Additionally, X-rays will have extremely shallow optical depth and thus the extent to which the high energy photons can influence the production of \Hep\ ionizing photons is unclear. 
However, X-ray photons would in principle degrade to lower energy photons and ionize \Hep\ over larger optical depths, which may be able to explain some of the spatial offsets. 
For our sample at $z\sim3$, limited by strong observational constraints of weak X-ray features at high-redshift we are unable to determine the role of X-ray binaries, if any, in producing \HeII.

\paragraph{Stripped stars:}\label{sec:stripped_stars}
Recent studies have shown that low mass stars in the presence of binary companions undergo stripping of (most but not all of) their hydrogen envelope exposing the very hot and compact helium core \citep{Kippenhahn1969,Podsiadlowski1992} capable of producing high energy \Hep\ photons \citep{Gotberg2017}. In contrast to W-R stars, these stars are exclusively produced in binary systems and are produced by low-mass sub-dwarf O and B type stars. 
\citet{Gotberg2017} finds that every interacting binary produces a hot stripped star, which is powered by core helium burning up to $\sim10\%$ of its total lifetime. 
The amount of which stripped stars contribute to the ionisation of \Hep\ is completely unconstrained with variations up to $\sim6$ orders of magnitude difference at higher metallicities (see Figure 11 of \citealt{Gotberg2017}). These variations, primarily driven by the uncertainties in the assumed mass loss rate is significantly reduced at extreme low metallicities ($Z\lesssim0.0002$), but is still in the order of $\sim1$ magnitude. 
However, given stripped stars are produced by low mass stars which are favoured by the IMF and their prolonged life-time in the helium burning phase compared to their W-R counterparts could lead them to be a significant \Hep\ photon emitting mechanism in our \HeII\ sample.
Additionally, transparent stellar atmospheres coupled with the longer time-delay between star-formation and stripped star production enables stripped stars to ionize far larger distances in the ISM/IGM \citep{Gotberg2017}, which could result in production of \He\ emission outside of central star-forming regions in galaxies.

\paragraph{IMF variations:}\label{sec:imf_variations} 
Recent studies have demonstrated the possibility of systematic variations in the IMF in star-forming galaxies at low \citep{Hoversten2008,Gunawardhana2011} and high \citep{Nanayakkara2017} redshifts. 
If star-forming galaxies do contain systematically higher amounts of massive stars compared to what is expected by  \citet{Salpeter1955} like IMFs, the higher abundance of massive stars would lead to the production of extra ionizing photons for a given SFR. 
In Figure \ref{fig:bpass_model_predictions}, we show the ionization photon production efficiency for H and \Hep\ in BPASS stellar population models for different IMFs. 
For clarity of the figure we remove effects of the upper mass IMF cutoff but we note that in both single and binary models it only has an influence between the first 1-3 Myrs of the star-burst and becomes negligible henceforth.
This is driven by the relatively short life-times of  100--300\msol\ stars.  
Changing the IMF slope shows a prominent effect even at later-times from the onset of the star-burst, with up to $\sim0.3-0.5$ dex higher H ionizing photon production rate predicted for binary models with the IMF slope change from $\Gamma=-1.35$ to $\Gamma=-1.00$. 
Additionally, at $\lambda<228$\AA, binary stars play a viral role in producing higher amounts of ionizing photons which could ionize \Hep. 
Thus, top heavy IMFs could be one possible contributer to \HeII\ emission in our sample \citep[also see][]{Kehrig2018}. 
However, as we show in Figure  \ref{fig:bpass_model_predictions}, the effect of IMF on $\xi_{ion}$ is negligible for \Hep\ and thus may not contribute significantly to increase the \HeII\ EW. 
Stacked galaxy analysis of $z\sim2$ rest-UV features by \citet{Steidel2016} have shown that IMF sensitive rest-UV features could be reproduced using BPASS binary models without invoking variations in the IMF and independently of the upper mass IMF cutoff (between 100--300\msol). 
However, their photo-ionisation models does not include dust depletion but includes dust physics (e.g., photoelectric heating), which would boost the strength of the coolant lines and result in strong emission lines. 
In order to completely rule out IMF variations to be a contributing \Hep\ emitting factor for our \HeII\ sample, we require higher S/N rest-UV/optical spectra with IMF sensitive features.


\section{Conclusions}
\label{sec:conclusions}

In this paper we explored deep spectroscopic observations from the VLT/MUSE integral field spectrograph to  compile a sample of strong \HeII\ nebular emitters between $1.93<z<4.67$. We have complemented our sample with other deep rest-UV spectroscopic surveys conducted around the same redshift as described in Appendix \ref{sec:other_surveys}.
Using custom built emission line fitting codes we obtained rest-UV emission line ratios to compare with expectations from photo-ionisation modeling. 

Our results are as summarized as follows:

\begin{itemize}
\item The MUSE \HeII\ sample comprises galaxies with multiple rest-UV emission line detections with a large range in \HeII\ EW ($\sim$ an order of magnitude) and $\mathrm{M_{UV}}$ ($\sim-19$ to $-23$) [Section \ref{sec:observed_sample}: Figure \ref{fig:HeII_flux_comparisions}].

\item Using photo-ionisation modeling from \citet{Gutkin2016}, we show that the observed emission line ratios of our \HeII\ sample can be reproduced primarily at sub-solar metallicities and high ionisation parameters [Sections \ref{sec:model_comp_HeII_individual}, \ref{sec:model_comp_stacked}: Figures \ref{fig:line_ratios_gutkin}, \ref{fig:line_ratios_gutkin_stacked}]. 

\item We use BPASS binary stellar population models from \citet{Xiao2018} to show that BPASS binary models are also able to reproduce the observed line ratios in the  \CIII/\HeII\ vs \OIII/\HeII\ diagnostic. However, when  effects of binaries are included, models become degenerate [Section \ref{sec:model_comp_HeII_Xiao}: Figure \ref{fig:line_ratios_xiao}].

\item We show that the dust attenuation law and assumption of dust sight-lines only have a negligible effect for our line-ratio analysis. However, photo-ionisation model assumptions of metal depletion and dust dissociation needs stronger constraints at higher redshifts [Section \ref{sec:uncerternities_dust}: Appendix \ref{appendix:dust_appendix}: Figure \ref{fig:line_ratios_dust}]. 

\item We show that BPASS models are able to re-produce the \CIII\ EWs but not \HeII\ and \OIII\ EWs. This is possibly driven by the lack of ionisation photons blue-ward of 228\AA\ (and also super solar O/Fe abundance) [Section \ref{sec:model_comp_HeII_EW}:  Figures \ref{fig:ews_xiao}, \ref{fig:bpass_Lyman_continuum_spectra}].

\item We show that observed \HeII\ luminosities could only be reproduced by BPASS models at $\sim1/200th$ solar metallicity, which is in contrast with gas phase metallicities inferred by rest-UV emission line diagnostics [Section \ref{sec:hep_production}: Figure \ref{fig:NHeII_def}]. 

\item We find the \HeII\ line widths to be $\sim200-400\ km/s$ and show the need for an emission line fitting algorithms that fit \HeII\ fitting independently of other emission lines to accurately constrain the shape of \HeII\ and measure line flux [Section \ref{sec:uncerternities_snr}: Figure \ref{fig:line_fits_difference}]. 

\item We compute ionising photon production efficiency and $\xi_{ion}$ for H  and \Hep\ ionising photons and find that binary stars and IMF have a significant impact [Section \ref{sec:hep_production}: Figure \ref{fig:bpass_model_predictions}]. 

\item We explore possible mechanisms of \HeII\ production:
	\begin{itemize}

		\item We rule out high-metallicity W-R stars as a possible mechanism, but we are unable to place any constraints on the contribution by low-metallicity W-R stars (e.g. WNh stars) [Section \ref{sec:WR-stars}].

		\item We note that binary stars can play a crucial role but the current binary stellar evolution implementations still lack \HeII\ ionising photons under the hypothesis that stars are the only ionizing source [Section \ref{sec:binary_effects}].

		\item Variations in the high-mass end IMF, with slopes steeper than the canonical \citet{Salpeter1955} slope could contribute to the missing \Hep\ ionising photons [Section \ref{sec:imf_variations}].  

		\item We rule out the contribution from strong AGN to the ionizing photon flux for our \HeII\ sample.  However, we cannot rule out a sub-dominant contribution from weak/low luminosity AGN and/or shocks [Section \ref{sec:AGN_contamination}].  

	\end{itemize}

\end{itemize}

Future work should focus on full spectral fitting analysis of individual rest-UV spectra of galaxies involving effects of binaries and varying IMF. 
More sophisticated binary stellar populations with realistic stellar atmosphere models including effects of low-Z very massive stars at Eddington limit are required to accurately determine the stellar and nebular production efficiencies of \HeII. 
Additionally, dust properties of galaxies at $z>2$ requires stronger constraints for photo-ionisation modeling. 
Future sub-mm observations of galaxies will allow dust temperatures and geometries to be constrained and linking with FIR emission lines would allow stronger constraints on the dust grain properties of galaxies in the early Universe. Links between gas metallicities and dust will be beneficial to understand the complicated processes the galaxies undergo in the epoch of the peak of the cosmic star-formation-rate density. 
The combined coverage of $z=2-4$ galaxies by MUSE XDF survey\footnote{A single 160h MUSE pointing in the UDF planned to be completed in 2019. $\sim2$ galaxies in the MUSE \HeII\ sample is within the FoV. } and \emph{JWST} will provide an ideal sample with high S/N to study stellar and ISM conditions of galaxies, to understand which is the dominant mechanisms for the \HeII\ emission at high redshift and to constrain the role of pristine stellar-populations in the production of \HeII.

\begin{acknowledgements}
The authors wish to thank the referee for constructive comments that improved the paper substantially. 
We thank the BPASS team to making the stellar population models available. We thank Elizabeth Stanway, Claus Leitherer, Daniel Schaerer,  Jorick Vink, and Nell Byler for insightful discussions.  
We thank the Lorentz Centre and the scientific organizers of the \emph{Characterizing galaxies with spectroscopy with a view for JWST} workshop held at the Lorentz Centre in 2017 October, which promoted useful discussions among the wider community.
TN, JB, and RB acknowledges the Nederlandse Organisatie voor Wetenschappelijk Onderzoek (NWO) top grant TOP1.16.057. 
AF acknowledges support from the ERC via an Advanced Grant under grant agreement no. 339659-MUSICOS.
JB acknowledges support by Funda{\c c}{\~a}o para a Ci{\^e}ncia e a Tecnologia (FCT) through national funds (UID/FIS/04434/2013) and Investigador FCT contract IF/01654/2014/CP1215/CT0003, and by FEDER through COMPETE2020 (POCI-01-0145-FEDER-007672).
JR acknowledges support from the ERC Starting grant 336736 (CALENDS).
This research made use of {\tt astropy}\footnote{http://www.astropy.org} a community-developed core Python package for Astronomy \citep{Astropy2013,Astropy2018} and {\tt pandas} \citep{Kinney2010}.
Figures were generated using {\tt matplotlib} \citep{Hunter2007} and {\tt seaborn}\footnote{https://seaborn.pydata.org}.
\end{acknowledgements}

Facilities: VLT (MUSE)

\bibliographystyle{aa}
\bibliography{HeII_cat.bbl}

\begin{thebibliography}{189}
\expandafter\ifx\csname natexlab\endcsname\relax\def\natexlab#1{#1}\fi

\bibitem[{{Agarwal} {et~al.}(2016){Agarwal}, {Johnson}, {Zackrisson}, {Labbe},
  {van den Bosch}, {Natarajan}, \& {Khochfar}}]{Agarwal2016}
{Agarwal}, B., {Johnson}, J.~L., {Zackrisson}, E., {et~al.} 2016, \mnras, 460,
  4003

\bibitem[{{Akerman} {et~al.}(2004){Akerman}, {Carigi}, {Nissen}, {Pettini}, \&
  {Asplund}}]{Akerman2004}
{Akerman}, C.~J., {Carigi}, L., {Nissen}, P.~E., {Pettini}, M., \& {Asplund},
  M. 2004, \aap, 414, 931

\bibitem[{{Allen} {et~al.}(1976){Allen}, {Wright}, \& {Goss}}]{Allen1976}
{Allen}, D.~A., {Wright}, A.~E., \& {Goss}, W.~M. 1976, \mnras, 177, 91

\bibitem[{{Allen} {et~al.}(2008){Allen}, {Groves}, {Dopita}, {Sutherland}, \&
  {Kewley}}]{Allen2008}
{Allen}, M.~G., {Groves}, B.~A., {Dopita}, M.~A., {Sutherland}, R.~S., \&
  {Kewley}, L.~J. 2008, \apjs, 178, 20

\bibitem[{{Amor{\'\i}n} {et~al.}(2017){Amor{\'\i}n}, {Fontana},
  {P{\'e}rez-Montero}, {Castellano}, {Guaita}, {Grazian}, {Le F{\`e}vre},
  {Ribeiro}, {Schaerer}, {Tasca}, {Thomas}, {Bardelli}, {Cassar{\`a}},
  {Cassata}, {Cimatti}, {Contini}, {de Barros}, {Garilli}, {Giavalisco},
  {Hathi}, {Koekemoer}, {Le Brun}, {Lemaux}, {Maccagni}, {Pentericci}, {Pforr},
  {Talia}, {Tresse}, {Vanzella}, {Vergani}, {Zamorani}, {Zucca}, \&
  {Merlin}}]{Amorin2017}
{Amor{\'\i}n}, R., {Fontana}, A., {P{\'e}rez-Montero}, E., {et~al.} 2017,
  Nature Astronomy, 1, 0052

\bibitem[{{Asano} {et~al.}(2013){Asano}, {Takeuchi}, {Hirashita}, \&
  {Nozawa}}]{Asano2013}
{Asano}, R.~S., {Takeuchi}, T.~T., {Hirashita}, H., \& {Nozawa}, T. 2013,
  \mnras, 432, 637

\bibitem[{{Astropy Collaboration} {et~al.}(2018){Astropy Collaboration},
  {Price-Whelan}, {Sip{\H o}cz}, {G{\"u}nther}, {Lim}, {Crawford}, {Conseil},
  {Shupe}, {Craig}, {Dencheva}, {Ginsburg}, {VanderPlas}, {Bradley},
  {P{\'e}rez-Su{\'a}rez}, {de Val-Borro}, {Aldcroft}, {Cruz}, {Robitaille},
  {Tollerud}, {Ardelean}, {Babej}, {Bach}, {Bachetti}, {Bakanov}, {Bamford},
  {Barentsen}, {Barmby}, {Baumbach}, {Berry}, {Biscani}, {Boquien}, {Bostroem},
  {Bouma}, {Brammer}, {Bray}, {Breytenbach}, {Buddelmeijer}, {Burke},
  {Calderone}, {Cano Rodr{\'{\i}}guez}, {Cara}, {Cardoso}, {Cheedella},
  {Copin}, {Corrales}, {Crichton}, {D'Avella}, {Deil}, {Depagne}, {Dietrich},
  {Donath}, {Droettboom}, {Earl}, {Erben}, {Fabbro}, {Ferreira}, {Finethy},
  {Fox}, {Garrison}, {Gibbons}, {Goldstein}, {Gommers}, {Greco}, {Greenfield},
  {Groener}, {Grollier}, {Hagen}, {Hirst}, {Homeier}, {Horton}, {Hosseinzadeh},
  {Hu}, {Hunkeler}, {Ivezi{\'c}}, {Jain}, {Jenness}, {Kanarek}, {Kendrew},
  {Kern}, {Kerzendorf}, {Khvalko}, {King}, {Kirkby}, {Kulkarni}, {Kumar},
  {Lee}, {Lenz}, {Littlefair}, {Ma}, {Macleod}, {Mastropietro}, {McCully},
  {Montagnac}, {Morris}, {Mueller}, {Mumford}, {Muna}, {Murphy}, {Nelson},
  {Nguyen}, {Ninan}, {N{\"o}the}, {Ogaz}, {Oh}, {Parejko}, {Parley}, {Pascual},
  {Patil}, {Patil}, {Plunkett}, {Prochaska}, {Rastogi}, {Reddy Janga},
  {Sabater}, {Sakurikar}, {Seifert}, {Sherbert}, {Sherwood-Taylor}, {Shih},
  {Sick}, {Silbiger}, {Singanamalla}, {Singer}, {Sladen}, {Sooley},
  {Sornarajah}, {Streicher}, {Teuben}, {Thomas}, {Tremblay}, {Turner},
  {Terr{\'o}n}, {van Kerkwijk}, {de la Vega}, {Watkins}, {Weaver}, {Whitmore},
  {Woillez}, {Zabalza}, \& {Astropy Contributors}}]{Astropy2018}
{Astropy Collaboration}, {Price-Whelan}, A.~M., {Sip{\H o}cz}, B.~M., {et~al.}
  2018, \aj, 156, 123

\bibitem[{{Astropy Collaboration} {et~al.}(2013){Astropy Collaboration},
  {Robitaille}, {Tollerud}, {Greenfield}, {Droettboom}, {Bray}, {Aldcroft},
  {Davis}, {Ginsburg}, {Price-Whelan}, {Kerzendorf}, {Conley}, {Crighton},
  {Barbary}, {Muna}, {Ferguson}, {Grollier}, {Parikh}, {Nair}, {Unther},
  {Deil}, {Woillez}, {Conseil}, {Kramer}, {Turner}, {Singer}, {Fox}, {Weaver},
  {Zabalza}, {Edwards}, {Azalee Bostroem}, {Burke}, {Casey}, {Crawford},
  {Dencheva}, {Ely}, {Jenness}, {Labrie}, {Lim}, {Pierfederici}, {Pontzen},
  {Ptak}, {Refsdal}, {Servillat}, \& {Streicher}}]{Astropy2013}
{Astropy Collaboration}, {Robitaille}, T.~P., {Tollerud}, E.~J., {et~al.} 2013,
  \aap, 558, A33

\bibitem[{{Bacon} {et~al.}(2010){Bacon}, {Accardo}, {Adjali}, {Anwand},
  {Bauer}, {Biswas}, {Blaizot}, {Boudon}, {Brau-Nogue}, {Brinchmann},
  {Caillier}, {Capoani}, {Carollo}, {Contini}, {Couderc}, {Daguis{\'e}},
  {Deiries}, {Delabre}, {Dreizler}, {Dubois}, {Dupieux}, {Dupuy}, {Emsellem},
  {Fechner}, {Fleischmann}, {Fran{\c c}ois}, {Gallou}, {Gharsa}, {Glindemann},
  {Gojak}, {Guiderdoni}, {Hansali}, {Hahn}, {Jarno}, {Kelz}, {Koehler},
  {Kosmalski}, {Laurent}, {Le Floch}, {Lilly}, {Lizon}, {Loupias}, {Manescau},
  {Monstein}, {Nicklas}, {Olaya}, {Pares}, {Pasquini}, {P{\'e}contal-Rousset},
  {Pell{\'o}}, {Petit}, {Popow}, {Reiss}, {Remillieux}, {Renault}, {Roth},
  {Rupprecht}, {Serre}, {Schaye}, {Soucail}, {Steinmetz}, {Streicher}, {Stuik},
  {Valentin}, {Vernet}, {Weilbacher}, {Wisotzki}, \& {Yerle}}]{Bacon2010}
{Bacon}, R., {Accardo}, M., {Adjali}, L., {et~al.} 2010, in \procspie, Vol.
  7735, Ground-based and Airborne Instrumentation for Astronomy III, 773508

\bibitem[{{Bacon} {et~al.}(2015){Bacon}, {Brinchmann}, {Richard}, {Contini},
  {Drake}, {Franx}, {Tacchella}, {Vernet}, {Wisotzki}, {Blaizot}, {Bouch{\'e}},
  {Bouwens}, {Cantalupo}, {Carollo}, {Carton}, {Caruana}, {Cl{\'e}ment},
  {Dreizler}, {Epinat}, {Guiderdoni}, {Herenz}, {Husser}, {Kamann}, {Kerutt},
  {Kollatschny}, {Krajnovic}, {Lilly}, {Martinsson}, {Michel-Dansac},
  {Patricio}, {Schaye}, {Shirazi}, {Soto}, {Soucail}, {Steinmetz}, {Urrutia},
  {Weilbacher}, \& {de Zeeuw}}]{Bacon2015}
{Bacon}, R., {Brinchmann}, J., {Richard}, J., {et~al.} 2015, \aap, 575, A75

\bibitem[{{Bacon} {et~al.}(2017){Bacon}, {Conseil}, {Mary}, {Brinchmann},
  {Shepherd}, {Akhlaghi}, {Weilbacher}, {Piqueras}, {Wisotzki}, {Lagattuta},
  {Epinat}, {Guerou}, {Inami}, {Cantalupo}, {Courbot}, {Contini}, {Richard},
  {Maseda}, {Bouwens}, {Bouch{\'e}}, {Kollatschny}, {Schaye}, {Marino},
  {Pello}, {Herenz}, {Guiderdoni}, \& {Carollo}}]{Bacon2017}
{Bacon}, R., {Conseil}, S., {Mary}, D., {et~al.} 2017, \aap, 608, A1

\bibitem[{{Balestra} {et~al.}(2010){Balestra}, {Mainieri}, {Popesso},
  {Dickinson}, {Nonino}, {Rosati}, {Teimoorinia}, {Vanzella}, {Cristiani},
  {Cesarsky}, {Fosbury}, {Kuntschner}, \& {Rettura}}]{Balestra2010}
{Balestra}, I., {Mainieri}, V., {Popesso}, P., {et~al.} 2010, \aap, 512, A12

\bibitem[{{Barkana} \& {Loeb}(2001)}]{Barkana2001}
{Barkana}, R. \& {Loeb}, A. 2001, \physrep, 349, 125

\bibitem[{Beckwith {et~al.}(2006)Beckwith, Stiavelli, Koekemoer, Caldwell,
  Ferguson, Hook, Lucas, Bergeron, Corbin, Jogee, Panagia, Robberto, Royle,
  Somerville, \& Sosey}]{Beckwith2006}
Beckwith, S. V.~W., Stiavelli, M., Koekemoer, A.~M., {et~al.} 2006, {AJ}, 132,
  1729

\bibitem[{{Berg} {et~al.}(2018){Berg}, {Erb}, {Auger}, {Pettini}, \&
  {Brammer}}]{Berg2018}
{Berg}, D.~A., {Erb}, D.~K., {Auger}, M.~W., {Pettini}, M., \& {Brammer}, G.~B.
  2018, ArXiv e-prints [\eprint[arXiv]{1803.02340}]

\bibitem[{{Berg} {et~al.}(2016){Berg}, {Skillman}, {Henry}, {Erb}, \&
  {Carigi}}]{Berg2016a}
{Berg}, D.~A., {Skillman}, E.~D., {Henry}, R. B.~C., {Erb}, D.~K., \& {Carigi},
  L. 2016, \apj, 827, 126

\bibitem[{{Borisova} {et~al.}(2016){Borisova}, {Cantalupo}, {Lilly}, {Marino},
  {Gallego}, {Bacon}, {Blaizot}, {Bouch{\'e}}, {Brinchmann}, {Carollo},
  {Caruana}, {Finley}, {Herenz}, {Richard}, {Schaye}, {Straka}, {Turner},
  {Urrutia}, {Verhamme}, \& {Wisotzki}}]{Borisova2016}
{Borisova}, E., {Cantalupo}, S., {Lilly}, S.~J., {et~al.} 2016, \apj, 831, 39

\bibitem[{{Bouwens} {et~al.}(2016){Bouwens}, {Aravena}, {Decarli}, {Walter},
  {da Cunha}, {Labb{\'e}}, {Bauer}, {Bertoldi}, {Carilli}, {Chapman}, {Daddi},
  {Hodge}, {Ivison}, {Karim}, {Le Fevre}, {Magnelli}, {Ota}, {Riechers},
  {Smail}, {van der Werf}, {Weiss}, {Cox}, {Elbaz}, {Gonzalez-Lopez},
  {Infante}, {Oesch}, {Wagg}, \& {Wilkins}}]{Bouwens2016b}
{Bouwens}, R.~J., {Aravena}, M., {Decarli}, R., {et~al.} 2016, \apj, 833, 72

\bibitem[{{Bouwens} {et~al.}(2010){Bouwens}, {Illingworth}, {Oesch},
  {Stiavelli}, {van Dokkum}, {Trenti}, {Magee}, {Labb{\'e}}, {Franx},
  {Carollo}, \& {Gonzalez}}]{Bouwens2010}
{Bouwens}, R.~J., {Illingworth}, G.~D., {Oesch}, P.~A., {et~al.} 2010, \apjl,
  709, L133

\bibitem[{{Bowler} {et~al.}(2017){Bowler}, {Dunlop}, {McLure}, \&
  {McLeod}}]{Bowler2017}
{Bowler}, R.~A.~A., {Dunlop}, J.~S., {McLure}, R.~J., \& {McLeod}, D.~J. 2017,
  \mnras, 466, 3612

\bibitem[{{Brinchmann} {et~al.}(2013){Brinchmann}, {Charlot}, {Kauffmann},
  {Heckman}, {White}, \& {Tremonti}}]{Brinchmann2013}
{Brinchmann}, J., {Charlot}, S., {Kauffmann}, G., {et~al.} 2013, \mnras, 432,
  2112

\bibitem[{{Brinchmann} {et~al.}(2008){Brinchmann}, {Kunth}, \&
  {Durret}}]{Brinchmann2008}
{Brinchmann}, J., {Kunth}, D., \& {Durret}, F. 2008, \aap, 485, 657

\bibitem[{{Bromm} \& {Larson}(2004)}]{Bromm2004}
{Bromm}, V. \& {Larson}, R.~B. 2004, \araa, 42, 79

\bibitem[{{Bromm} \& {Yoshida}(2011)}]{Bromm2011}
{Bromm}, V. \& {Yoshida}, N. 2011, \araa, 49, 373

\bibitem[{{Brorby} {et~al.}(2016){Brorby}, {Kaaret}, {Prestwich}, \&
  {Mirabel}}]{Brorby2016}
{Brorby}, M., {Kaaret}, P., {Prestwich}, A., \& {Mirabel}, I.~F. 2016, \mnras,
  457, 4081

\bibitem[{{Bruzual} \& {Charlot}(2003)}]{Bruzual2003}
{Bruzual}, G. \& {Charlot}, S. 2003, \mnras, 344, 1000

\bibitem[{{Buat} {et~al.}(2011){Buat}, {Giovannoli}, {Heinis}, {Charmandaris},
  {Coia}, {Daddi}, {Dickinson}, {Elbaz}, {Hwang}, {Morrison}, {Dasyra},
  {Aussel}, {Altieri}, {Dannerbauer}, {Kartaltepe}, {Leiton}, {Magdis},
  {Magnelli}, \& {Popesso}}]{Buat2011}
{Buat}, V., {Giovannoli}, E., {Heinis}, S., {et~al.} 2011, \aap, 533, A93

\bibitem[{{Calzetti}(2001)}]{Calzetti2001}
{Calzetti}, D. 2001, \pasp, 113, 1449

\bibitem[{{Calzetti} {et~al.}(2000){Calzetti}, {Armus}, {Bohlin}, {Kinney},
  {Koornneef}, \& {Storchi-Bergmann}}]{Calzetti2000}
{Calzetti}, D., {Armus}, L., {Bohlin}, R.~C., {et~al.} 2000, \apj, 533, 682

\bibitem[{{Calzetti} {et~al.}(1994){Calzetti}, {Kinney}, \&
  {Storchi-Bergmann}}]{Calzetti1994}
{Calzetti}, D., {Kinney}, A.~L., \& {Storchi-Bergmann}, T. 1994, \apj, 429, 582

\bibitem[{{Cantalupo} {et~al.}(2019){Cantalupo}, {Pezzulli}, {Lilly}, {Marino},
  {Gallego}, {Schaye}, {Bacon}, {Feltre}, {Kollatschny}, {Nanayakkara},
  {Richard}, {Wendt}, {Wisotzki}, \& {Prochaska}}]{Cantalupo2019}
{Cantalupo}, S., {Pezzulli}, G., {Lilly}, S.~J., {et~al.} 2019, \mnras, 483,
  5188

\bibitem[{{Cardelli} {et~al.}(1989){Cardelli}, {Clayton}, \&
  {Mathis}}]{Cardelli1989}
{Cardelli}, J.~A., {Clayton}, G.~C., \& {Mathis}, J.~S. 1989, \apj, 345, 245

\bibitem[{{Casares} {et~al.}(2017){Casares}, {Jonker}, \&
  {Israelian}}]{Casares2017}
{Casares}, J., {Jonker}, P.~G., \& {Israelian}, G. 2017, ArXiv e-prints
  [\eprint[arXiv]{1701.07450}]

\bibitem[{{Cassata} {et~al.}(2013){Cassata}, {Le F{\`e}vre}, {Charlot},
  {Contini}, {Cucciati}, {Garilli}, {Zamorani}, {Adami}, {Bardelli}, {Le Brun},
  {Lemaux}, {Maccagni}, {Pollo}, {Pozzetti}, {Tresse}, {Vergani}, {Zanichelli},
  \& {Zucca}}]{Cassata2013}
{Cassata}, P., {Le F{\`e}vre}, O., {Charlot}, S., {et~al.} 2013, \aap, 556, A68

\bibitem[{Chabrier(2003)}]{Chabrier2003}
Chabrier, G. 2003, Publications of the Astronomical Society of the Pacific,
  115, pp. 763

\bibitem[{{Charlot} \& {Longhetti}(2001)}]{Charlot2001}
{Charlot}, S. \& {Longhetti}, M. 2001, \mnras, 323, 887

\bibitem[{{Chevallard} \& {Charlot}(2016)}]{Chevallard2016}
{Chevallard}, J. \& {Charlot}, S. 2016, \mnras, 462, 1415

\bibitem[{{Chiappini} {et~al.}(2003){Chiappini}, {Romano}, \&
  {Matteucci}}]{Chiappini2003}
{Chiappini}, C., {Romano}, D., \& {Matteucci}, F. 2003, \mnras, 339, 63

\bibitem[{{Cimatti} {et~al.}(2002){Cimatti}, {Daddi}, {Mignoli}, {Pozzetti},
  {Renzini}, {Zamorani}, {Broadhurst}, {Fontana}, {Saracco}, {Poli},
  {Cristiani}, {D'Odorico}, {Giallongo}, {Gilmozzi}, \& {Menci}}]{Cimatti2002}
{Cimatti}, A., {Daddi}, E., {Mignoli}, M., {et~al.} 2002, \aap, 381, L68

\bibitem[{{Conroy}(2013)}]{Conroy2013}
{Conroy}, C. 2013, \araa, 51, 393

\bibitem[{{Cooke} {et~al.}(2011){Cooke}, {Pettini}, {Steidel}, {Rudie}, \&
  {Nissen}}]{Cooke2011}
{Cooke}, R., {Pettini}, M., {Steidel}, C.~C., {Rudie}, G.~C., \& {Nissen},
  P.~E. 2011, \mnras, 417, 1534

\bibitem[{{Crowther}(2007)}]{Crowther2007}
{Crowther}, P.~A. 2007, \araa, 45, 177

\bibitem[{{Crowther} \& {Hadfield}(2006)}]{Crowther2006}
{Crowther}, P.~A. \& {Hadfield}, L.~J. 2006, \aap, 449, 711

\bibitem[{{Cullen} {et~al.}(2018){Cullen}, {McLure}, {Khochfar}, {Dunlop},
  {Dalla Vecchia}, {Carnall}, {Bourne}, {Castellano}, {Cimatti}, {Cirasuolo},
  {Elbaz}, {Fynbo}, {Garilli}, {Koekemoer}, {Marchi}, {Pentericci}, {Talia}, \&
  {Zamorani}}]{Cullen2018}
{Cullen}, F., {McLure}, R.~J., {Khochfar}, S., {et~al.} 2018, \mnras, 476, 3218

\bibitem[{{De Cia}(2018)}]{DeCia2018}
{De Cia}, A. 2018, ArXiv e-prints [\eprint[arXiv]{1805.05365}]

\bibitem[{{de Mink} {et~al.}(2013){de Mink}, {Langer}, {Izzard}, {Sana}, \& {de
  Koter}}]{deMink2013}
{de Mink}, S.~E., {Langer}, N., {Izzard}, R.~G., {Sana}, H., \& {de Koter}, A.
  2013, \apj, 764, 166

\bibitem[{Draine(2003)}]{Draine2003}
Draine, B. 2003, Annual Review of Astronomy and Astrophysics, 41, 241

\bibitem[{{Draine} {et~al.}(2007){Draine}, {Dale}, {Bendo}, {Gordon}, {Smith},
  {Armus}, {Engelbracht}, {Helou}, {Kennicutt}, {Li}, {Roussel}, {Walter},
  {Calzetti}, {Moustakas}, {Murphy}, {Rieke}, {Bot}, {Hollenbach}, {Sheth}, \&
  {Teplitz}}]{Draine2007}
{Draine}, B.~T., {Dale}, D.~A., {Bendo}, G., {et~al.} 2007, \apj, 663, 866

\bibitem[{{Draine} \& {Salpeter}(1979)}]{Draine1979}
{Draine}, B.~T. \& {Salpeter}, E.~E. 1979, \apj, 231, 77

\bibitem[{{Eldridge} {et~al.}(2008){Eldridge}, {Izzard}, \&
  {Tout}}]{Eldridge2008}
{Eldridge}, J.~J., {Izzard}, R.~G., \& {Tout}, C.~A. 2008, \mnras, 384, 1109

\bibitem[{{Eldridge} \& {Stanway}(2009)}]{Eldridge2009}
{Eldridge}, J.~J. \& {Stanway}, E.~R. 2009, \mnras, 400, 1019

\bibitem[{{Eldridge} \& {Stanway}(2012)}]{Eldridge2012a}
{Eldridge}, J.~J. \& {Stanway}, E.~R. 2012, \mnras, 419, 479

\bibitem[{{Eldridge} {et~al.}(2017){Eldridge}, {Stanway}, {Xiao}, {McClelland},
  {Taylor}, {Ng}, {Greis}, \& {Bray}}]{Eldridge2017}
{Eldridge}, J.~J., {Stanway}, E.~R., {Xiao}, L., {et~al.} 2017, PASA, 34, e058

\bibitem[{{Epinat} {et~al.}(2018){Epinat}, {Contini}, {Finley}, {Boogaard},
  {Gu{\'e}rou}, {Brinchmann}, {Carton}, {Michel-Dansac}, {Bacon}, {Cantalupo},
  {Carollo}, {Hamer}, {Kollatschny}, {Krajnovi{\'c}}, {Marino}, {Richard},
  {Soucail}, {Weilbacher}, \& {Wisotzki}}]{Epinat2018}
{Epinat}, B., {Contini}, T., {Finley}, H., {et~al.} 2018, \aap, 609, A40

\bibitem[{{Erb} {et~al.}(2010){Erb}, {Pettini}, {Shapley}, {Steidel}, {Law}, \&
  {Reddy}}]{Erb2010}
{Erb}, D.~K., {Pettini}, M., {Shapley}, A.~E., {et~al.} 2010, \apj, 719, 1168

\bibitem[{{Erben} {et~al.}(2013){Erben}, {Hildebrandt}, {Miller}, {van
  Waerbeke}, {Heymans}, {Hoekstra}, {Kitching}, {Mellier}, {Benjamin}, {Blake},
  {Bonnett}, {Cordes}, {Coupon}, {Fu}, {Gavazzi}, {Gillis}, {Grocutt}, {Gwyn},
  {Holhjem}, {Hudson}, {Kilbinger}, {Kuijken}, {Milkeraitis}, {Rowe},
  {Schrabback}, {Semboloni}, {Simon}, {Smit}, {Toader}, {Vafaei}, {van Uitert},
  \& {Velander}}]{Erben2013}
{Erben}, T., {Hildebrandt}, H., {Miller}, L., {et~al.} 2013, \mnras, 433, 2545

\bibitem[{{Fardal} {et~al.}(2001){Fardal}, {Katz}, {Gardner}, {Hernquist},
  {Weinberg}, \& {Dav{\'e}}}]{Fardal2001}
{Fardal}, M.~A., {Katz}, N., {Gardner}, J.~P., {et~al.} 2001, \apj, 562, 605

\bibitem[{{Feltre} {et~al.}(2016){Feltre}, {Charlot}, \& {Gutkin}}]{Feltre2016}
{Feltre}, A., {Charlot}, S., \& {Gutkin}, J. 2016, \mnras, 456, 3354

\bibitem[{{Ferland} {et~al.}(2013){Ferland}, {Porter}, {van Hoof}, {Williams},
  {Abel}, {Lykins}, {Shaw}, {Henney}, \& {Stancil}}]{Ferland2013}
{Ferland}, G.~J., {Porter}, R.~L., {van Hoof}, P.~A.~M., {et~al.} 2013, \rmxaa,
  49, 137

\bibitem[{{Freeman} {et~al.}(2017){Freeman}, {Siana}, {Kriek}, {Shapley},
  {Reddy}, {Coil}, {Mobasher}, {Muratov}, {Azadi}, {Leung}, {Sanders},
  {Shivaei}, {Price}, {DeGroot}, \& {Kere{\v s}}}]{Freeman2017}
{Freeman}, W.~R., {Siana}, B., {Kriek}, M., {et~al.} 2017, ArXiv e-prints
  [\eprint[arXiv]{1710.03230}]

\bibitem[{{Garc{\'{\i}}a-Vergara} {et~al.}(2017){Garc{\'{\i}}a-Vergara},
  {Hennawi}, {Barrientos}, \& {Rix}}]{Garcia-Vergara2017}
{Garc{\'{\i}}a-Vergara}, C., {Hennawi}, J.~F., {Barrientos}, L.~F., \& {Rix},
  H.-W. 2017, \apj, 848, 7

\bibitem[{{Garilli} {et~al.}(2014){Garilli}, {Guzzo}, {Scodeggio},
  {Bolzonella}, {Abbas}, {Adami}, {Arnouts}, {Bel}, {Bottini}, {Branchini},
  {Cappi}, {Coupon}, {Cucciati}, {Davidzon}, {De Lucia}, {de la Torre},
  {Franzetti}, {Fritz}, {Fumana}, {Granett}, {Ilbert}, {Iovino}, {Krywult}, {Le
  Brun}, {Le F{\`e}vre}, {Maccagni}, {Ma{\l}ek}, {Marulli}, {McCracken},
  {Paioro}, {Polletta}, {Pollo}, {Schlagenhaufer}, {Tasca}, {Tojeiro},
  {Vergani}, {Zamorani}, {Zanichelli}, {Burden}, {Di Porto}, {Marchetti},
  {Marinoni}, {Mellier}, {Moscardini}, {Nichol}, {Peacock}, {Percival},
  {Phleps}, \& {Wolk}}]{Garilli2014}
{Garilli}, B., {Guzzo}, L., {Scodeggio}, M., {et~al.} 2014, \aap, 562, A23

\bibitem[{{Garnett} {et~al.}(1991){Garnett}, {Kennicutt}, {Chu}, \&
  {Skillman}}]{Garnett1991}
{Garnett}, D.~R., {Kennicutt}, Jr., R.~C., {Chu}, Y.-H., \& {Skillman}, E.~D.
  1991, \apj, 373, 458

\bibitem[{{Gioannini} {et~al.}(2017){Gioannini}, {Matteucci}, {Vladilo}, \&
  {Calura}}]{Gioannini2017}
{Gioannini}, L., {Matteucci}, F., {Vladilo}, G., \& {Calura}, F. 2017, \mnras,
  464, 985

\bibitem[{{G{\"o}tberg} {et~al.}(2017){G{\"o}tberg}, {de Mink}, \&
  {Groh}}]{Gotberg2017}
{G{\"o}tberg}, Y., {de Mink}, S.~E., \& {Groh}, J.~H. 2017, \aap, 608, A11

\bibitem[{{Gr{\"a}fener} \& {Vink}(2015)}]{Grafener2015}
{Gr{\"a}fener}, G. \& {Vink}, J.~S. 2015, \aap, 578, L2

\bibitem[{{Grogin} {et~al.}(2011){Grogin}, {Kocevski}, {Faber}, {Ferguson},
  {Koekemoer}, {Riess}, {Acquaviva}, {Alexander}, {Almaini}, {Ashby}, {Barden},
  {Bell}, {Bournaud}, {Brown}, {Caputi}, {Casertano}, {Cassata}, {Castellano},
  {Challis}, {Chary}, {Cheung}, {Cirasuolo}, {Conselice}, {Roshan Cooray},
  {Croton}, {Daddi}, {Dahlen}, {Dav{\'e}}, {de Mello}, {Dekel}, {Dickinson},
  {Dolch}, {Donley}, {Dunlop}, {Dutton}, {Elbaz}, {Fazio}, {Filippenko},
  {Finkelstein}, {Fontana}, {Gardner}, {Garnavich}, {Gawiser}, {Giavalisco},
  {Grazian}, {Guo}, {Hathi}, {H{\"a}ussler}, {Hopkins}, {Huang}, {Huang},
  {Jha}, {Kartaltepe}, {Kirshner}, {Koo}, {Lai}, {Lee}, {Li}, {Lotz}, {Lucas},
  {Madau}, {McCarthy}, {McGrath}, {McIntosh}, {McLure}, {Mobasher},
  {Moustakas}, {Mozena}, {Nandra}, {Newman}, {Niemi}, {Noeske}, {Papovich},
  {Pentericci}, {Pope}, {Primack}, {Rajan}, {Ravindranath}, {Reddy}, {Renzini},
  {Rix}, {Robaina}, {Rodney}, {Rosario}, {Rosati}, {Salimbeni}, {Scarlata},
  {Siana}, {Simard}, {Smidt}, {Somerville}, {Spinrad}, {Straughn}, {Strolger},
  {Telford}, {Teplitz}, {Trump}, {van der Wel}, {Villforth}, {Wechsler},
  {Weiner}, {Wiklind}, {Wild}, {Wilson}, {Wuyts}, {Yan}, \& {Yun}}]{Grogin2011}
{Grogin}, N.~A., {Kocevski}, D.~D., {Faber}, S.~M., {et~al.} 2011, \apjs, 197,
  35

\bibitem[{{Groves} {et~al.}(2004{\natexlab{a}}){Groves}, {Dopita}, \&
  {Sutherland}}]{Groves2004a}
{Groves}, B.~A., {Dopita}, M.~A., \& {Sutherland}, R.~S. 2004{\natexlab{a}},
  \apjs, 153, 9

\bibitem[{{Groves} {et~al.}(2004{\natexlab{b}}){Groves}, {Dopita}, \&
  {Sutherland}}]{Groves2004b}
{Groves}, B.~A., {Dopita}, M.~A., \& {Sutherland}, R.~S. 2004{\natexlab{b}},
  \apjs, 153, 75

\bibitem[{{Gunawardhana} {et~al.}(2011){Gunawardhana}, {Hopkins}, {Sharp},
  {Brough}, {Taylor}, {Bland-Hawthorn}, {Maraston}, {Tuffs}, {Popescu},
  {Wijesinghe}, {Jones}, {Croom}, {Sadler}, {Wilkins}, {Driver}, {Liske},
  {Norberg}, {Baldry}, {Bamford}, {Loveday}, {Peacock}, {Robotham}, {Zucker},
  {Parker}, {Conselice}, {Cameron}, {Frenk}, {Hill}, {Kelvin}, {Kuijken},
  {Madore}, {Nichol}, {Parkinson}, {Pimbblet}, {Prescott}, {Sutherland},
  {Thomas}, \& {van Kampen}}]{Gunawardhana2011}
{Gunawardhana}, M.~L.~P., {Hopkins}, A.~M., {Sharp}, R.~G., {et~al.} 2011,
  \mnras, 415, 1647

\bibitem[{{Gutkin} {et~al.}(2016){Gutkin}, {Charlot}, \&
  {Bruzual}}]{Gutkin2016}
{Gutkin}, J., {Charlot}, S., \& {Bruzual}, G. 2016, \mnras, 462, 1757

\bibitem[{{Heger} \& {Woosley}(2002)}]{Heger2002}
{Heger}, A. \& {Woosley}, S.~E. 2002, \apj, 567, 532

\bibitem[{{Henry} {et~al.}(2000){Henry}, {Edmunds}, \&
  {K{\"o}ppen}}]{Henry2000}
{Henry}, R.~B.~C., {Edmunds}, M.~G., \& {K{\"o}ppen}, J. 2000, \apj, 541, 660

\bibitem[{{Heymans} {et~al.}(2012){Heymans}, {Van Waerbeke}, {Miller}, {Erben},
  {Hildebrandt}, {Hoekstra}, {Kitching}, {Mellier}, {Simon}, {Bonnett},
  {Coupon}, {Fu}, {Harnois D{\'e}raps}, {Hudson}, {Kilbinger}, {Kuijken},
  {Rowe}, {Schrabback}, {Semboloni}, {van Uitert}, {Vafaei}, \&
  {Velander}}]{Heymans2012}
{Heymans}, C., {Van Waerbeke}, L., {Miller}, L., {et~al.} 2012, \mnras, 427,
  146

\bibitem[{{Hoversten} \& {Glazebrook}(2008)}]{Hoversten2008}
{Hoversten}, E.~A. \& {Glazebrook}, K. 2008, \apj, 675, 163

\bibitem[{{Hu} {et~al.}(2004){Hu}, {Cowie}, {Capak}, {McMahon}, {Hayashino}, \&
  {Komiyama}}]{Hu2004}
{Hu}, E.~M., {Cowie}, L.~L., {Capak}, P., {et~al.} 2004, \aj, 127, 563

\bibitem[{Hunter(2007)}]{Hunter2007}
Hunter, J.~D. 2007, Computing In Science \& Engineering, 9, 90

\bibitem[{{Inami} {et~al.}(2017){Inami}, {Bacon}, {Brinchmann}, {Richard},
  {Contini}, {Conseil}, {Hamer}, {Akhlaghi}, {Bouch{\'e}}, {Cl{\'e}ment},
  {Desprez}, {Drake}, {Hashimoto}, {Leclercq}, {Maseda}, {Michel-Dansac},
  {Paalvast}, {Tresse}, {Ventou}, {Kollatschny}, {Boogaard}, {Finley},
  {Marino}, {Schaye}, \& {Wisotzki}}]{Inami2017}
{Inami}, H., {Bacon}, R., {Brinchmann}, J., {et~al.} 2017, \aap, 608, A2

\bibitem[{{Izotov} {et~al.}(2012){Izotov}, {Thuan}, \& {Privon}}]{Izotov2012}
{Izotov}, Y.~I., {Thuan}, T.~X., \& {Privon}, G. 2012, \mnras, 427, 1229

\bibitem[{{Jaskot} \& {Ravindranath}(2016)}]{Jaskot2016}
{Jaskot}, A.~E. \& {Ravindranath}, S. 2016, \apj, 833, 136

\bibitem[{{Jeans}(1902)}]{Jeans1902}
{Jeans}, J.~H. 1902, Philosophical Transactions of the Royal Society of London
  Series A, 199, 1

\bibitem[{{Jenkins}(2009)}]{Jenkins2009}
{Jenkins}, E.~B. 2009, \apj, 700, 1299

\bibitem[{{Kacprzak} {et~al.}(2016){Kacprzak}, {van de Voort}, {Glazebrook},
  {Tran}, {Yuan}, {Nanayakkara}, {Allen}, {Alcorn}, {Cowley}, {Labb{\'e}},
  {Spitler}, {Straatman}, \& {Tomczak}}]{Kacprzak2016}
{Kacprzak}, G.~G., {van de Voort}, F., {Glazebrook}, K., {et~al.} 2016, \apjl,
  826, L11

\bibitem[{{Kacprzak} {et~al.}(2015){Kacprzak}, {Yuan}, {Nanayakkara},
  {Kobayashi}, {Tran}, {Kewley}, {Glazebrook}, {Spitler}, {Taylor}, {Cowley},
  {Labbe}, {Straatman}, \& {Tomczak}}]{Kacprzak2015}
{Kacprzak}, G.~G., {Yuan}, T., {Nanayakkara}, T., {et~al.} 2015, \apjl, 802,
  L26

\bibitem[{{Kehrig} {et~al.}(2011){Kehrig}, {Oey}, {Crowther}, {Fogel},
  {Pellegrini}, {Schnurr}, {Schaerer}, {Massey}, \& {Roth}}]{Kehrig2011}
{Kehrig}, C., {Oey}, M.~S., {Crowther}, P.~A., {et~al.} 2011, \aap, 526, A128

\bibitem[{{Kehrig} {et~al.}(2018){Kehrig}, {V{\'{\i}}lchez}, {Guerrero},
  {Iglesias-P{\'a}ramo}, {Hunt}, {Duarte-Puertas}, \&
  {Ramos-Larios}}]{Kehrig2018}
{Kehrig}, C., {V{\'{\i}}lchez}, J.~M., {Guerrero}, M.~A., {et~al.} 2018,
  \mnras, 480, 1081

\bibitem[{{Kehrig} {et~al.}(2015){Kehrig}, {V{\'{\i}}lchez},
  {P{\'e}rez-Montero}, {Iglesias-P{\'a}ramo}, {Brinchmann}, {Kunth}, {Durret},
  \& {Bayo}}]{Kehrig2015}
{Kehrig}, C., {V{\'{\i}}lchez}, J.~M., {P{\'e}rez-Montero}, E., {et~al.} 2015,
  \apjl, 801, L28

\bibitem[{{Kewley} {et~al.}(2016){Kewley}, {Yuan}, {Nanayakkara}, {Kacprzak},
  {Tran}, {Glazebrook}, {Spitler}, {Cowley}, {Dopita}, {Straatman},
  {Labb{\'e}}, \& {Tomczak}}]{Kewley2016}
{Kewley}, L.~J., {Yuan}, T., {Nanayakkara}, T., {et~al.} 2016, \apj, 819, 100

\bibitem[{{Kippenhahn}(1969)}]{Kippenhahn1969}
{Kippenhahn}, R. 1969, \aap, 3, 83

\bibitem[{{Knobel} {et~al.}(2012){Knobel}, {Lilly}, {Iovino}, {Kova{\v c}},
  {Bschorr}, {Presotto}, {Oesch}, {Kampczyk}, {Carollo}, {Contini}, {Kneib},
  {Le Fevre}, {Mainieri}, {Renzini}, {Scodeggio}, {Zamorani}, {Bardelli},
  {Bolzonella}, {Bongiorno}, {Caputi}, {Cucciati}, {de la Torre}, {de Ravel},
  {Franzetti}, {Garilli}, {Lamareille}, {Le Borgne}, {Le Brun}, {Maier},
  {Mignoli}, {Pello}, {Peng}, {Perez Montero}, {Silverman}, {Tanaka}, {Tasca},
  {Tresse}, {Vergani}, {Zucca}, {Barnes}, {Bordoloi}, {Cappi}, {Cimatti},
  {Coppa}, {Koekemoer}, {L{\'o}pez-Sanjuan}, {McCracken}, {Moresco}, {Nair},
  {Pozzetti}, \& {Welikala}}]{Knobel2012}
{Knobel}, C., {Lilly}, S.~J., {Iovino}, A., {et~al.} 2012, \apj, 753, 121

\bibitem[{{Koekemoer} {et~al.}(2011){Koekemoer}, {Faber}, {Ferguson}, {Grogin},
  {Kocevski}, {Koo}, {Lai}, {Lotz}, {Lucas}, {McGrath}, {Ogaz}, {Rajan},
  {Riess}, {Rodney}, {Strolger}, {Casertano}, {Castellano}, {Dahlen},
  {Dickinson}, {Dolch}, {Fontana}, {Giavalisco}, {Grazian}, {Guo}, {Hathi},
  {Huang}, {van der Wel}, {Yan}, {Acquaviva}, {Alexander}, {Almaini}, {Ashby},
  {Barden}, {Bell}, {Bournaud}, {Brown}, {Caputi}, {Cassata}, {Challis},
  {Chary}, {Cheung}, {Cirasuolo}, {Conselice}, {Roshan Cooray}, {Croton},
  {Daddi}, {Dav{\'e}}, {de Mello}, {de Ravel}, {Dekel}, {Donley}, {Dunlop},
  {Dutton}, {Elbaz}, {Fazio}, {Filippenko}, {Finkelstein}, {Frazer}, {Gardner},
  {Garnavich}, {Gawiser}, {Gruetzbauch}, {Hartley}, {H{\"a}ussler},
  {Herrington}, {Hopkins}, {Huang}, {Jha}, {Johnson}, {Kartaltepe},
  {Khostovan}, {Kirshner}, {Lani}, {Lee}, {Li}, {Madau}, {McCarthy},
  {McIntosh}, {McLure}, {McPartland}, {Mobasher}, {Moreira}, {Mortlock},
  {Moustakas}, {Mozena}, {Nandra}, {Newman}, {Nielsen}, {Niemi}, {Noeske},
  {Papovich}, {Pentericci}, {Pope}, {Primack}, {Ravindranath}, {Reddy},
  {Renzini}, {Rix}, {Robaina}, {Rosario}, {Rosati}, {Salimbeni}, {Scarlata},
  {Siana}, {Simard}, {Smidt}, {Snyder}, {Somerville}, {Spinrad}, {Straughn},
  {Telford}, {Teplitz}, {Trump}, {Vargas}, {Villforth}, {Wagner}, {Wandro},
  {Wechsler}, {Weiner}, {Wiklind}, {Wild}, {Wilson}, {Wuyts}, \&
  {Yun}}]{Koekemoer2011}
{Koekemoer}, A.~M., {Faber}, S.~M., {Ferguson}, H.~C., {et~al.} 2011, \apjs,
  197, 36

\bibitem[{{Kormendy} \& {Richstone}(1995)}]{Kormendy1995}
{Kormendy}, J. \& {Richstone}, D. 1995, Annual Review of Astronomy and
  Astrophysics, 33, 581

\bibitem[{{Kriek} {et~al.}(2009){Kriek}, {van Dokkum}, {Labb{\'e}}, {Franx},
  {Illingworth}, {Marchesini}, \& {Quadri}}]{Kriek2009}
{Kriek}, M., {van Dokkum}, P.~G., {Labb{\'e}}, I., {et~al.} 2009, \apj, 700,
  221

\bibitem[{{Lagache} {et~al.}(2005){Lagache}, {Puget}, \& {Dole}}]{Lagache2005}
{Lagache}, G., {Puget}, J.-L., \& {Dole}, H. 2005, \araa, 43, 727

\bibitem[{{Langer}(2012)}]{Langer2012}
{Langer}, N. 2012, \araa, 50, 107

\bibitem[{{Le F{\`e}vre} {et~al.}(2013){Le F{\`e}vre}, {Cassata}, {Cucciati},
  {Garilli}, {Ilbert}, {Le Brun}, {Maccagni}, {Moreau}, {Scodeggio}, {Tresse},
  {Zamorani}, {Adami}, {Arnouts}, {Bardelli}, {Bolzonella}, {Bondi},
  {Bongiorno}, {Bottini}, {Cappi}, {Charlot}, {Ciliegi}, {Contini}, {de la
  Torre}, {Foucaud}, {Franzetti}, {Gavignaud}, {Guzzo}, {Iovino}, {Lemaux},
  {L{\'o}pez-Sanjuan}, {McCracken}, {Marano}, {Marinoni}, {Mazure}, {Mellier},
  {Merighi}, {Merluzzi}, {Paltani}, {Pell{\`o}}, {Pollo}, {Pozzetti},
  {Scaramella}, {Tasca}, {Vergani}, {Vettolani}, {Zanichelli}, \&
  {Zucca}}]{LeFevre2013}
{Le F{\`e}vre}, O., {Cassata}, P., {Cucciati}, O., {et~al.} 2013, \aap, 559,
  A14

\bibitem[{{Le F{\`e}vre} {et~al.}(2015){Le F{\`e}vre}, {Tasca}, {Cassata},
  {Garilli}, {Le Brun}, {Maccagni}, {Pentericci}, {Thomas}, {Vanzella},
  {Zamorani}, {Zucca}, {Amorin}, {Bardelli}, {Capak}, {Cassar{\`a}},
  {Castellano}, {Cimatti}, {Cuby}, {Cucciati}, {de la Torre}, {Durkalec},
  {Fontana}, {Giavalisco}, {Grazian}, {Hathi}, {Ilbert}, {Lemaux}, {Moreau},
  {Paltani}, {Ribeiro}, {Salvato}, {Schaerer}, {Scodeggio}, {Sommariva},
  {Talia}, {Taniguchi}, {Tresse}, {Vergani}, {Wang}, {Charlot}, {Contini},
  {Fotopoulou}, {L{\'o}pez-Sanjuan}, {Mellier}, \& {Scoville}}]{LeFevre2015}
{Le F{\`e}vre}, O., {Tasca}, L.~A.~M., {Cassata}, P., {et~al.} 2015, \aap, 576,
  A79

\bibitem[{{Leitherer} {et~al.}(2014){Leitherer}, {Ekstr{\"o}m}, {Meynet},
  {Schaerer}, {Agienko}, \& {Levesque}}]{Leitherer2014}
{Leitherer}, C., {Ekstr{\"o}m}, S., {Meynet}, G., {et~al.} 2014, \apjs, 212, 14

\bibitem[{{Leitherer} {et~al.}(1999){Leitherer}, {Schaerer}, {Goldader},
  {Delgado}, {Robert}, {Kune}, {de Mello}, {Devost}, \&
  {Heckman}}]{Leitherer1999}
{Leitherer}, C., {Schaerer}, D., {Goldader}, J.~D., {et~al.} 1999, \apjs, 123,
  3

\bibitem[{{Leja} {et~al.}(2016){Leja}, {Johnson}, {Conroy}, {van Dokkum}, \&
  {Byler}}]{Leja2017}
{Leja}, J., {Johnson}, B.~D., {Conroy}, C., {van Dokkum}, P.~G., \& {Byler}, N.
  2016, ArXiv e-prints [\eprint[arXiv]{1609.09073}]

\bibitem[{{Lilly} {et~al.}(2007){Lilly}, {Le F{\`e}vre}, {Renzini}, {Zamorani},
  {Scodeggio}, {Contini}, {Carollo}, {Hasinger}, {Kneib}, {Iovino}, {Le Brun},
  {Maier}, {Mainieri}, {Mignoli}, {Silverman}, {Tasca}, {Bolzonella},
  {Bongiorno}, {Bottini}, {Capak}, {Caputi}, {Cimatti}, {Cucciati}, {Daddi},
  {Feldmann}, {Franzetti}, {Garilli}, {Guzzo}, {Ilbert}, {Kampczyk}, {Kovac},
  {Lamareille}, {Leauthaud}, {Borgne}, {McCracken}, {Marinoni}, {Pello},
  {Ricciardelli}, {Scarlata}, {Vergani}, {Sanders}, {Schinnerer}, {Scoville},
  {Taniguchi}, {Arnouts}, {Aussel}, {Bardelli}, {Brusa}, {Cappi}, {Ciliegi},
  {Finoguenov}, {Foucaud}, {Franceschini}, {Halliday}, {Impey}, {Knobel},
  {Koekemoer}, {Kurk}, {Maccagni}, {Maddox}, {Marano}, {Marconi}, {Meneux},
  {Mobasher}, {Moreau}, {Peacock}, {Porciani}, {Pozzetti}, {Scaramella},
  {Schiminovich}, {Shopbell}, {Smail}, {Thompson}, {Tresse}, {Vettolani},
  {Zanichelli}, \& {Zucca}}]{Lilly2007}
{Lilly}, S.~J., {Le F{\`e}vre}, O., {Renzini}, A., {et~al.} 2007, \apjs, 172,
  70

\bibitem[{{Lo Faro} {et~al.}(2017){Lo Faro}, {Buat}, {Roehlly},
  {Alvarez-Marquez}, {Burgarella}, {Silva}, \& {Efstathiou}}]{Faro2017}
{Lo Faro}, B., {Buat}, V., {Roehlly}, Y., {et~al.} 2017, \mnras, 472, 1372

\bibitem[{{Luo} {et~al.}(2017){Luo}, {Brandt}, {Xue}, {Lehmer}, {Alexander},
  {Bauer}, {Vito}, {Yang}, {Basu-Zych}, {Comastri}, {Gilli}, {Gu},
  {Hornschemeier}, {Koekemoer}, {Liu}, {Mainieri}, {Paolillo}, {Ranalli},
  {Rosati}, {Schneider}, {Shemmer}, {Smail}, {Sun}, {Tozzi}, {Vignali}, \&
  {Wang}}]{Luo2017}
{Luo}, B., {Brandt}, W.~N., {Xue}, Y.~Q., {et~al.} 2017, The Astrophysical
  Journal Supplement Series, 228, 2

\bibitem[{{Madau} \& {Dickinson}(2014)}]{Madau2014}
{Madau}, P. \& {Dickinson}, M. 2014, \araa, 52, 415

\bibitem[{{Magorrian} {et~al.}(1998){Magorrian}, {Tremaine}, {Richstone},
  {Bender}, {Bower}, {Dressler}, {Faber}, {Gebhardt}, {Green}, {Grillmair},
  {Kormendy}, \& {Lauer}}]{Magorrian1998}
{Magorrian}, J., {Tremaine}, S., {Richstone}, D., {et~al.} 1998, \aj, 115, 2285

\bibitem[{{Marino} {et~al.}(2018){Marino}, {Cantalupo}, {Lilly}, {Gallego},
  {Straka}, {Borisova}, {Pezzulli}, {Bacon}, {Brinchmann}, {Carollo},
  {Caruana}, {Conseil}, {Contini}, {Diener}, {Finley}, {Inami}, {Leclercq},
  {Muzahid}, {Richard}, {Schaye}, {Wendt}, \& {Wisotzki}}]{Marino2018}
{Marino}, R.~A., {Cantalupo}, S., {Lilly}, S.~J., {et~al.} 2018, \apj, 859, 53

\bibitem[{{Maseda} {et~al.}(2017){Maseda}, {Brinchmann}, {Franx}, {Bacon},
  {Bouwens}, {Schmidt}, {Boogaard}, {Contini}, {Feltre}, {Inami},
  {Kollatschny}, {Marino}, {Richard}, {Verhamme}, \& {Wisotzki}}]{Maseda2017}
{Maseda}, M.~V., {Brinchmann}, J., {Franx}, M., {et~al.} 2017, ArXiv e-prints
  [\eprint[arXiv]{1710.06432}]

\bibitem[{{Mathis}(1990)}]{Mathis1990}
{Mathis}, J.~S. 1990, \araa, 28, 37

\bibitem[{{Matthee} \& {Schaye}(2018)}]{Matthee2018a}
{Matthee}, J. \& {Schaye}, J. 2018, \mnras, 479, L34

\bibitem[{{Matthee} {et~al.}(2017){Matthee}, {Sobral}, {Boone},
  {R{\"o}ttgering}, {Schaerer}, {Girard}, {Pallottini}, {Vallini}, {Ferrara},
  {Darvish}, \& {Mobasher}}]{Matthee2017}
{Matthee}, J., {Sobral}, D., {Boone}, F., {et~al.} 2017, \apj, 851, 145

\bibitem[{{McClelland} \& {Eldridge}(2016)}]{McClelland2016}
{McClelland}, L.~A.~S. \& {Eldridge}, J.~J. 2016, \mnras, 459, 1505

\bibitem[{McKinney(2010)}]{Kinney2010}
McKinney, W. 2010, in Proceedings of the 9th Python in Science Conference, ed.
  S.~van~der Walt \& J.~Millman, 51 -- 56

\bibitem[{{McLure} {et~al.}(2018){McLure}, {Dunlop}, {Cullen}, {Bourne},
  {Best}, {Khochfar}, {Bowler}, {Biggs}, {Geach}, {Scott}, {Micha{\l}owski},
  {Rujopakarn}, {van Kampen}, {Kirkpatrick}, \& {Pope}}]{McLure2018}
{McLure}, R.~J., {Dunlop}, J.~S., {Cullen}, F., {et~al.} 2018, \mnras, 476,
  3991

\bibitem[{{McLure} {et~al.}(2011){McLure}, {Dunlop}, {de Ravel}, {Cirasuolo},
  {Ellis}, {Schenker}, {Robertson}, {Koekemoer}, {Stark}, \&
  {Bowler}}]{McLure2011}
{McLure}, R.~J., {Dunlop}, J.~S., {de Ravel}, L., {et~al.} 2011, \mnras, 418,
  2074

\bibitem[{{Meurer} {et~al.}(1999){Meurer}, {Heckman}, \&
  {Calzetti}}]{Meurer1999}
{Meurer}, G.~R., {Heckman}, T.~M., \& {Calzetti}, D. 1999, \apj, 521, 64

\bibitem[{{Miralles-Caballero} {et~al.}(2016){Miralles-Caballero},
  {D{\'{\i}}az}, {L{\'o}pez-S{\'a}nchez}, {Rosales-Ortega}, {Monreal-Ibero},
  {P{\'e}rez-Montero}, {Kehrig}, {Garc{\'{\i}}a-Benito}, {S{\'a}nchez},
  {Walcher}, {Galbany}, {Iglesias-P{\'a}ramo}, {V{\'{\i}}lchez}, {Gonz{\'a}lez
  Delgado}, {van de Ven}, {Barrera-Ballesteros}, {Lyubenova}, {Meidt},
  {Falcon-Barroso}, {Mast}, {Mendoza}, \& {Califa
  Collaboration}}]{Miralles-Caballero2016}
{Miralles-Caballero}, D., {D{\'{\i}}az}, A.~I., {L{\'o}pez-S{\'a}nchez},
  {\'A}.~R., {et~al.} 2016, \aap, 592, A105

\bibitem[{{Muratov} {et~al.}(2015){Muratov}, {Kere{\v s}},
  {Faucher-Gigu{\`e}re}, {Hopkins}, {Quataert}, \& {Murray}}]{Muratov2015}
{Muratov}, A.~L., {Kere{\v s}}, D., {Faucher-Gigu{\`e}re}, C.-A., {et~al.}
  2015, \mnras, 454, 2691

\bibitem[{{Murayama} {et~al.}(2007){Murayama}, {Taniguchi}, {Scoville},
  {Ajiki}, {Sanders}, {Mobasher}, {Aussel}, {Capak}, {Koekemoer}, {Shioya},
  {Nagao}, {Carilli}, {Ellis}, {Garilli}, {Giavalisco}, {Kitzbichler}, {Le
  F{\`e}vre}, {Maccagni}, {Schinnerer}, {Smol{\v c}i{\'c}}, {Tribiano},
  {Cimatti}, {Komiyama}, {Miyazaki}, {Sasaki}, {Koda}, \&
  {Karoji}}]{Murayama2007}
{Murayama}, T., {Taniguchi}, Y., {Scoville}, N.~Z., {et~al.} 2007, \apjs, 172,
  523

\bibitem[{{Naidu} {et~al.}(2017){Naidu}, {Oesch}, {Reddy}, {Holden}, {Steidel},
  {Montes}, {Atek}, {Bouwens}, {Carollo}, {Cibinel}, {Illingworth},
  {Labb{\'e}}, {Magee}, {Morselli}, {Nelson}, {van Dokkum}, \&
  {Wilkins}}]{Naidu2017}
{Naidu}, R.~P., {Oesch}, P.~A., {Reddy}, N., {et~al.} 2017, \apj, 847, 12

\bibitem[{{Nakajima} {et~al.}(2018){Nakajima}, {Fletcher}, {Ellis},
  {Robertson}, \& {Iwata}}]{Nakajima2018}
{Nakajima}, K., {Fletcher}, T., {Ellis}, R.~S., {Robertson}, B.~E., \& {Iwata},
  I. 2018, \mnras, 477, 2098

\bibitem[{{Nanayakkara} {et~al.}(2017){Nanayakkara}, {Glazebrook}, {Kacprzak},
  {Yuan}, {Fisher}, {Tran}, {Kewley}, {Spitler}, {Alcorn}, {Cowley}, {Labbe},
  {Straatman}, \& {Tomczak}}]{Nanayakkara2017}
{Nanayakkara}, T., {Glazebrook}, K., {Kacprzak}, G.~G., {et~al.} 2017, \mnras,
  468, 3071

\bibitem[{{Narayanan} {et~al.}(2018){Narayanan}, {Conroy}, {Dave}, {Johnson},
  \& {Popping}}]{Narayanan2018}
{Narayanan}, D., {Conroy}, C., {Dave}, R., {Johnson}, B., \& {Popping}, G.
  2018, ArXiv e-prints [\eprint[arXiv]{1805.06905}]

\bibitem[{{Nozawa} {et~al.}(2015){Nozawa}, {Asano}, {Hirashita}, \&
  {Takeuchi}}]{Nozawa2015}
{Nozawa}, T., {Asano}, R.~S., {Hirashita}, H., \& {Takeuchi}, T.~T. 2015,
  \mnras, 447, L16

\bibitem[{{Oke} \& {Gunn}(1983)}]{Oke1983}
{Oke}, J.~B. \& {Gunn}, J.~E. 1983, \apj, 266, 713

\bibitem[{{Ono} {et~al.}(2017){Ono}, {Ouchi}, {Harikane}, {Toshikawa}, {Rauch},
  {Yuma}, {Sawicki}, {Shibuya}, {Shimasaku}, {Oguri}, {Willott}, {Akhlaghi},
  {Akiyama}, {Coupon}, {Kashikawa}, {Komiyama}, {Konno}, {Lin}, {Matsuoka},
  {Miyazaki}, {Nagao}, {Nakajima}, {Silverman}, {Tanaka}, {Taniguchi}, \&
  {Wang}}]{Ono2017}
{Ono}, Y., {Ouchi}, M., {Harikane}, Y., {et~al.} 2017, \pasj
  [\eprint[arXiv]{1704.06004}]

\bibitem[{{Osterbrock} \& {Cohen}(1982)}]{Osterbrock1982}
{Osterbrock}, D.~E. \& {Cohen}, R.~D. 1982, \apj, 261, 64

\bibitem[{{Ouchi} {et~al.}(2017){Ouchi}, {Harikane}, {Shibuya}, {Shimasaku},
  {Taniguchi}, {Konno}, {Kobayashi}, {Kajisawa}, {Nagao}, {Ono}, {Inoue},
  {Umemura}, {Mori}, {Hasegawa}, {Higuchi}, {Komiyama}, {Matsuda}, {Nakajima},
  {Saito}, \& {Wang}}]{Ouchi2017}
{Ouchi}, M., {Harikane}, Y., {Shibuya}, T., {et~al.} 2017, ArXiv e-prints
  [\eprint[arXiv]{1704.07455}]

\bibitem[{{Pacifici} {et~al.}(2015){Pacifici}, {da Cunha}, {Charlot}, {Rix},
  {Fumagalli}, {Wel}, {Franx}, {Maseda}, {van Dokkum}, {Brammer}, {Momcheva},
  {Skelton}, {Whitaker}, {Leja}, {Lundgren}, {Kassin}, \& {Yi}}]{Pacifici2015}
{Pacifici}, C., {da Cunha}, E., {Charlot}, S., {et~al.} 2015, \mnras, 447, 786

\bibitem[{{Patr{\'{\i}}cio} {et~al.}(2016){Patr{\'{\i}}cio}, {Richard},
  {Verhamme}, {Wisotzki}, {Brinchmann}, {Turner}, {Christensen}, {Weilbacher},
  {Blaizot}, {Bacon}, {Contini}, {Lagattuta}, {Cantalupo}, {Cl{\'e}ment}, \&
  {Soucail}}]{Patricio2017}
{Patr{\'{\i}}cio}, V., {Richard}, J., {Verhamme}, A., {et~al.} 2016, \mnras,
  456, 4191

\bibitem[{{Podsiadlowski} {et~al.}(1992){Podsiadlowski}, {Joss}, \&
  {Hsu}}]{Podsiadlowski1992}
{Podsiadlowski}, P., {Joss}, P.~C., \& {Hsu}, J.~J.~L. 1992, \apj, 391, 246

\bibitem[{{Raiter} {et~al.}(2010){Raiter}, {Schaerer}, \&
  {Fosbury}}]{Raiter2010}
{Raiter}, A., {Schaerer}, D., \& {Fosbury}, R.~A.~E. 2010, \aap, 523, A64

\bibitem[{{Reddy} {et~al.}(2015){Reddy}, {Kriek}, {Shapley}, {Freeman},
  {Siana}, {Coil}, {Mobasher}, {Price}, {Sanders}, \& {Shivaei}}]{Reddy2015}
{Reddy}, N.~A., {Kriek}, M., {Shapley}, A.~E., {et~al.} 2015, \apj, 806, 259

\bibitem[{{Reddy} {et~al.}(2018){Reddy}, {Oesch}, {Bouwens}, {Montes},
  {Illingworth}, {Steidel}, {van Dokkum}, {Atek}, {Carollo}, {Cibinel},
  {Holden}, {Labb{\'e}}, {Magee}, {Morselli}, {Nelson}, \&
  {Wilkins}}]{Reddy2018}
{Reddy}, N.~A., {Oesch}, P.~A., {Bouwens}, R.~J., {et~al.} 2018, \apj, 853, 56

\bibitem[{{Reddy} {et~al.}(2016{\natexlab{a}}){Reddy}, {Steidel}, {Pettini}, \&
  {Bogosavljevi{\'c}}}]{Reddy2016}
{Reddy}, N.~A., {Steidel}, C.~C., {Pettini}, M., \& {Bogosavljevi{\'c}}, M.
  2016{\natexlab{a}}, \apj, 828, 107

\bibitem[{{Reddy} {et~al.}(2016{\natexlab{b}}){Reddy}, {Steidel}, {Pettini},
  {Bogosavljevi{\'c}}, \& {Shapley}}]{Reddy2016b}
{Reddy}, N.~A., {Steidel}, C.~C., {Pettini}, M., {Bogosavljevi{\'c}}, M., \&
  {Shapley}, A.~E. 2016{\natexlab{b}}, \apj, 828, 108

\bibitem[{{Rigby} {et~al.}(2018){Rigby}, {Bayliss}, {Chisholm}, {Bordoloi},
  {Sharon}, {Gladders}, {Johnson}, {Paterno-Mahler}, {Wuyts}, {Dahle}, \&
  {Acharyya}}]{Rigby2018b}
{Rigby}, J.~R., {Bayliss}, M.~B., {Chisholm}, J., {et~al.} 2018, \apj, 853, 87

\bibitem[{{Salmon} {et~al.}(2016){Salmon}, {Papovich}, {Long}, {Willner},
  {Finkelstein}, {Ferguson}, {Dickinson}, {Duncan}, {Faber}, {Hathi},
  {Koekemoer}, {Kurczynski}, {Newman}, {Pacifici}, {P{\'e}rez-Gonz{\'a}lez}, \&
  {Pforr}}]{Salmon2016}
{Salmon}, B., {Papovich}, C., {Long}, J., {et~al.} 2016, \apj, 827, 20

\bibitem[{{Salpeter}(1955)}]{Salpeter1955}
{Salpeter}, E.~E. 1955, \apj, 121, 161

\bibitem[{{Sana} {et~al.}(2013){Sana}, {de Koter}, {de Mink}, {Dunstall},
  {Evans}, {H{\'e}nault-Brunet}, {Ma{\'{\i}}z Apell{\'a}niz},
  {Ram{\'{\i}}rez-Agudelo}, {Taylor}, {Walborn}, {Clark}, {Crowther},
  {Herrero}, {Gieles}, {Langer}, {Lennon}, \& {Vink}}]{Sana2013}
{Sana}, H., {de Koter}, A., {de Mink}, S.~E., {et~al.} 2013, \aap, 550, A107

\bibitem[{{Sana} {et~al.}(2012){Sana}, {de Mink}, {de Koter}, {Langer},
  {Evans}, {Gieles}, {Gosset}, {Izzard}, {Le Bouquin}, \&
  {Schneider}}]{Sana2012}
{Sana}, H., {de Mink}, S.~E., {de Koter}, A., {et~al.} 2012, Science, 337, 444

\bibitem[{{Sanders} {et~al.}(2015{\natexlab{a}}){Sanders}, {Shapley}, {Kriek},
  {Reddy}, {Freeman}, {Coil}, {Siana}, {Mobasher}, {Shivaei}, {Price}, \& {de
  Groot}}]{Sanders2015b}
{Sanders}, R.~L., {Shapley}, A.~E., {Kriek}, M., {et~al.} 2015{\natexlab{a}},
  ArXiv e-prints [\eprint[arXiv]{1509.03636}]

\bibitem[{{Sanders} {et~al.}(2015{\natexlab{b}}){Sanders}, {Shapley}, {Kriek},
  {Reddy}, {Freeman}, {Coil}, {Siana}, {Mobasher}, {Shivaei}, {Price}, \& {de
  Groot}}]{Sanders2015}
{Sanders}, R.~L., {Shapley}, A.~E., {Kriek}, M., {et~al.} 2015{\natexlab{b}},
  \apj, 799, 138

\bibitem[{{Savage} \& {Sembach}(1996)}]{Savage1996}
{Savage}, B.~D. \& {Sembach}, K.~R. 1996, \apj, 470, 893

\bibitem[{{Schaerer}(2002)}]{Schaerer2002}
{Schaerer}, D. 2002, \aap, 382, 28

\bibitem[{{Schaerer}(2003)}]{Schaerer2003}
{Schaerer}, D. 2003, \aap, 397, 527

\bibitem[{{Schaerer} {et~al.}(2019){Schaerer}, {Fragos}, \&
  {Izotov}}]{Schaerer2019}
{Schaerer}, D., {Fragos}, T., \& {Izotov}, Y.~I. 2019, \aap, 622, L10

\bibitem[{{Schmutz} {et~al.}(1992){Schmutz}, {Leitherer}, \&
  {Gruenwald}}]{Schmutz1992}
{Schmutz}, W., {Leitherer}, C., \& {Gruenwald}, R. 1992, \pasp, 104, 1164

\bibitem[{{Scodeggio} {et~al.}(2016){Scodeggio}, {Guzzo}, {Garilli}, {Granett},
  {Bolzonella}, {de la Torre}, {Abbas}, {Adami}, {Arnouts}, {Bottini}, {Cappi},
  {Coupon}, {Cucciati}, {Davidzon}, {Franzetti}, {Fritz}, {Iovino}, {Krywult},
  {Le Brun}, {Le F{\'e}vre}, {Maccagni}, {Malek}, {Marchetti}, {Marulli},
  {Polletta}, {Pollo}, {Tasca}, {Tojeiro}, {Vergani}, {Zanichelli}, {Bel},
  {Branchini}, {De Lucia}, {Ilbert}, {McCracken}, {Moutard}, {Peacock},
  {Zamorani}, {Burden}, {Fumana}, {Jullo}, {Marinoni}, {Mellier}, {Moscardini},
  \& {Percival}}]{Scodeggio2016}
{Scodeggio}, M., {Guzzo}, L., {Garilli}, B., {et~al.} 2016, ArXiv e-prints
  [\eprint[arXiv]{1611.07048}]

\bibitem[{{Scoville} {et~al.}(2007){Scoville}, {Aussel}, {Brusa}, {Capak},
  {Carollo}, {Elvis}, {Giavalisco}, {Guzzo}, {Hasinger}, {Impey}, {Kneib},
  {LeFevre}, {Lilly}, {Mobasher}, {Renzini}, {Rich}, {Sanders}, {Schinnerer},
  {Schminovich}, {Shopbell}, {Taniguchi}, \& {Tyson}}]{Scoville2007}
{Scoville}, N., {Aussel}, H., {Brusa}, M., {et~al.} 2007, \apjs, 172, 1

\bibitem[{{Scoville} {et~al.}(2015){Scoville}, {Faisst}, {Capak}, {Kakazu},
  {Li}, \& {Steinhardt}}]{Scoville2015}
{Scoville}, N., {Faisst}, A., {Capak}, P., {et~al.} 2015, \apj, 800, 108

\bibitem[{{Senchyna} {et~al.}(2017){Senchyna}, {Stark}, {Vidal-Garc{\'{\i}}a},
  {Chevallard}, {Charlot}, {Mainali}, {Jones}, {Wofford}, {Feltre}, \&
  {Gutkin}}]{Senchyna2017}
{Senchyna}, P., {Stark}, D.~P., {Vidal-Garc{\'{\i}}a}, A., {et~al.} 2017, ArXiv
  e-prints [\eprint[arXiv]{1706.00881}]

\bibitem[{{Shapley} {et~al.}(2003){Shapley}, {Steidel}, {Pettini}, \&
  {Adelberger}}]{Shapley2003}
{Shapley}, A.~E., {Steidel}, C.~C., {Pettini}, M., \& {Adelberger}, K.~L. 2003,
  \apj, 588, 65

\bibitem[{{Shibuya} {et~al.}(2017){Shibuya}, {Ouchi}, {Harikane}, {Rauch},
  {Ono}, {Mukae}, {Higuchi}, {Kojima}, {Yuma}, {Lee}, {Furusawa}, {Konno},
  {Martin}, {Shimasaku}, {Taniguchi}, {Kobayashi}, {Kajisawa}, {Nagao}, {Goto},
  {Kashikawa}, {Komiyama}, {Kusakabe}, {Momose}, {Nakajima}, {Tanaka}, \&
  {Wang}}]{Shibuya2017}
{Shibuya}, T., {Ouchi}, M., {Harikane}, Y., {et~al.} 2017, ArXiv e-prints
  [\eprint[arXiv]{1705.00733}]

\bibitem[{{Shields} \& {Kennicutt}(1995)}]{Shields1995}
{Shields}, J.~C. \& {Kennicutt}, Jr., R.~C. 1995, \apj, 454, 807

\bibitem[{{Shirazi} \& {Brinchmann}(2012)}]{Shirazi2012}
{Shirazi}, M. \& {Brinchmann}, J. 2012, \mnras, 421, 1043

\bibitem[{{Smith} {et~al.}(2017){Smith}, {Gotberg}, \& {de Mink}}]{Smith2017}
{Smith}, N., {Gotberg}, Y., \& {de Mink}, S.~E. 2017, ArXiv e-prints
  [\eprint[arXiv]{1704.03516}]

\bibitem[{{Sobral} {et~al.}(2018){Sobral}, {Matthee}, {Brammer}, {Ferrara},
  {Alegre}, {R{\"o}ttgering}, {Schaerer}, {Mobasher}, \&
  {Darvish}}]{Sobral2018b}
{Sobral}, D., {Matthee}, J., {Brammer}, G., {et~al.} 2018, \mnras, 2683

\bibitem[{{Sobral} {et~al.}(2015){Sobral}, {Matthee}, {Darvish}, {Schaerer},
  {Mobasher}, {R{\"o}ttgering}, {Santos}, \& {Hemmati}}]{Sobral2015}
{Sobral}, D., {Matthee}, J., {Darvish}, B., {et~al.} 2015, \apj, 808, 139

\bibitem[{{Spitzer}(1978)}]{Spitzer1978}
{Spitzer}, L. 1978, {Physical processes in the interstellar medium}

\bibitem[{{Stanway} {et~al.}(2016){Stanway}, {Eldridge}, \&
  {Becker}}]{Stanway2016}
{Stanway}, E.~R., {Eldridge}, J.~J., \& {Becker}, G.~D. 2016, \mnras, 456, 485

\bibitem[{{Steidel} {et~al.}(2003){Steidel}, {Adelberger}, {Shapley},
  {Pettini}, {Dickinson}, \& {Giavalisco}}]{Steidel2003}
{Steidel}, C.~C., {Adelberger}, K.~L., {Shapley}, A.~E., {et~al.} 2003, \apj,
  592, 728

\bibitem[{{Steidel} {et~al.}(2014){Steidel}, {Rudie}, {Strom}, {Pettini},
  {Reddy}, {Shapley}, {Trainor}, {Erb}, {Turner}, {Konidaris}, {Kulas}, {Mace},
  {Matthews}, \& {McLean}}]{Steidel2014}
{Steidel}, C.~C., {Rudie}, G.~C., {Strom}, A.~L., {et~al.} 2014, \apj, 795, 165

\bibitem[{{Steidel} {et~al.}(2016){Steidel}, {Strom}, {Pettini}, {Rudie},
  {Reddy}, \& {Trainor}}]{Steidel2016}
{Steidel}, C.~C., {Strom}, A.~L., {Pettini}, M., {et~al.} 2016, \apj, 826, 159

\bibitem[{{Strom} {et~al.}(2017){Strom}, {Steidel}, {Rudie}, {Trainor},
  {Pettini}, \& {Reddy}}]{Strom2017}
{Strom}, A.~L., {Steidel}, C.~C., {Rudie}, G.~C., {et~al.} 2017, \apj, 836, 164

\bibitem[{{Sz{\'e}csi} {et~al.}(2015){Sz{\'e}csi}, {Langer}, {Yoon}, {Sanyal},
  {de Mink}, {Evans}, \& {Dermine}}]{Szecsi2015}
{Sz{\'e}csi}, D., {Langer}, N., {Yoon}, S.-C., {et~al.} 2015, \aap, 581, A15

\bibitem[{{Tapken} {et~al.}(2006){Tapken}, {Appenzeller}, {Gabasch}, {Heidt},
  {Hopp}, {Bender}, {Mehlert}, {Noll}, {Seitz}, \& {Seifert}}]{Tapken2006}
{Tapken}, C., {Appenzeller}, I., {Gabasch}, A., {et~al.} 2006, \aap, 455, 145

\bibitem[{{Tasca} {et~al.}(2017){Tasca}, {Le F{\`e}vre}, {Ribeiro}, {Thomas},
  {Moreau}, {Cassata}, {Garilli}, {Le Brun}, {Lemaux}, {Maccagni},
  {Pentericci}, {Schaerer}, {Vanzella}, {Zamorani}, {Zucca}, {Amorin},
  {Bardelli}, {Cassar{\`a}}, {Castellano}, {Cimatti}, {Cucciati}, {Durkalec},
  {Fontana}, {Giavalisco}, {Grazian}, {Hathi}, {Ilbert}, {Paltani}, {Pforr},
  {Scodeggio}, {Sommariva}, {Talia}, {Tresse}, {Vergani}, {Capak}, {Charlot},
  {Contini}, {de la Torre}, {Dunlop}, {Fotopoulou}, {Guaita}, {Koekemoer},
  {L{\'o}pez-Sanjuan}, {Mellier}, {Salvato}, {Scoville}, {Taniguchi}, \&
  {Wang}}]{Tasca2017}
{Tasca}, L.~A.~M., {Le F{\`e}vre}, O., {Ribeiro}, B., {et~al.} 2017, \aap, 600,
  A110

\bibitem[{{Theios} {et~al.}(2018){Theios}, {Steidel}, {Strom}, {Rudie},
  {Trainor}, \& {Reddy}}]{Theios2018}
{Theios}, R.~L., {Steidel}, C.~C., {Strom}, A.~L., {et~al.} 2018, ArXiv
  e-prints [\eprint[arXiv]{1805.00016}]

\bibitem[{{Thuan} \& {Izotov}(2005)}]{Thuan2005}
{Thuan}, T.~X. \& {Izotov}, Y.~I. 2005, \apjs, 161, 240

\bibitem[{{Tremonti} {et~al.}(2004){Tremonti}, {Heckman}, {Kauffmann},
  {Brinchmann}, {Charlot}, {White}, {Seibert}, {Peng}, {Schlegel}, {Uomoto},
  {Fukugita}, \& {Brinkmann}}]{Tremonti2004}
{Tremonti}, C.~A., {Heckman}, T.~M., {Kauffmann}, G., {et~al.} 2004, \apj, 613,
  898

\bibitem[{{Tumlinson} {et~al.}(2001){Tumlinson}, {Giroux}, \&
  {Shull}}]{Tumlinson2001}
{Tumlinson}, J., {Giroux}, M.~L., \& {Shull}, J.~M. 2001, \apjl, 550, L1

\bibitem[{{Tumlinson} \& {Shull}(2000)}]{Tumlinson2000}
{Tumlinson}, J. \& {Shull}, J.~M. 2000, \apjl, 528, L65

\bibitem[{{Tumlinson} {et~al.}(2003){Tumlinson}, {Shull}, \&
  {Venkatesan}}]{Tumlinson2003}
{Tumlinson}, J., {Shull}, J.~M., \& {Venkatesan}, A. 2003, \apj, 584, 608

\bibitem[{{Vanzella} {et~al.}(2008){Vanzella}, {Cristiani}, {Dickinson},
  {Giavalisco}, {Kuntschner}, {Haase}, {Nonino}, {Rosati}, {Cesarsky},
  {Ferguson}, {Fosbury}, {Grazian}, {Moustakas}, {Rettura}, {Popesso},
  {Renzini}, {Stern}, \& {GOODS Team}}]{Vanzella2008}
{Vanzella}, E., {Cristiani}, S., {Dickinson}, M., {et~al.} 2008, \aap, 478, 83

\bibitem[{{Vanzella} {et~al.}(2016){Vanzella}, {de Barros}, {Vasei}, {Alavi},
  {Giavalisco}, {Siana}, {Grazian}, {Hasinger}, {Suh}, {Cappelluti}, {Vito},
  {Amorin}, {Balestra}, {Brusa}, {Calura}, {Castellano}, {Comastri}, {Fontana},
  {Gilli}, {Mignoli}, {Pentericci}, {Vignali}, \& {Zamorani}}]{Vanzella2016}
{Vanzella}, E., {de Barros}, S., {Vasei}, K., {et~al.} 2016, \apj, 825, 41

\bibitem[{{Vidal-Garc{\'\i}a} {et~al.}(2017){Vidal-Garc{\'\i}a}, {Charlot},
  {Bruzual}, \& {Hubeny}}]{Vidal-Garca2017}
{Vidal-Garc{\'\i}a}, A., {Charlot}, S., {Bruzual}, G., \& {Hubeny}, I. 2017,
  \mnras, 470, 3532

\bibitem[{{Wilkins} {et~al.}(2013){Wilkins}, {Coulton}, {Caruana}, {Croft}, {di
  Matteo}, {Khandai}, {Feng}, {Bunker}, \& {Elbert}}]{Wilkins2013}
{Wilkins}, S.~M., {Coulton}, W., {Caruana}, J., {et~al.} 2013, \mnras, 435,
  2885

\bibitem[{{Williams} {et~al.}(1996){Williams}, {Blacker}, {Dickinson}, {Dixon},
  {Ferguson}, {Fruchter}, {Giavalisco}, {Gilliland}, {Heyer}, {Katsanis},
  {Levay}, {Lucas}, {McElroy}, {Petro}, {Postman}, {Adorf}, \&
  {Hook}}]{Williams1996}
{Williams}, R.~E., {Blacker}, B., {Dickinson}, M., {et~al.} 1996, \aj, 112,
  1335

\bibitem[{{Wirth} {et~al.}(2015){Wirth}, {Trump}, {Barro}, {Guo}, {Koo}, {Liu},
  {Kassis}, {Lyke}, {Rizzi}, {Campbell}, {Goodrich}, \& {Faber}}]{Wirth2015}
{Wirth}, G.~D., {Trump}, J.~R., {Barro}, G., {et~al.} 2015, \aj, 150, 153

\bibitem[{{Wise} {et~al.}(2014){Wise}, {Demchenko}, {Halicek}, {Norman},
  {Turk}, {Abel}, \& {Smith}}]{Wise2014}
{Wise}, J.~H., {Demchenko}, V.~G., {Halicek}, M.~T., {et~al.} 2014, \mnras,
  442, 2560

\bibitem[{{Wise} {et~al.}(2012){Wise}, {Turk}, {Norman}, \& {Abel}}]{Wise2012}
{Wise}, J.~H., {Turk}, M.~J., {Norman}, M.~L., \& {Abel}, T. 2012, \apj, 745,
  50

\bibitem[{{Wiseman} {et~al.}(2017){Wiseman}, {Schady}, {Bolmer}, {Kr{\"u}hler},
  {Yates}, {Greiner}, \& {Fynbo}}]{Wiseman2017}
{Wiseman}, P., {Schady}, P., {Bolmer}, J., {et~al.} 2017, \aap, 599, A24

\bibitem[{{Wofford} {et~al.}(2016){Wofford}, {Charlot}, {Bruzual}, {Eldridge},
  {Calzetti}, {Adamo}, {Cignoni}, {de Mink}, {Gouliermis}, {Grasha}, {Grebel},
  {Lee}, {{\"O}stlin}, {Smith}, {Ubeda}, \& {Zackrisson}}]{Wofford2016}
{Wofford}, A., {Charlot}, S., {Bruzual}, G., {et~al.} 2016, \mnras, 457, 4296

\bibitem[{{Woods} \& {Gilfanov}(2016)}]{Woods2016}
{Woods}, T.~E. \& {Gilfanov}, M. 2016, \mnras, 455, 1770

\bibitem[{{Xiao} {et~al.}(2018){Xiao}, {Stanway}, \& {Eldridge}}]{Xiao2018}
{Xiao}, L., {Stanway}, E.~R., \& {Eldridge}, J.~J. 2018, \mnras, 477, 904

\bibitem[{{Yamasawa} {et~al.}(2011){Yamasawa}, {Habe}, {Kozasa}, {Nozawa},
  {Hirashita}, {Umeda}, \& {Nomoto}}]{Yamasawa2011}
{Yamasawa}, D., {Habe}, A., {Kozasa}, T., {et~al.} 2011, \apj, 735, 44

\bibitem[{{Yang} {et~al.}(2006){Yang}, {Zabludoff}, {Dav{\'e}}, {Eisenstein},
  {Pinto}, {Katz}, {Weinberg}, \& {Barton}}]{Yang2006}
{Yang}, Y., {Zabludoff}, A.~I., {Dav{\'e}}, R., {et~al.} 2006, \apj, 640, 539

\bibitem[{{Zafar} {et~al.}(2015){Zafar}, {M{\o}ller}, {Watson}, {Fynbo},
  {Krogager}, {Zafar}, {Saturni}, {Geier}, \& {Venemans}}]{Zafar2015}
{Zafar}, T., {M{\o}ller}, P., {Watson}, D., {et~al.} 2015, \aap, 584, A100

\bibitem[{{Zagury}(2017)}]{Zagury2017}
{Zagury}, F. 2017, Astronomische Nachrichten, 338, 807

\end{thebibliography}

\begin{appendix}

\section{\HeII\ detections from other public surveys}
\label{appendix:other_surveys}

We present a brief description of the public surveys exploited to investigate \HeII\ detections below and a summary is presented in Table \ref{tab:summary_table_others} accompanied by individual sources in Table \ref{tab:full_data_table_others}.

\paragraph{GOODS FORS2} 
The Great Observatories Origins Deep Survey (GOODS) FORS2 sample \citep{Vanzella2008} comprise 1225 spectra of individual galaxies in the GOODS South field with a $R=660$, which corresponds to 13 \AA\ at 8600 \AA. Out of 1166 galaxies in the GOODS FORS2 catalogue with a redshift quality flag of $A$, $B$, or $C$, we select 131 galaxies with spectroscopic redshifts between $2.65<z<5.71$.
However, due to variations in the slit-position in the masks 33 galaxies fall outside the desired spectral coverage and are removed from the sample. 
In the remaining 98 galaxies, we remove galaxies contaminated with large continuum noise and broad line AGN to select three galaxies with possible \HeII\ detections via visual examination.

\paragraph{GOODS VIMOS}
VIMOS ESO/GOODS spectroscopic survey was designed to complement the GOODS FORS2 survey by increasing the optical completeness and sky coverage over the GOODS-South field \citep{Balestra2010}. The final catalogue\footnote{Vizier ID: J/A+A/512/A12/catalogC} comprise of 4602 unique object spectra obtained via the VIMOS spectrograph with two resolution gratings; a low resolution mode with a nominal $R=180$ covering $3500-6900$ \AA\ and a medium resolution mode with $R=580$ covering $4000-10000$ \AA. Here, we combine galaxies observed via both modes to investigate \HeII\ detections within the redshift range $1.13<z<5.10$ and obtain 854 galaxies with with a redshift quality flag of $A$ or $B$. 
We visually examine the spectra of the 854 galaxies and select five spectra with tentative \HeII\ detections.

\paragraph{VANDELS}
VANDELS survey \citep{McLure2018} exploits multi-wavelength imaging and \emph{HST} grism coverage in CANDELS UDS and CDFS footprints footprints \citep{Grogin2011,Koekemoer2011} to investigate the properties of high redshift galaxies via emission and absorption line spectroscopy using the VLT/VIMOS. The survey primarily targets bright ($H_\mathrm{AB}<27$) star-forming galaxies between $2.5<z<5.5$. 
The observations for the VANDELS survey has been carried out using the  CG475 filter with the medium resolution grism, and thus closely spaced lines are blended.
VANDELS UDS catalogue contains 464 targets out of which 292 has a {\tt zflg}$>1$ and fall within $1.9\lesssim z \lesssim 5$,  the \HeII\ wavelength coverage of the CG475 filter. We visually inspect the spectra of these 292 galaxies. Similarly, the CDFS catalogue contains 415 galaxies out of which 285 galaxies are selected for visual inspection. After removing broad-line AGN, we select six and two galaxies respectively from the UDS and CDFS fields with tentative \HeII\ detections.

\paragraph{VIPERS}
The VIMOS Public Extragalactic Redshift Survey \citep[VIPERS;][]{Garilli2014,Scodeggio2016} comprise 91,507 galaxies obtained by the VLT/VIMOS spectrograph using the $R=220$ grism covering a wavelength range of $5500-9500$ \AA. Spectroscopic follow up targets were selected from the CFHTLS survey \citep{Heymans2012,Erben2013} covering a total FoV of $\mathrm{\sim224arcmin^2}$. In the two VIPERS fields, we find 50 galaxies between $3.0<z<4.5$ (the wavelength coverage for \HeII\ detections in the VIMOS mode implemented by VIPERS) with a redshift quality flag of  {\tt zflg-10}$>1$. We visually inspect all spectra and find eight galaxies with tentative \HeII\ detections, however, all of these galaxies show evidence for broad-line AGN.

\paragraph{VUDS}
The VIMOS Ultra-Deep Survey \citep[VUDS;][]{LeFevre2015} obtained spectra of $\sim10,000$ galaxies up to $I_{AB}\sim27$ using VLT/VIMOS. The Data Release 1 \citep{Tasca2017} contains all data obtained in the CANDLES COSMOS and CDFS fields using the low resolution grism and comprise of 677 galaxy redshifts. VUDS observations were carried out to obtain the maximum wavelength coverage and hence observations comprise of $\sim14$ hours per pointing per grism using the LRBLUE ($3600-6700$ \AA) and LRRED ($5500-9350$ \AA) grisms providing a total wavelength coverage between $3600-9350$ \AA. 
Therefore, VUDS obtained \HeII\ coverage between $1.2<z<5.1$ and comprise of 132 and 122 galaxies each in COSMOS and CDFS fields with a redshift quality flag {\tt zflags}$>1$. We visually inspect all spectra and identify three galaxies with tentative \HeII\ detections. We note that the full data set of VUDS contains additional \HeII\ detections as shown by \citet{Amorin2017}.

\paragraph{zCOSMOS Bright survey}
zCOSMOS survey \citep{Lilly2007} utilized the VLT/VIMOS over 600 hours between $2005-2010$ to obtain rest-frame UV and optical spectra of galaxies in the COSMOS field. The survey was designed in two parts: zCOSMOS-bright to obtain spectra of of galaxies with $I_{AB}<22.5$ using the medium resolution grism ($R\sim600$) with a spectral coverage between $5550-9650$ \AA and zCOSMOS-deep to obtain preferentially high redshift sources with a magnitude limit of $B_{AB}\sim25$ using the low resolution ($R\sim200$) blue grism covering a spectral range of $3600-6800$ \AA. 
The public zCOSMOS bright catalogue contains 20,689 galaxy spectra out of which 48 fall within $2.4<z<4.8$, the wavelength range to obtain \HeII\ in the observed VIMOS configuration. We visually inspect all spectra and identify 13 galaxies with tentative \HeII\ detections, however, 11 galaxies selected show either broad \CIII\ or \CIV\ emission lines suggesting the presence of an AGN, thus the final sample presented contains two galaxies. 
The zCOSMOS deep survey would provide a better selection of higher-$z$ targets, however, the data is yet to be publicly released.

Additionally, we explored the K20 survey \citep{Cimatti2002}, and did not find any convincing \HeII\ emitters.

\begin{table*}
\tiny
\caption{Catalogue of other \HeII\ emitters. [Appendix \ref{appendix:other_surveys}]
\label{tab:summary_table_others}}
\begin{tabular}{  r r r r r r  r r r }
\hline\hline
{Survey name}    					&
{Field}    							&
{Instrument}             			&
{$\Delta \lambda$ (\AA)}          	&
{Resolution}     					&
{Quality flag}       				& 
{$\mathrm{N_{\HeII\ coverage}}$}    &
{$\mathrm{N_{\HeII\ tentative}}$} 	\\
\hline \hline
GOODS FORS2			& GOODS    	& FORS2   & 6000--10000  & 100        & A, B, C       	& 131  &  3   \\
GOODS VIMOS		    & GOODS    	& VIMOS   & 4000--10000  & 180/580    & A, B       		& 854  &  5   \\
VANDELS				& CDFS    	& VIMOS   & 4800--10000	 & 580    & \tt{zflg$>$1} 		& 285  &  2  \\
VANDELS				& UDS    	& VIMOS   & 4800--10000	 & 580    & \tt{zflg$>$1} 		& 293  &  6   \\
VIPERS 				& W1    	& VIMOS   & 4800--10000	 & 220    & \tt{zflg-10$\geq$2} &  24  &  0   \\
VIPERS 				& W4    	& VIMOS   & 4800--10000	 & 220    & \tt{zflg-10$\geq$2} &  26  &  0  \\
VUDS				& CDFS    	& VIMOS   & 3600--9350   & 230/580& \tt{zflg$>$2}		& 122  &  0    \\
VUDS				& COSMOS   	& VIMOS   & 3600--9350   & 230    & \tt{zflg$>$2}		& 132  &  3   \\
ZCOSMOS	BRIGHT		& COSMOS   	& VIMOS   & 5550--9450	 & 600    & None		  		&  48  &  2     \\
\hline
\end{tabular}
\end{table*}

\begin{table}
\small
\caption{Non-MUSE tentative \HeII\ sample.  [Appendix \ref{appendix:other_surveys}]
\label{tab:full_data_table_others}}
\begin{tabular}{  r r r r r r  r r r }
\hline\hline
{ID}    &
{Field} &
{RA}    &
{Dec}   &
{zspec} \\
\hline\hline
779  & goods-fors2 & 3.0: 32.0: 29    & -27: 42: 34 & 3.59 \\
1673 & goods-fors2 & 3.0: 32.0: 52    & -27: 52: 37 & 3.47 \\
1701 & goods-fors2 & 3.0: 32.0: 55    & -27: 54: 14 & 4.72 \\
40 & vuds-cosmos & 10.0: 00.0: 27      & 2: 24:     33 & 3.25 \\
66 & vuds-cosmos & 10.0: 00.0: 27      & 2: 32:     54    & 2.47 \\
109 & vuds-cosmos & 10.0: 00.0: 46     & 2: 19:      05 & 4.38 \\
1524 & zcosmos & 9.0: 59.0: 57        & 1: 45:     36   & 3.18 \\
20603 & zcosmos & 9.0: 59.0: 01        & 2: 44:      19 & 3.52 \\
001765 & vandels-cdfs & 3.0: 32.0: 43 & -27: 54:   07 & 3.77 \\
126819 & vandels-cdfs & 3.0: 31.0: 56 & -27: 45:  33 & 2.82 \\
016296 & vandels-uds & 2.0: 17.0: 47  & -5: 11: 08 & 3.71 \\
017893 & vandels-uds & 2.0: 17.0: 17  & -5: 10: 36 & 4.14 \\
020721 & vandels-uds & 2.0: 17.0: 38  & -5: 9: 47 & 2.52 \\
145830 & vandels-uds & 2.0: 18.0: 14  & -5: 20: 11 & 3.21 \\
281893 & vandels-uds & 2.0: 17.0: 11  & -5: 22: 18 & 2.70 \\
287621 & vandels-uds & 2.0: 16.0: 52  & -5: 21: 26 & 2.88 \\
128 & vimos goodss & 22.0: 08.0:  55  & 52: 57: 19 & 2.34 \\
490 & vimos goodss & 22.0: 08.0:  19  & 53: 00: 36 & 2.02\\
1319 & vimos goodss & 22.0: 09.0: 08    & 53: 5: 51 & 2.15 \\
1947 & vimos goodss & 22.0: 08.0: 31   & 53: 9: 27 & 1.61 \\
4050 & vimos goodss & 22.0: 09.0: 12   & 53: 8: 15 & 2.45 \\
\hline
\end{tabular}
\end{table}

\section{Analysis of Gutkin et al. (2016) rest-UV emission line ratio models used in our analysis}
\label{appendix:model_comp_models}

\citet{Gutkin2016} models are parameterized by the following six parameters: the interstellar metallicity ($Z_{ISM}$), the zero-age ionization parameter at the Stromgren radius ($U_{s}$), the dust-to-metal ratio ($\xi_{d}$), the carbon-to-Oxygen abundance ratio (C/O), the hydrogen gas density ($n_{H}$), and the upper mass cut-off of the IMF ($m_{up}$). 
We use the following emission lines for our analysis: \HeII, \CIII=(\CIII$\lambda1907$+\CIII$\lambda1909$), \OIII(=\OIII$\lambda1661$+\OIII$\lambda1666$), \SiIII(=\SiIII$\lambda1883$+\SiIII$\lambda1892$). 
For each emission line ratio diagnostic, we select a subsample of galaxies with S/N$\geq3$ for the emission lines considered in that specific diagnostic.

\begin{figure*}[h!]
\includegraphics[trim = 10 10 10 0, clip, scale=0.625]{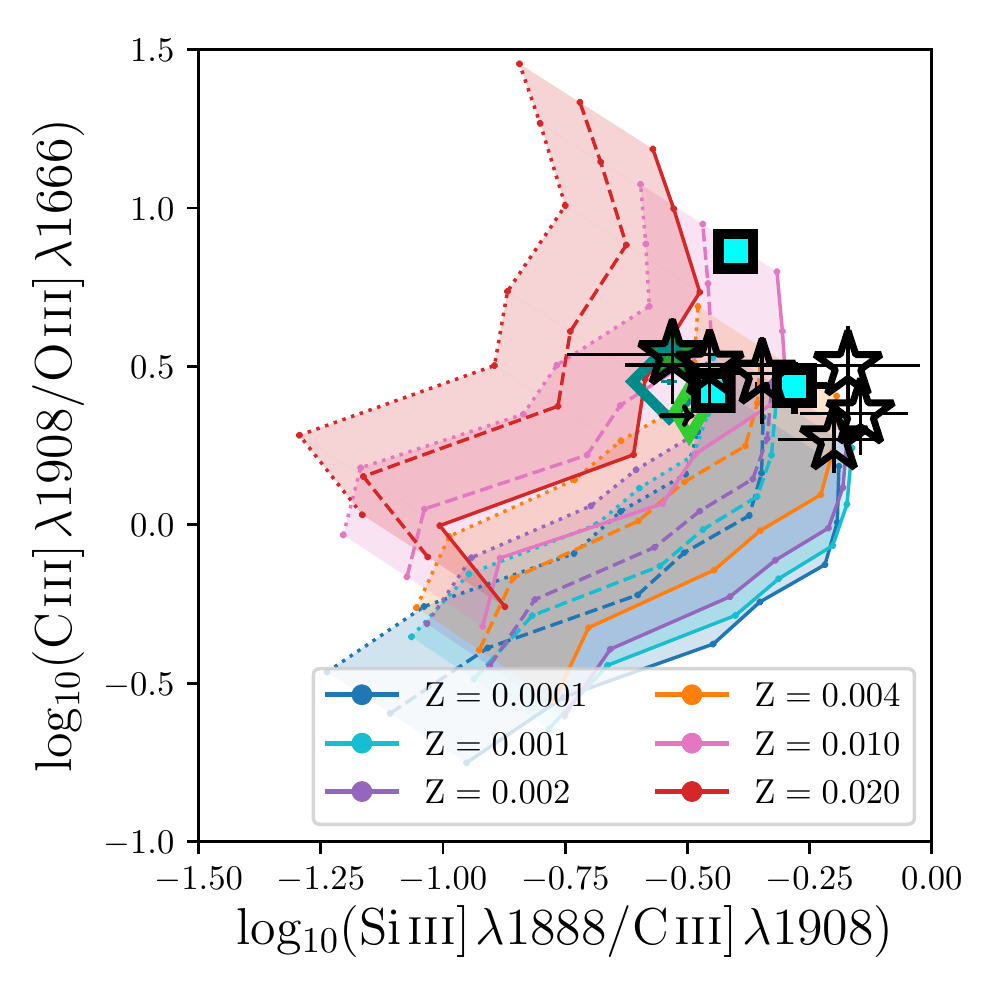}
\includegraphics[trim = 10 10 10 0, clip, scale=0.625]{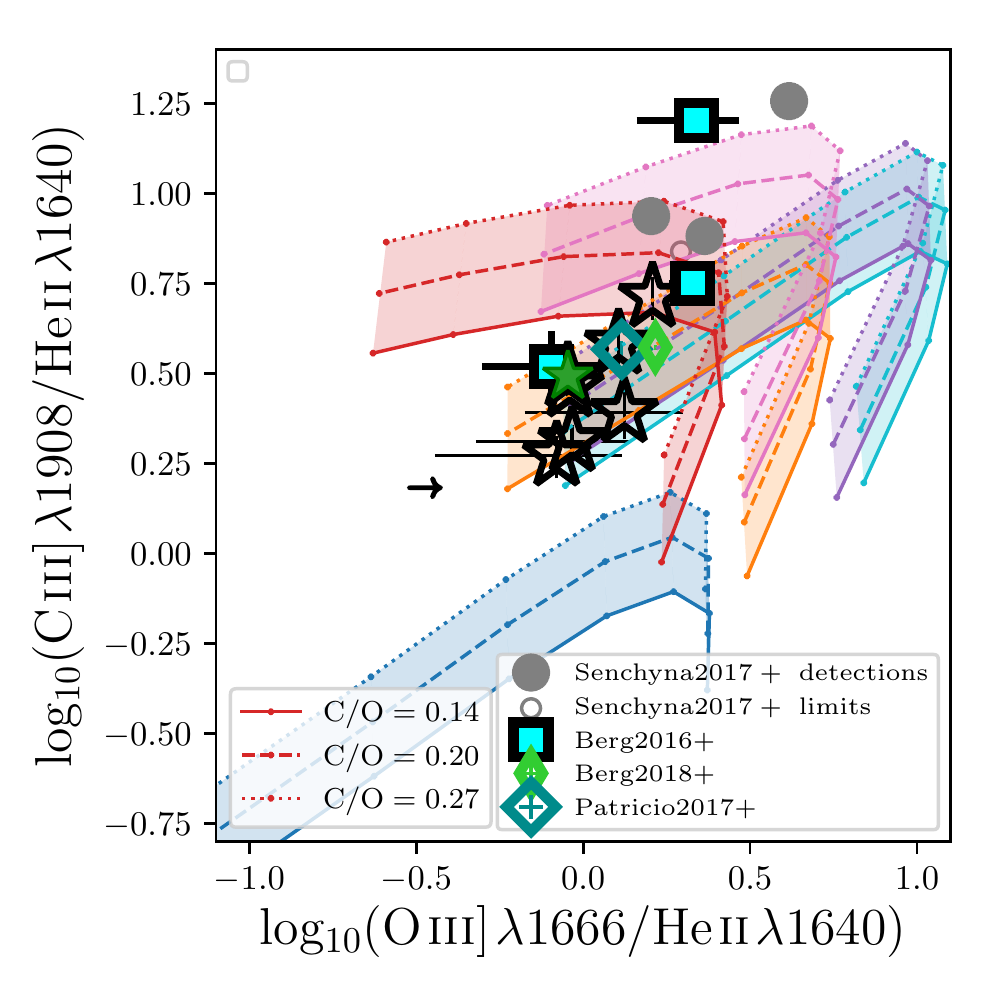}
\includegraphics[trim = 10 10 10 0, clip, scale=0.625]{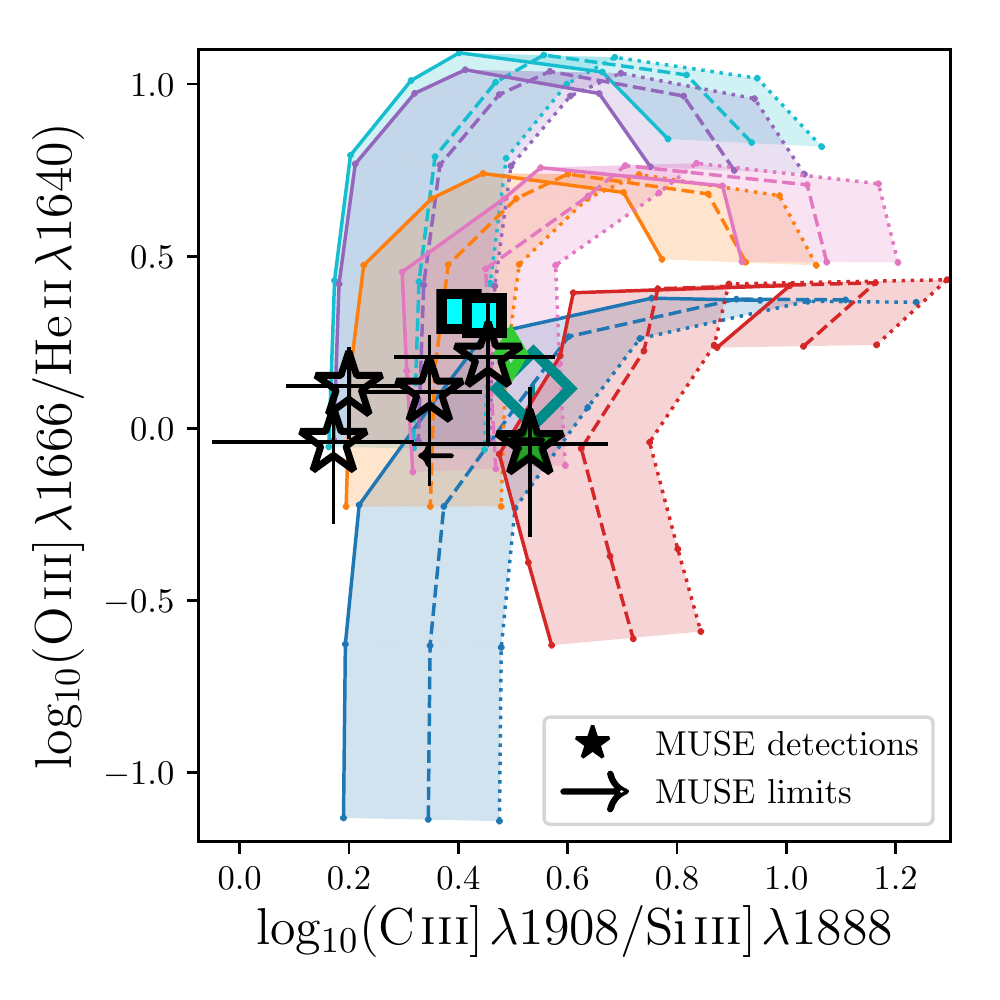}
\caption{Similar to  \ref{fig:line_ratios_gutkin} but the emission line ratio diagnostic plots have been zoomed out to demonstrate the dependence of emission line ratios with model parameters. [Appendix \ref{appendix:model_comp_models}]
\label{fig:similar_to_figline_ratios_gutkin_but_the_emission_line_ratio_diagnostic_plots_have_been_zoomed_out_to_demonstrate_the_dependence_of_emission_line_ratios_with_model_parameters_appendix_appendixmodel_comp_models_figline_ratios_gutkin_model}
\label{fig:line_ratios_gutkin_model}
}
\end{figure*}

In Figure \ref{fig:line_ratios_gutkin_model} we show the model distribution in  \CIII/\OIII\ vs \SiIII/\CIII, \CIII/\HeII\ vs \OIII/\HeII, and \OIII/\HeII\ vs \CIII/\SiIII\ emission line ratio diagnostics.  
The line ratios of the model tracks from \citet{Gutkin2016} computed with $n_{H}=100cm^{-3}$, $\xi_{d}=0.5$, $m_{up}=300M_\odot$ for $(C/O)/(C/O)_\odot$ ratios of 0.14, 0.20, and 0.27, $Z_{ISM}$ of 0.004, 0.008, 0.017, 0.02, 0.04 and  $log_{10}(U_{s})$ between $-1.0$ to $-4.0$ in increments of 0.5 are shown in this figure.   
Here we briefly discuss the influence of models parameters on the emission line ratios.

Increasing the C/O ratio results in the increase of C abundance compared to other heavy elements, thus we observe an increase in the \CIII/\OIII\ ratio and a decrease in the \SiIII/\CIII\ ratio. 
When all other parameters are fixed, the tracks move towards the upper left as a function of C/O ratio in the \CIII/\OIII\ vs \SiIII/\CIII\ line ratio diagram.
Similarly, in the \CIII/\HeII\ vs \OIII/\HeII\ line ratio diagram, the prominent rise in \CIII\ results in higher \CIII/\HeII\ moving the tracks upwards, while tracks move leftwards in  \OIII/\HeII\ vs \CIII/\SiIII\ line ratio diagram. 
However, due to C being a prominent coolant it has been shown that \CIII\ emission does not linearly correlate with the C/O ratio \citep[e.g.,][]{Jaskot2016}.

In the \CIII/\OIII\ vs \SiIII/\CIII\ line ratio diagram, increasing $n_{H}$ while keeping the other parameters fixed move the tracks towards the lower right, with the largest shift visible at higher metallicities and higher ionization parameters (which increase from top to bottom).
At higher $n_{H}$, excited atoms will favour collisional de-excitation rather than radiative de-excitation. Therefore, increasing the $n_{H}$ will result in stronger collisional emission lines. 
However, the infrared fine structure transitions have lower critical density compared to UV/optical transitions and thus at higher densities cooling through UV/optical line transitions become prominent.  
At higher metallicities infrared transitions dominate the cooling process, hence the shift in change is higher for rest-UV lines at higher metallicities. 
Similarly, the \CIII/\HeII\ vs \OIII/\HeII\ and  \OIII/\HeII\ vs \CIII/\SiIII\ line ratio diagrams, which are relatively insensitive to changes in $n_{H}$ at lower metallicities, shows some dependence at higher metallicities driven by the increase in strength of UV cooling lines that compensate for the lack of infrared transitions at higher $n_{H}$.

At higher $U_{s}$, \HII\ regions becomes more compact and closer to the ionization inner boundaries favouring high ionization lines. Therefore, the \OIII\ line gets stronger compared to \CIII\ and \CIII\ gets stronger by a lesser amount compared to \SiIII, resulting in a larger vertical shift and a slight horizontal shift to lower values with increasing $U_{s}$ when other parameters are fixed.
Given the increase in UV line transitions and increase in high ionization line strength, at higher $U_{s}$  the change of line ratios are more prominent as a function of $n_{H}$. 
With increasing  $U_{s}$, the models in  \CIII/\HeII\ vs \OIII/\HeII\ line ratio diagram show a curved shape where the tracks reach a maximum \CIII/\HeII\ and \OIII/\HeII\ value, after which all line fluxes reduce, with \CIII\ showing the largest drop due to the lower ionisation potential of C$^{++}$ (24.38 eV) compared to O$^{++}$ (35.12 eV). 
In \OIII/\HeII\ vs \CIII/\SiIII\ line ratio diagram, the higher metallicity models show a plateauing  towards higher $U_{s}$.

Increasing  $\xi_{d}$ with other parameters fixed moves the tracks towards the left as a result of a balance between depletion of metals and electron temperature. 
An increase of $\xi_{d}$ results in less amount of coolants in the gas phase, thus the electron temperature and cooling through collisionally excited lines increases. However, in \citet{Gutkin2016} models the metals have relative differences in depletion (see Table 1 of \citet{Gutkin2016}) and thus at higher metallicities the rise of electron temperature due to depletion of metal coolants is more prominent. 
In \CIII/\OIII\ vs \SiIII/\CIII\ line ratio diagram, given that Si has the highest amount of depletion compared to C and O, the \SiIII\ intensity drops more prominently with the increase of $\xi_{d}$ resulting in the observed shift of the tracks to the right. 
Similarly, in \CIII/\HeII\ vs \OIII/\HeII\ line ratio diagram \CIII\ has the highest depletion and the highest metallicity models show the lowest line ratios at high $U_{s}$. 
At low $\xi_{d}$, high metallicity models show a lower drop in line ratios at extreme $U_{s}$ driven by the rise in electron temperature due to depletion of high order metals. 
This is also seen in \OIII/\HeII\ vs \CIII/\SiIII\ line ratio diagram, where only higher metallicity models show evolution with $\xi_{d}$ in the \OIII/\HeII\ ratio. 
Models at all metallicities show strong evolution in the \CIII/\SiIII\ ratio, given $\sim\times2$ depletion factor of Si compared to C.

\section{Effects of dust in the rest-UV}
\label{appendix:dust_appendix}

Here we provide a detailed discussion on the role of dust in photo-ionisation modeling and interpreting observed spectra.  
To the first order, dust grains primarily absorb UV light and re-emit the absorbed energy in as IR emission \citep{Spitzer1978}. The shorter wavelength light have higher optical depth, thus the rest-UV regime probed in our study undergoes the highest amount of extinction leading to necessity in accurate calibration and correction of dust attenuation to probe the underlying properties of the ISM of our galaxies. 
However, our understanding of dust at high redshift is limited and variations in dust grain size distribution and variations in spatial geometry of stars and dust have shown to contribute significantly to the rest-UV dust attenuation properties \citep[e.g.,][]{Reddy2015,Reddy2016}.

Interstellar dust originate through natural condensation of stellar atmospheres and stellar winds and supernovae that release heavy elements produced by massive stars to the ISM \citep{Lagache2005}, and $\sim30-50\%$ of released metals condense into dust grains \citep{Draine2007}.  
Dust grain models of local star-forming regions have been calibrated from abundances of the heavy elements, absorption and scattering properties from rest-UV to rest-FIR, IR emission properties, and polarization of absorbed and emitted light from local star-forming regions.  Results have shown that local regions primarily comprise of a mixture of amorphous silicate and carbonaceous grains with a distribution in grain size \citep{Draine2003}.  
Given observational constraints to perform such a diverse analysis of emission line properties, our understanding of dust grain properties of high redshift galaxies is limited and studies are yet to break the degeneracies between larger covering fraction of dust and differences in dust grain size distributions \citep[e.g.,][]{Reddy2015}. \\

Uncertainties in dust grain models lead to three interdependent problems in our analysis.

\paragraph{Attenuation}
Firstly, large dust grains lead to high scattering cross sections, thus dust composition results in complexities for dust absorption and scattering of rest-UV light. 
However, the shattering, grain growth, and coagulation of dust grains is a strong function of the star-formation history and IMF \citep{Asano2013,Nozawa2015}, and thus requires strong constraints on cosmic galaxy evolution models to fully comprehend the physics of dust grains at high-redshift.
To first order, large dust grains which undergo efficient dissociation through strong UV radiation and supernova shocks and overabundance of silicate grains \citep{Zafar2015} results in smaller grain sizes, primarily in the presence of young O and B stars.
Smaller grains have lower scattering cross sections, thus steepens the attenuation law in the rest-UV. 
Therefore, in the context of rest-UV emission lines, uncertainty in the wavelength dependent attenuation contribute to differences in the dust corrected emission line ratios to the intrinsic emission line ratios. 
In Section \ref{sec:uncerternities_dust} we have discussed the implication of the dust law on the rest-UV emission line ratios.

\paragraph{Metal depletion}
Secondly, limited understanding of depletion of metals onto dust grains adds complexity to photo-ionization modeling.
Dust grains, which comprise of metals depleted from the gas phase, affect the UV radiation produced by the stars via absorption and scattering of incident radiation and influence the radiation pressure, photoelectric heating, and collisional cooling within the ISM \citep{Shields1995,Groves2004a}.  
Therefore, understanding the processes and rates that lead metals to deplete to dust is necessary to constrain heating and cooling mechanisms of the ISM.  
Different elements deplete into dust at varying rates \citep{Savage1996,Jenkins2009}, thus constraining relative abundances of refractory (elements that deplete rapidly to dust, e.g. Fe, Ni) to volatile elements (elements with lower depletion rates, e.g. Zn, P, S) is necessary to probe fractions of metals that are locked within dust, but is observationally challenging. 
Dustier systems have been shown to have a significantly higher fraction of metals depleted onto dust grains with a universal sequence observed at low and high-$z$ systems \citep[e.g.,][]{Brinchmann2013,Steidel2016,Wiseman2017,DeCia2018} 
At high redshift, studies using  gamma-ray burst damped Lyman-alpha absorbers have shown the dust-to-metal ratios to positively correlate with ISM metallicity, but with a significant low dust-metal ratios compared to the Milky-Way \citep{Wiseman2017}. 
However, at high-$z$ intrinsic biases in selecting damped Lyman-alpha absorbers to probe dust properties, and the confusion between circum-galactic dust absorption to ISM dust absorption introduce additional uncertainties.  
Due to such complications, photo-ionisation models make simplifying assumptions to compute heavy element abundances and their depletion factors, such as that the total mass of carbon depleted in dust produce the carbonaceous grains and most other heavy elements produce varieties of silicate grains \citep[e.g.,][]{Groves2004a}. 
However, the validity and consequences of such effects are yet to be explored.

\paragraph{Dust dissociation}
Energetic events that occur within galaxies, such as star formation provide high energy photons that have been shown to destroy dust grains \citep[e.g.,][]{Draine1979,Yamasawa2011}. 
When the collisional and photo-electrical charging rates balance in a dust grain, the full radiation pressure acting on the grain would rapidly accelerate the grain to its terminal velocity, thus increasing the probability of shattering through high velocity grain-grain collisions. 
However, the equilibrium of charging rate and shattering cross sections are a function of grain size and composition \citep{Groves2004a}, and the process leading to larger grains shattering to smaller ones until the sizes are small enough to dissociate through stochastic heating is yet to be incorporated into photo-ionisation models. 
The extent to which such processes would affect our analysis is unclear.\\

Quantifying the amount of metals depleted from the gas to the dust-phase and the dust dissociation is imperative to probe the influence of young stars on the ISM. 
The balance between these processes influence the gas phase metallicity of the ISM. Metal coolants are important in regulating the temperature structure of the ISM. 
With the increase of dust-to-metal ratio, the dust optical depth increases reducing the electron temperature through increased absorption of high energy photons \citep{Gutkin2016}. 
Thus, the change of gas-to-dust ratio of galaxies result in multiple effects thats influence the strength of the rest-UV emission line features, and accurate constraints of metal depletion and dust dissociation is important to explore the properties of the ISM through emission line ratios.

Only a handful of evolutionary models account for the role of dust in chemical evolution \citep[e.g.,][]{Gioannini2017}.
In the \citet{Gutkin2016} photo-ionization models considered in our analysis, the metal depletion uses the default {\tt CLOUDY} values with updates from \citet{Groves2004b}. 
In the three emission line ratios we explored in our analysis,  \CIII/\OIII\ vs \SiIII/\CIII, \CIII/\HeII\ vs \OIII/\HeII, and \OIII/\HeII\ vs \CIII/\SiIII, the limited S/N of our sample, and degeneracies between the variable parameters constraints our understanding of accurately interpreting rest-UV emission line diagnostics. 
Additionally, the dust attenuation law of high-$z$ galaxies also play a role in obtaining accurate intrinsic emission line ratios, specially for lines that are not close apart in wavelength space.

Several studies have explored the dust attenuation law of high redshift star-forming galaxies \citep[e.g.,][]{Wilkins2013,Reddy2015,Reddy2016,Reddy2016b,Reddy2018,Scoville2015,Bouwens2016b,Salmon2016,Faro2017,Cullen2018} with some results favouring star-burst like attenuation curves similar to \citet{Calzetti2000} and others suggesting the need for flexible attenuation laws to constrain the dust of these systems. 
Dust attenuation from dust-to-metal ratios have shown to introduce significant discrepancies compared to SED derived attenuation estimates \citep{Wiseman2017}, thus has been shown not to be a good proxy to estimate dust attenuation.

Intrinsic galaxy spectra for our analysis are obtained using extinction values derived from FAST (except for the quasar field where UV slope \AvBeta\ is used due to the lack of broadband photometry, however, we show in Figure \ref{fig:AvSED_AvBeta_comp} that \AvBeta\ shows good agreement with \AvSED)  assuming a \citet{Calzetti2000} dust attenuation law. 
If the physics of dust grains in the early Universe is quite different, variations in the total-to-selective dust extinction will introduce discrepancies from the $z\sim0$ calibrated attenuation laws. 
Additionally, variations with dust geometries have shown that nebular emission lines originating from the dense \HII\ regions in the vicinity of young O,B stars have higher optical depths \citep{Calzetti2000,Reddy2015,Theios2018}, with $z\sim0$ star-forming galaxies showing $\sim\times2$ more attenuation for nebular emission lines compared to the continuum \citep{Calzetti2000}. 
However, in the context of our work, we have ruled out effects of the considered dust law or dust sightlines (see Figure \ref{fig:line_ratios_dust}) to be of significant impact to the analysis of our observed emission line ratios. 

\end{appendix}

\end{document}